\documentclass[prb,  a4paper,twocolumn,longbibliography]{revtex4}
\usepackage{amssymb}
\usepackage{amsmath}
\usepackage{bbm}
\usepackage{amsfonts}
\usepackage{graphicx,epsfig,psfrag,xcolor,subcaption}
\usepackage{bm,comment}
\usepackage{color}
\usepackage[english]{babel}
\usepackage{hyperref}
\usepackage{cancel}

\usepackage{ragged2e} 

\usepackage{float}
\graphicspath{{figures/}}

\def\ve{\varepsilon}

\def\vf{\varphi}
\def\de{\partial}

\def\mE{\mathcal{E}}
\def\arg{{\rm arg}}

\newcommand{\ignore}[1]{} 
\newcommand{\figref}[1]{Fig.~\ref{#1}}

\begin{document}

\captionsetup{
    justification=justified,
    singlelinecheck=false
}

\title{Quantum oscillations of helical edge states of periodically deformed 2D topological insulator in magnetic field}

 \author{A.~V.~Tsvetkova}
  \affiliation{National University of Science and Technology MISIS, Moscow 119049, Russia}
  
  \author{P.~D.~Grigoriev}
  \affiliation{L.D. Landau Institute for Theoretical Physics, RAS, Chernogolovka 142432, Russia}

\author{Ya.~I.~Rodionov}
  \affiliation{Dukhov Research Institute of Automatics (VNIIA), Moscow 127055, Russia}
  \affiliation{National Research University Higher School of Economics, Moscow 101000, Russia}
\begin{abstract}

We study edge-state transport in a two-dimensional topological insulator with a periodically deformed edge subjected to a uniform magnetic field. Zeeman coupling breaks time-reversal symmetry and enables elastic backscattering, producing oscillations of the forbidden-band widths. In the strong-field regime, the gaps can close completely at discrete field values. In the weak-field regime, we identify an important class of periodic deformations for which the dominant semiclassical scattering is controlled by complex infinity rather than by the nearest turning points. We develop a semiclassical treatment of this process and establish its agreement with perturbation theory and direct numerical calculations. The gap modulation should produce observable oscillations of the edge conductance. Unlike conventional magnetic quantum oscillations, which are periodic in inverse field, the predicted oscillations are periodic in the magnetic field itself, with a period determined by the Fermi velocity and effective \(g\)-factor.

\end{abstract}

\maketitle

\section{Introduction}

Two-dimensional topological insulators (TI) are characterized by a full insulating gap in the bulk and gapless edge states which are topologically protected. The paradigmatic example of topological protection is the quantum Hall effect, where the Hall conductance is determined by a Chern number and its precise quantization underlies the electrical resistance standard~\cite{Yennie1987,Hatsugai1997}. Other systems with topological protection were proposed, and after the experimental observation of 2D TI in a HgTe/CdTe quantum well (QW) \cite{konig2007quantum} in the context of quantum spin Hall (QSH) effect, the topology became an important area of condensed matter physics \cite{TopInsRevModPhys2010,TopInsRevModPhys2011}. The important distinction of these TIs is their pronounced spin-orbit interaction (SOI)\cite{hinz2006, yang2014}, resulting in band inversion. Below we consider only such SOI-induced, or band-inverted, 2D TIs.

Topological boundary states are either 1D edge states at the boundaries of 2D TIs (e.g., HgTe quantum wells) or 2D surface states at the boundaries of 3D TIs, e.g., \(\mathrm{Bi}_2\mathrm{Se}_3\)~\cite{zhang2009}. In most cases 2D and 3D TI samples are made from different compounds, the only exception being HgTe \cite{kvon2020}. Other realizations of 1D topologically protected states include those on the edges between surfaces of 3D TI \cite{deb2014} and states that occur on step edges \cite{herath2013,fedotov2017}. Scattering of 1D edge states in a 2D TI can be either forward or backward. The former is only a phase shift and does not contribute to the transport scattering rate. Hence, only backscattering contributes to the transport relaxation of the edge states.

The edge states in such 2D TIs, such as HgTe/CdTe QWs, exhibit spin-momentum locking, i.e. a backscattering event necessarily entails a spin flip. Therefore, in the absence of magnetic impurities or a magnetic field, both of which break time-reversal (TR) symmetry, elastic backscattering of the edge states is forbidden. This suppression of elastic backscattering is a hallmark of the topological protection provided by TR symmetry~\cite{Kane2005}. Following the experimental observation of transport by edge states in HgTe quantum wells~\cite{konig2007} and by surface states in \(\mathrm{Bi}_2\mathrm{Se}_3\) crystals~\cite{hsieh2008}, the properties of TIs attracted considerable interest from the scientific community.

The measured mobility of edge states, although being remarkably high, turned out to be smaller than expected according to the topological protection \cite{konig2007,daumer2003,nowack2013,gusev2014,roth2009,kvon2020}. A multitude of theoretical explanations \cite{hsu2021,yevtushenko2022} for the scattering mechanisms of helical edge states and topological surface states have been developed in the last decade. They include the scattering from random fluctuations of the band gap \cite{tkachov2011}, the elastic scattering in charge puddles close to the edges \cite{vayrynen2014,essert2015,kurilovich2019prb,kurilovich2019prl}, the scattering from edge irregularities described by a certain coupling function \cite{magarill2015}, the scattering of counter-propagating states \cite{li2017}, etc. The interaction-induced backscattering mechanism for helical edge states of a 2D TI due to tunnel coupling to puddles located near the edge channel was also considered \cite{Krainov2025}. The influence of specific defects on topological-state transport has also been investigated theoretically~\cite{zhangting2012,herath2013} and directly probed experimentally at individual defects~\cite{lupke2017}. In spite of extensive theoretical work, the origin of the unexpectedly low mobility of edge states in TIs with strong SOI and band inversion remains unresolved.

The topological protection of these edge states is broken by any perturbation violating the time-reversal symmetry, e.g., by a magnetic field.
Edge states in such systems in a transverse magnetic field were studied in Ref.~\cite{Durnev}, although the influence of SOI and edge deformations was neglected. The scattering by single edge deformation of a 2D TI in a magnetic field was recently studied~\cite{dotdaev2024}. The resulting reflection probability exhibits an interesting and even nonmonotonic dependence on magnetic field.

In this paper, building on the model Hamiltonian developed in Ref.~\cite{dotdaev2024} for an isolated edge bend in the presence of strong SOI, we extend the analysis to a periodic sequence of edge deformations and study the resulting scattering and band formation of the edge states in a 2D TI. The edge bending is described by the deformation angle profile (see Fig. \ref{fig:bend}). The elastic scattering and the band formation become possible in the presence of a uniform magnetic field orthogonal to the plane of a 2D TI. The external magnetic field breaks TR symmetry, allowing backscattering and opening forbidden gaps in the edge-state band structure.
Typically, the edge deformation is smooth on the scale of the carriers' de Broglie wavelength, so that a semiclassical treatment is applicable. We therefore perform a full semiclassical analysis using the powerful Pokrovsky--Khalatnikov--Dykhne (PKhD) method and obtain analytical expressions for the energy-band widths with pre-exponential accuracy.
We analyze two complementary regimes: the strong-field limit, \(\mu\sim\varepsilon\), and the weak-field limit, \(\mu\ll\varepsilon\), where \(\mu\) is the Zeeman energy and \(\varepsilon\) is the edge-state energy. In the latter regime, the semiclassical analysis is supplemented by perturbation theory in the subregime \(\mu L/(\hbar v_F)\ll1\), where \(L\) is the deformation period and \(v_F\) is the Fermi velocity.

In the strong field limit, we show that the forbidden-band widths $\Delta$ oscillate as a function of magnetic field, resulting in a novel type of magnetic  quantum oscillations (MQO) inherent to topological insulators:
\begin{gather}
    \Delta \sim \cos{[\mu L]}.
\end{gather}
These oscillations of the band widths, and particularly of the gap at the Fermi level, should manifest themselves in thermodynamic and transport observables.

Similarly to ordinary MQO in normal metals \cite{Shoenberg2009Sep}, the proposed effect 
allows for measuring the microscopic properties of charge carriers. In contrast to conventional bulk MQO, which requires 2D or 3D carrier motion, the proposed MQO arises from 1D edge states. They are also distinct from previously discussed MQO involving TI edge states~\cite{WangMQO2010,Alisultanov2023}. 
Although the frequency of the proposed MQO contains the period $L$ of the edge bending profile, these MQO completely differ from Sondheimer oscillations 
the commensurability between the sample thickness and the average distance the electrons on extremal orbits travel during the cyclotron period. At present, ARPES provides one of the principal direct probes of the microscopic structure of edge states, but its accuracy and availability are limited.   

In the weak-field limit, we identify an important class of periodic deformation profiles exhibiting an unusual phenomenon: the leading backscattering amplitude, and hence the band-gap formation, is controlled by a Stokes transition at complex infinity, \(z\to i\infty\), rather than by finite complex turning points.
To treat this situation semiclassically, we develop an extension of the PKhD method. Finally, we compare all our analytical results with direct numerical integration of the corresponding Dirac equation and find remarkably good agreement. 

The paper is organized as follows. Section II introduces the model and the methods used throughout the paper. In Sec. III, we develop the band theory in the strong-magnetic-field regime. Section IV is devoted to the weak-field limit for a general class of periodic deformation profiles. In Sec. V, we consider an important subclass of profiles that appears to lead to an internal inconsistency of the general weak-field theory and show how this apparent paradox is resolved. Section VI develops the corresponding nonperturbative semiclassical description of scattering controlled by complex infinity. Finally, in Sec. VII we discuss the applicability of our results.

\section{Model and methods}
\subsection{Model and Hamiltonian}
\begin{figure}[t!]
	\centering
\includegraphics[width=1\columnwidth]{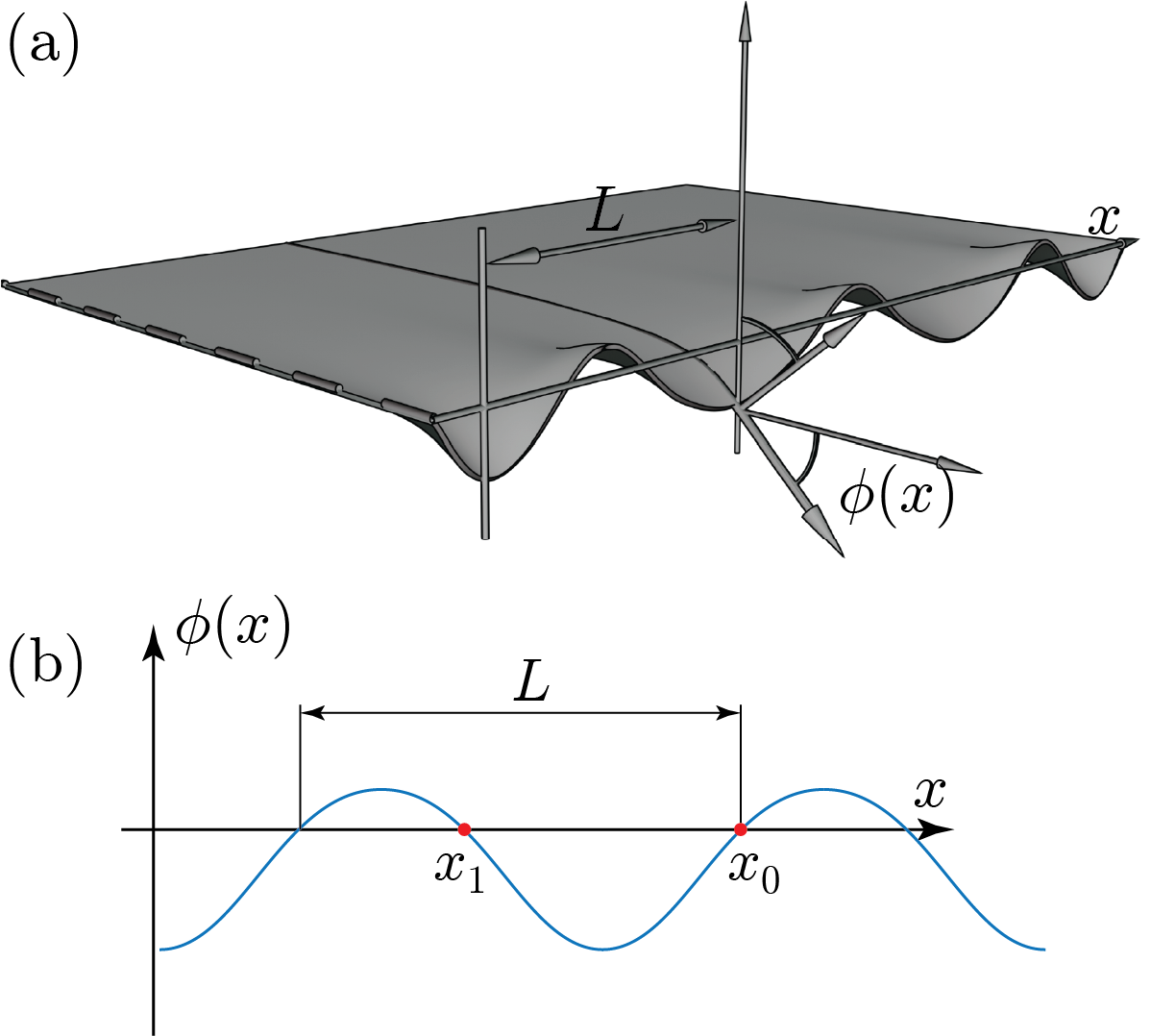}
	\caption{\justifying (a) A schematic illustration of a periodic sign-changing geometric undulation of  the edge of a 2D topological insulator sample.  (b) The red dots (zeroes of the deformation correspond to two successive LZ-transitions)}
\label{fig:bend}
\end{figure}
We employ the effective edge-state model introduced in Ref.~\onlinecite{dotdaev2024}, where scattering from a single smooth out-of-plane deformation was considered. Here we extend the same model to a periodic deformation profile, $\phi(x+L)=\phi(x)$. For completeness, we briefly summarize the resulting one-dimensional Hamiltonian.

We consider the effective 1D Hamiltonian of the helical edge states in the form
\begin{gather}
\label{eq:h1}
\hat H=H_0(\mathbf p)+\alpha(\boldsymbol{\sigma}\times\mathbf p)\cdot\boldsymbol{\nu},
\end{gather}
where \(h(\mathbf p)\) describes the unperturbed propagating edge states, while the second term is the standard Bychkov–Rashba spin-orbit interaction~\cite{bychkov1984}. Here, \(\boldsymbol{\nu}\) is the local unit vector normal to the two-dimensional topological-insulator surface, $\boldsymbol{\sigma} = (\sigma_x,\ \sigma_y,\ \sigma_z)$ are Pauli matrices  and \(\alpha\) is the Rashba coupling constant. The dependence of this term on \(\boldsymbol{\nu}\) is essential for the present problem, since an out-of-plane deformation makes the local normal coordinate dependent.

For a straight edge directed along the \(x\) axis, the corresponding low-energy Hamiltonian of a single helical pair can be written as
\begin{gather}
\label{ham0}
\hat{H}_0=v_F\hat{p}_x\hat\sigma_y.
\end{gather}
where \(v_F\) is the spin-orbit-renormalized Fermi velocity.

An out-of-plane deformation rotates the local normal \(\boldsymbol{\nu}\) and therefore modifies the spin-orbit term. We parameterize the local inclination of the edge by \(\vf(x)\), as illustrated in \figref{fig:bend}. For the geometry considered here,

\begin{gather} 
\alpha(\boldsymbol{\sigma}\times\mathbf p)\cdot\boldsymbol{\nu} =-\alpha p_x\sigma_y+\alpha p_x\sigma_z\sin\vf(x). 
\end{gather}
The first term merely renormalizes \(v_F\), whereas the deformation-dependent contribution must be symmetrized to preserve hermiticity. Substituting \(\alpha\sin\vf = \phi\) we obtain the potential field due to Rashba interaction in the following form
\begin{gather}
\label{interaction}
\hat U(x)=\frac12[\hat p_x\phi(x)+\phi(x)\hat p_x]\sigma_z.
\end{gather}
In the present work, in contrast to Ref. \onlinecite{dotdaev2024}, the profile is periodic, \(\phi(x+L)=\phi(x)\).

At zero magnetic field the deformation preserves time-reversal symmetry and cannot produce elastic backscattering within a single helical Kramers pair. To break time-reversal symmetry, we apply a magnetic field perpendicular to the undeformed TI plane, \(\mathbf H=H\hat{\mathbf z}\). Choosing the gauge \(\mathbf A=(Hy,0,0)\), the orbital contribution can be removed from the effective one-dimensional edge problem by a gauge transformation, while the Zeeman coupling remains. The resulting Hamiltonian is
\begin{gather}
\label{ham1}
\hat H^{1D}(x)=v_F\hat p_x\sigma_y+\mu\sigma_z+\hat U(x), \qquad \mu=\frac{g\mu_B}{2}H,
\end{gather}
where  $\mu_B$ is Bohr magneton and $g$ is a Lande factor for edge electrons \cite{kernreiter2016}.
Therefore, we need to solve the scattering problem for the following Dirac equation:
\begin{gather}
\label{dirac:main}
\left[v_F\hat{p}_x\sigma_y+\mu\sigma_z+\hat U(x)\right]\psi(x) = \ve\psi(x)
\end{gather}
In the absence of deformation potential $\hat{U}(x)$ the Zeeman term $\mu\sigma_z$ opens a gap in the spectrum of edge state of width $2\mu$: $\ve_\pm(p)=\pm\sqrt{v_F^2p^2+\mu^2}$. We consider propagating states, for which
\begin{gather}
    \label{zeeman}
    |\ve|>\mu.
\end{gather}

\subsection{PKhD method}
\subsubsection{Semiclassical approximation}
To apply the semiclassical framework to Eq.~\eqref{dirac:main}, one must first identify a small parameter. Physically, the semiclassical approximation is applicable when the spatial variation of the effective potential is sufficiently smooth. In the present problem, the deformation profile \(\phi(x)\) varies on the characteristic scale of its period \(L\). Accordingly, the reduced de Broglie wavelength must be much smaller than \(L\),
\begin{gather}
\label{semi_cond}
\frac{\lambda}{L} = \frac{\hbar v_F}{\ve L} \ll 1.
\end{gather}
As we shall see, condition~\eqref{zeeman}  ensures that the semiclassical momentum derived below remains nonzero everywhere on the real axis. The scattering therefore belongs to the over-barrier regime, for which the Pokrovsky–Khalatnikov–Dykhne method ~\cite{dykhne1961,pokrovskii1958} is applicable. Henceforth we set \(\hbar=v_F=1\), restoring them when needed for physical interpretation.

\subsubsection{Semiclassical treatment of periodic band structure}
Applying the Pokrovsky–Khalatnikov method~\cite{pokrovskii1958}, Dykhne~\cite{dykhne1961} obtained the band structure of a particle in a periodic potential \(U(x)\) in the over-barrier regime, \(E>U_0\), where $U_0=\max_x U(x).$ We refer the reader to Appendix~\ref{Dykhne} for details. Here, we summarize the central equations.

For a one-dimensional \(L\)-periodic potential, the centers \(E_n\) of the forbidden bands are determined semiclassically by
\begin{gather}
\label{action0}
S_L(E_n)=\int_0^L\pi(x)\,dx=\pi n\hbar, 
\end{gather}
where \(\pi(x)\) is the semiclassical momentum and \(S_L\) is the action accumulated over one period.
The width of the forbidden band (band gap) is given by equation
\begin{gather}
\label{dykhne2}
    \Delta = \frac{2 \sqrt{R}}{T},\quad
    T = \int\limits_{0}^L\frac{d\pi(x)}{d\ve}dx,
\end{gather}
where $T$ is the semiclassical oscillation time (the time of passage of a classical particle with the energy $E$ through one period of the potential) and $R$ is the over barrier reflection coefficient of the periodic potential. The computation of the semiclassical reflection coefficient is in turn obtained by PKh method.
Let us briefly overview its main steps:
\begin{enumerate}
\item Analytically continue the semiclassical solution into the complex plane along an anti-Stokes line, defined by
$\operatorname{Im}\int_{z_0}^{z}\pi(z')\,dz'=0,$
where \(z_0\) is a point at which the semiclassical approximation breaks down.
\item Construct the exact local solution in the vicinity of \(z_0\), where the deformation profile \(\phi(z)\) can be expanded, and determine its asymptotic behavior along the anti-Stokes lines corresponding to the incident and reflected waves.
\item Match the local asymptotics to the semiclassical solutions in their common domain of validity.
\item Analytically continue the matched solution back to the anti-Stokes line containing the real axis.
\item Use Dykhne's construction~\cite{dykhne1961} to determine the centers \(E_n\) and widths \(\Delta_n\) of the forbidden bands.
\end{enumerate}

\subsubsection{General analysis of the Dirac equation}
\paragraph{Single component equation.}

Equation~\eqref{dirac:main} is a pair of coupled first-order differential equations for the two-component spinor \(\psi(x)=(\psi_1(x),\psi_2(x))^T\). Eliminating one of the components yields the following second-order equation for \(\psi_1(x)\)\cite{dotdaev2024}:
\begin{gather}
\label{Dirac2}
\begin{split}
    &2(1+\phi^2)\alpha \psi''_1 + 2i\left[(1+\phi^2)\phi '' + \phi \alpha(2\mu - 3i\phi')\right]\psi'_1\\
    &+\left[-\frac{1}{2}\alpha\beta(\alpha-2 i\phi') + 4\ve \phi \phi '' \right]\psi_1 = 0,
    \\
    &\psi_2 = \frac{2(1+\phi^2)\,\psi_1' + i\psi_1 \phi \beta}{\alpha}.
\end{split}
\end{gather}
where $\alpha(x) = 2(\mu + \ve) - i \phi'$, $\beta(x) = 2(\mu - \ve) - i\phi'$.
Eq.~\eqref{Dirac2} will be referred to as Dirac's equation as well. 
\paragraph{Semiclassical solution.}
The basic semiclassical construction was developed for a single edge deformation in Ref. \onlinecite{dotdaev2024}. For completeness, we summarize below the elements required for the periodic problem and for the extensions developed in the subsequent sections.

The cumbersomeness of the single-component Dirac equation (11) is almost grotesque. Remarkably, however, as was shown in Ref.~\onlinecite{dotdaev2024}, it admits an elegant semiclassical solution. Discarding entirely second derivatives of the deformation potential $\phi(x)$ and expanding the equation in $1/(\ve L)$ we obtain the following formulae for semiclassical wave functions:
\begin{gather}
\begin{split}
    &\psi_{\pm}(x) = \sqrt{\pm q_{\pm} \left[ 1 \pm \frac{\phi \ve}{p}  \right]} \exp \left(i\int^x q_{\pm}(x')dx'\right), \\
    &q_{\pm} = \frac{-\mu \phi \pm p}{\phi^2 + 1}, \; p = \sqrt{\ve^2(\phi^2 + 1) - \mu^2}.
\label{semi0}
\end{split}
\end{gather}
Hereafter we suppress the component index and write \(\psi_1\equiv\psi\). 
A notable property of the semiclassical solutions~\eqref{semi0} is that they become exact solutions of the original equation at zero magnetic field, \(\mu=0\). As we show below, this property leads to an unusual asymptotic regime in which the resulting analytical expressions exhibit remarkably high numerical accuracy.

\paragraph{Schr\"{o}dinger representation.}
A useful representation of Eq.~\eqref{Dirac2} follows from the standard transformation that eliminates the first-derivative term and reduces a general second-order equation to Schr\"{o}dinger form:
\begin{gather}
\begin{split}
&\psi''(x) + \eta(x)\psi'(x) + \kappa(x)\psi(x) = 0
\;\Rightarrow\\
&\theta''(x) + \pi^2(x)\theta(x) = 0
\quad (\text{Schrödinger equation})\\
&\theta(x) = e^{\frac{1}{2}\int^x \eta(t)\,dt}\,\psi(x),\\
&\pi^2(x) = \kappa(x) - \frac{1}{2}\eta'(x) - \frac{1}{4}\eta^2(x).
\end{split}
\end{gather}
At first sight, these expressions are not particularly illuminating. However, as we are going to see below, \(\eta(x)\) plays the role of a Berry connection: the seemingly obscure integral \(\int \eta(x)\,dx\) becomes the familiar Berry phase. At the same time, the quantity \(\pi(x)\) appearing in the resulting Schr\"{o}dinger equation is precisely the semiclassical momentum.
To make this interpretation more transparent, we evaluate \(\pi(x)\) and \(\eta(x)\) semiclassically, treating derivatives of \(\phi(x)\) as small parameters. This yields
\begin{gather}
    \begin{split}
        \eta(x) = \frac{2i\mu \varphi(x)}{\varphi^2(x) + 1},
\qquad
\pi^2(x) = \frac{\varepsilon^2(\varphi^2 + 1) - \mu^2}{(\varphi^2 + 1)^2},\\
\theta_{\pm}(z) = \frac{1}{\sqrt{\pi(z)}}
\exp\!\left( \pm i \int^{z} \pi(t)\,dt \right).
    \end{split}
\label{semi2}
\end{gather}
which immediately reproduces the semiclassical solutions~\eqref{semi0}.

\paragraph{Types of the deformation potential.}
\label{types}
In the semiclassical analysis of scattering, the deformation profile is analytically continued to the complex plane. It is  enough to mention two broad classes of periodic profiles \(\phi(z)\): (i) profiles with singularities at finite complex \(z\), a canonical example being meromorphic functions constructed from trigonometric functions; and (ii) entire periodic profiles, such as trigonometric polynomials. As we show below, the latter class leads to the most unusual consequences for the band structure of the edge states.

\paragraph{Singular and turning points.}
Within the semiclassical paradigm, the relevant complex points are those at which the semiclassical approximation breaks down. In their vicinity, an exact local solution can be constructed and used to extract the reflected wave. These points were analyzed in detail in Ref.~\onlinecite{dotdaev2024}. Here we only summarize the results, using the momentum \(\pi(z)\) defined in Eq.~\ref{semi2}:
\begin{itemize}
  \item Poles $z_p$ of $\phi(z)$  are  zeros of $\pi(z)$.
  \item Branch points $z_\pm$ of $\pi(z)$ satisfy $\phi(z_{\pm})= \pm i\sqrt{1-\mu^2/\ve^2}$.
  \item Poles $z_1$ of $\pi(z)$ satisfy $\phi^2(z_1) = -1,\ \ \mu\neq 0$.     
\end{itemize}
Below we address the two general types of deformation potential outlined in Sec.~\ref{types}. 

\subsubsection{General solution for an arbitrary magnetic field}
Before proceeding, it is useful to recall the general semiclassical result for an arbitrary magnetic field.
As follows from the Dykhne formula~\eqref{dykhne2}, the gap widths $\Delta$ are determined by the semiclassical reflection coefficient $R$. 
The semiclassical reflection of an edge carrier from a smooth deformation was analyzed for arbitrary magnetic-field strength in Ref.~\onlinecite{dotdaev2024}. For a general ratio \(\varepsilon/\mu\), the reflection coefficient follows from the standard semiclassical expressions ~\eqref{semi0} and \eqref{semi2}, yielding the following compact result for the band gap:
\begin{gather}
\label{band_general}
    \Delta\equiv\frac{2\sqrt{R}}{T}=\frac{2}{T} e^{-2{\rm Im} S},\quad S = \int_{x_0}^{z_0}\pi(z)\,dz
\end{gather}
where momentum $\pi(z)$ is given by Eq.~\eqref{semi2}, and $z_0$ is its turning point in the upper complex plane. For the magnetic field of an arbitrary strength, there is only single turning point $z_0$ per deformation period. 
Equation~\eqref{band_general} therefore predicts a nonvanishing and, as it turns out, monotonic reflection probability as a function of magnetic field. Consequently, the forbidden-band widths are also predicted to vary monotonically with field. Apart from being rather unexciting, this result is problematic from a physical point of view.

Before confronting the reader with the full glory of complex analysis, let us first expose the limitation of Eq.~\eqref{band_general}.
As discussed above, TR symmetry strictly forbids elastic backscattering in the absence of a magnetic field, so that the reflection probability must vanish. Consequently, the forbidden bands must close as \(H\to0\).
Equation~\eqref{band_general}, however, fails to capture this constraint and predicts a finite reflection probability even in the zero-field limit, \(\mu\to0\). The origin of this failure becomes apparent from the semiclassical momentum ~\eqref{semi2}.
For \(\mu/\varepsilon\ll1\), the turning point—the branch point of the semiclassical momentum \(\pi(z)\)—approaches a pole of \(\pi(z)\), located at \(\phi^2(z)+1=0\). The scattering is therefore governed by the coherent interplay between the nearby turning point and pole. The pole qualitatively modifies the semiclassical connection problem and restores the vanishing of the reflection amplitude as \(\mu\to0\).

A related, but analytically distinct, situation arises in the strong-field limit, $ \varepsilon-\mu\ll\varepsilon$. In this regime the two turning points of \(\pi(z)\), $ \phi=\pm i\sqrt{2(\varepsilon-\mu)/\mu}, $
nearly coalesce in the vicinity of a zero of the deformation profile. This leads to two qualitatively different cases.

\begin{enumerate}
    \item \textbf{Sign-changing deformation}\\ 
    The periodic deformation profile has a real zero \(x_0\). Since the profile is periodic and changes sign, continuity requires another zero \(x_1\) within the same period (see Fig.\ref{fig:bend}). The two zeros generate two successive scattering events, whose amplitudes interfere coherently. As a result, the reflection probability oscillates with the semiclassical phase accumulated between the two turning-point regions. These oscillations are directly inherited by the forbidden-band widths. 
    \begin{gather}
        \Delta \propto |\cos S_{01}|,
    \end{gather}
    with $ S_{01} = \int_{x_0}^{x_1}\pi(x)\,dx$.
    \item \textbf{Sign-definite deformation}\\ 
    
    The periodic deformation profile does not vanish on the real axis. Geometrically, this corresponds to an edge that remains bent predominantly in one direction, with the amplitude of the periodic modulation smaller than the overall bending (see Fig.~\ref{fig:bend2}). The zeros of the analytically continued function \(\phi(z)\) move off the real axis into the complex plane. The relevant turning points therefore approach a complex zero rather than a real one. The question is then whether the same interference mechanism survives and how it manifests itself in the band spectrum.
\end{enumerate}
In this case, the over-barrier scattering occurs in the vicinity of a pair of nearly coalescing turning points associated with a complex zero of the deformation profile. Both types of deformation lead to magnetic-field oscillations of the band gaps. We begin our analysis with a regular deformation profile in the strong-field regime.

\section{Strong magnetic field}
We begin with the strong-field regime, in which the Zeeman energy approaches the carrier energy, \(\mu\simeq\varepsilon\). The corresponding group velocity becomes small, and the semiclassical condition~\eqref{semi_cond} reduces to
\begin{gather}
\label{semi_cond2}
\mu L\gg 1, 
\end{gather}
where \(L\) is the deformation period. 
The strong-field regime gives rise to quantum oscillations. As we show below, the underlying mechanism is closely analogous to Landau–Zener–Stückelberg (LZS) interferometry, except that the interfering transitions now occur in the complex plane and are shaped by the peculiar analytic structure of the TI edge-state problem.

\subsection{Sign-changing edge deformation as an LZS interferometer}
\begin{figure}[t!]
	\centering
\includegraphics[width=1\columnwidth]{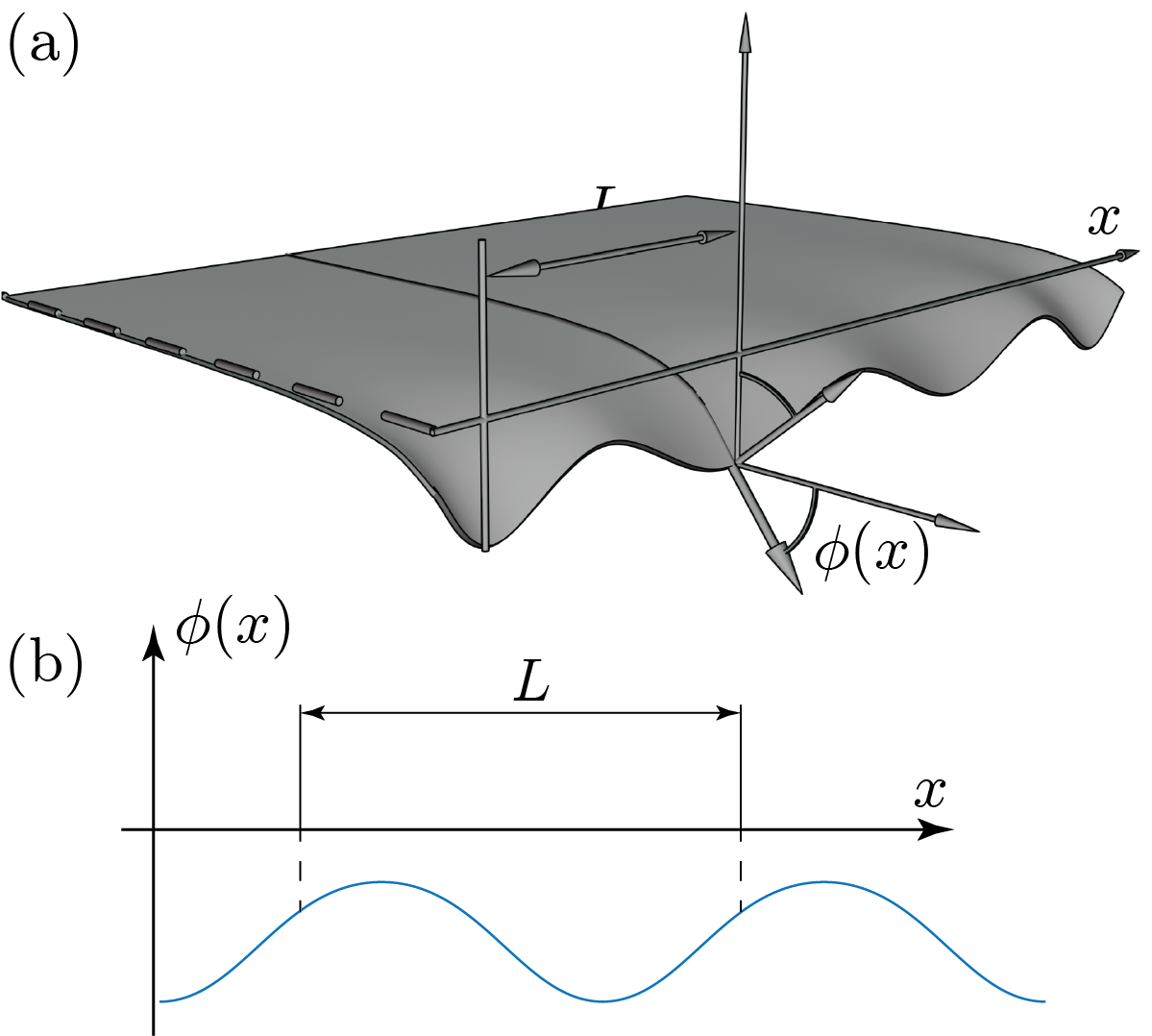}
	\caption{(a) A schematic illustration of a periodic sign-definite geometric undulation of  the edge of a 2D topological insulator sample. 
    (b) The corresponding deformation profile \(\phi(x)\).}
\label{fig:bend2}
\end{figure}
We first consider a sign-changing periodic deformation of the TI edge (see Fig.~\ref{fig:bend}(a)). We start the analysis by locating the branch points 
of the semiclassical momentum.
Introducing the energy offset
\begin{gather}
\begin{split}
    \ve = \mu + \delta \ve, \quad\frac{\delta \ve}{\mu} \ll 1
\label{eq:energy-offset}
\end{split}
\end{gather}
we obtain  from the semiclassical momentum expression~\eqref{semi2} the branch point equation: 
\begin{gather}
\label{roots}
\phi^2 + \frac{2\delta\ve}{\ve} = 0\ \Rightarrow \phi = \pm i\sqrt{\frac{2\delta\ve}{\ve}}
\end{gather}
Equation ~\ref{roots} shows that in the strong-field limit, \(\delta\varepsilon/\varepsilon\ll1\), the two branch points approach a zero \(x_0\) of the deformation profile. For a sign-changing profile [Fig.~\ref{fig:bend}], such zeros lie on the real axis and, by periodicity and continuity, occur in pairs within each period. Due to proximity of branch points to $x_0$, it is more natural to expand the deformation profile near \(x_0\) than the momentum near either branch point
\begin{gather}
    \phi(z) = \frac{z - x_0}{a} + ... ,
\label{sub1}
\end{gather}
where \(a\) is of the order of the deformation period \(L\), and its sign may be chosen positive without loss of generality.

\paragraph{The exact equation near the coalescing turning points.}
Substituting Eq.~\eqref{sub1} into Eq.~\eqref{Dirac2} and retaining the leading terms in \(1/(\mu a)\) [(cf. Eq.~\eqref{semi_cond2}], we obtain
\begin{gather}
\begin{split}
    &\psi''_1 + 2i\mu \frac{\zeta}{a} \psi'_1 + \mu\left[ 2\delta \ve + \frac{i}{a}  \right] \psi_1 = 0 \\
    &\zeta = z - x_0,
\label{exact_eq1}
\end{split}
\end{gather}
here $\zeta$ is the coordinate offset from $z_0$.
The corresponding exact solution can be represented as a Laplace integral
\begin{gather}\label{sol1}
  \psi(\zeta) = A\int\limits_{C}e^{\zeta t -i\frac{at^2}{4\mu}} t^{-ia\delta\ve-\frac{1}{2}}\,dt,
\end{gather}
where \(C\) runs along the negative real semiaxis (see Appendix~\ref{app:indef} for details). Equation~\ref{exact_eq1} is immediately recognized as the Hermite equation characteristic of the Landau–Zener (LZ) problem. This is not accidental. To make the connection with the LZ transition explicit, we now examine the corresponding semiclassical structure.

\subsubsection{The semiclassical analysis near the turning point}
\begin{figure}[t]
    \includegraphics[width=1\linewidth]{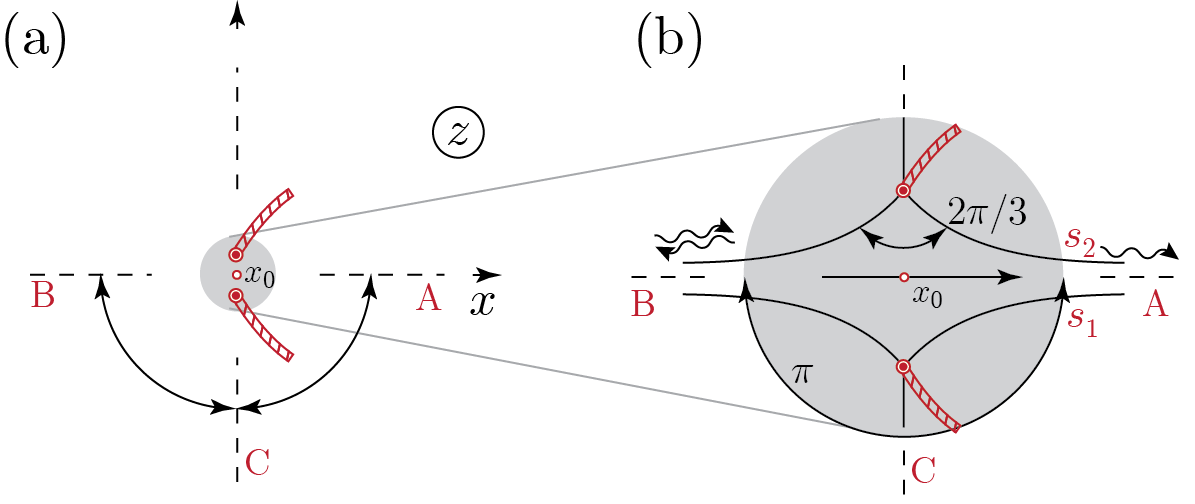}
    \caption{\justifying
    Sign - changing undulation of the edge of TI.
    (a)  Large scale pattern of anti-Stokes lines emanating from the branch points of the semiclassical momentum $\pi(z)$, $x_0$ is the root of the deformation potential $\phi$. The anti-Stokes lines are separated by $\pi/2$ angle. The ambiguity of the analytical continuation is resolved via detailed pattern of the semiclassical phase $\int \pi(\zeta)d\zeta$  presented in subplot (b).\\
    (b) Two anti-Stokes lines $s_1$ and $s_2$ merge into the horizontal \textit{transmission} line A. The incident and reflected waves propagate along line B, rotated with respect to the \textit{transmission} line by angle $\pi$. Anti-Stokes line C proves to be irrelevant.}
    \label{fig:antistokes01}
\end{figure}

The semiclassical structure of the Dirac equation is encoded in the pattern of its anti-Stokes lines. To determine them, we use Eq.~\eqref{semi2} together with the local expansion~\eqref{sub1} for the semiclassical momentum $\pi(z)$, which gives
\begin{gather}
\label{moment1}
    \begin{split}
    \pi(\zeta) &= \mu \sqrt{\frac{2\delta\ve}{\mu}+\frac{\zeta^2}{a^2}},\quad \frac{|\zeta|}{a}\ll1,\\
    S(z) &=\!\!\!\!\!\!\!\!\int\limits_{\pm ai\sqrt{2\delta\ve/\mu}}^z\!\!\!\!\!\!\pi(\zeta)d\zeta
    \end{split}
\end{gather}
The phase \(S(z)\) in Eq.~\eqref{moment1} is immediately recognized as the dynamical phase associated with the adiabatic eigenenergies of the Landau–Zener problem with Hamiltonian $H(t) =  \sigma_z \alpha t +\Delta\sigma_x$,
whose diabatic levels are \(\pm\alpha t\) and the gap is \(\Delta\).
For $|\zeta|/a\gg \delta\ve/\mu$
the asymptotic directions of the anti-Stokes lines follow immediately from
\begin{gather}\label{antiStokes1}
    \begin{split}
       {\rm Im}\,S(z) &= 0\ {\Rightarrow}\ 
         \arg\zeta = \frac{\pi n}{2},
    \end{split}
\end{gather}
The respective lines are depicted in Fig.~\ref{fig:antistokes01}(a).
Equation~\eqref{antiStokes1} yields four asymptotic directions, two of which coincide with the real axis and are therefore the natural directions for analytic continuation. An important subtlety should be emphasized here. These four asymptotic directions originate from six anti-Stokes lines emanating from the two nearly coalescing branch points of the semiclassical momentum [Fig.  Fig.~\ref{fig:antistokes01}(b)]. Two pairs of these lines merge asymptotically into horizontal directions, providing the natural matching channels for the semiclassical solutions. To transform the asymptotic transmitted wave (line A in Fig.~\ref{fig:antistokes01}) into the incident and reflected waves, one therefore analytically continues the solution through an angle \(-\pi\) in the complex \(z\) plane.

The semiclassical solutions follow from Eq.~\eqref{semi0}. Matching them to the asymptotics of the exact solution~\eqref{sol1} yields the transfer matrix (see Appendix~\ref{app:indef} for details):
\begin{gather}
\label{LZ}
\begin{split}
    T_{\rm LZ} &= \begin{pmatrix}
    \sqrt{1+P}e^{i\vf}\ &\ i\sqrt{P}\\
    -i\sqrt{P}\ &\ \sqrt{1+P}e^{-i\vf}
    \end{pmatrix},\\
    P &= e^{-2\pi s},\quad s = \delta\ve|a|  \\
  \vf &= -s\ln s + s - \arg\,\Gamma(1/2-is),
\end{split}
\end{gather}
where $P$  is the LZ transition probability and  $\ \vf$ is the corresponding LZ Stokes phase~\cite{stokesPhase}. A single deformation period contains two such LZ transitions, located near the consecutive zeros \(x_0\) and \(x_1\) of the deformation profile. Between them the evolution is adiabatic and amounts only to the accumulation of the semiclassical phases. The monodromy matrix over one complete period is therefore:
\begin{gather}
\label{mon1}
    T = U_1T^{(1)}_{\rm LZ}U_0T^{(0)}_{\rm LZ},
\end{gather}
where $T^{(0,1)}_{\rm LZ}$ are the LZ transfer matrices for the transitions in the vicinity of points $x_0$ and $x_1$ (see Fig.\ref{fig:bend}(b)) and matrices $U_{1,2} = {\rm diag}(e^{i\Phi_{1,2}},\ e^{-i\Phi_{1,2}})$ are the diagonal propagation matrices describing adiabatic evolution between successive LZ transitions 
\begin{gather}
\label{phases}
    \Phi_0= \int\limits_{x_1}^{x_0}\pi(x)\,dx,\quad
    \Phi_1 = \int\limits_{x_0-L}^{x_1}\pi(x)\,dx.
\end{gather}
\begin{figure}
    \centering
    \includegraphics[width=1.0\linewidth]{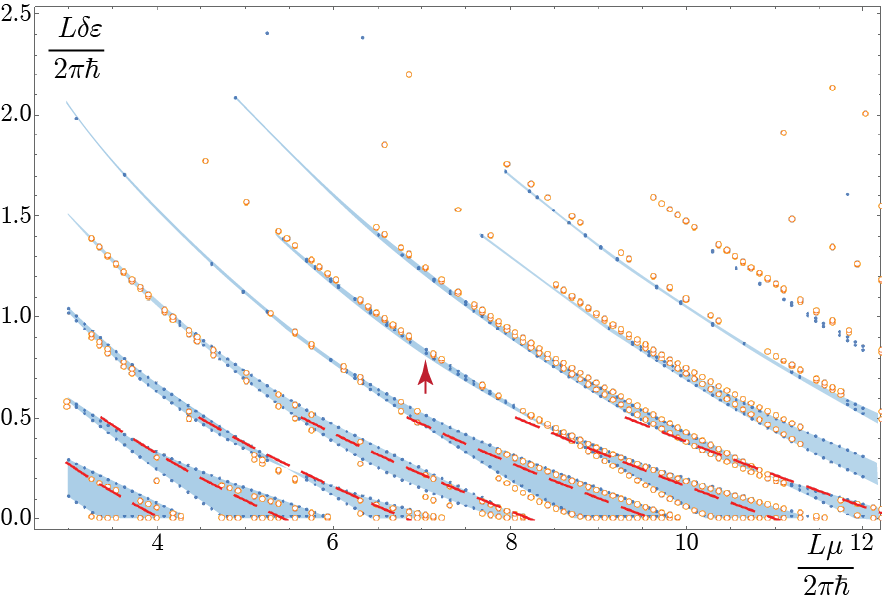}
    \caption{\justifying The pattern  of  forbidden bands ($\delta\ve = \ve-\mu$) for the potential $\phi(x) = 0.5+\cos (2\pi x/L)$,  obtained with semiclassical expression~\eqref{band01} (orange dots) and numerical solution of the original Dirac equation~\eqref{Dirac2} (blue dots with light blue filling)  as a function of the magnetic field $\mu$. The  band centers are obtained with analytic expression~\eqref{band-edges}  for small $\delta\ve$ are shown with red dashed lines. The width of the band highlighted by the red arrow $n=8$ is depicted in detail below in Fig.~\ref{fig:pLZ01b}.}
    \label{fig:pLZ01a}
\end{figure}
Equations ~\eqref{LZ} -~\eqref{phases} therefore map one period of the deformed TI edge directly onto a Landau–Zener–Stückelberg interferometer: the two zeros of \(\phi(x)\) play the role of successive LZ transitions, while \(\Phi_0\) and \(\Phi_1\) are the dynamical phases accumulated between them.
\subsubsection{Band positions}
Periodic repetition of this elementary interferometer converts the scattering problem into a Bloch problem. Since \(T\) is the transfer matrix over one period, forbidden bands correspond to real Bloch exponents and are determined by
\begin{gather}
\label{Bloch}
    |{\rm tr}T|\geq 2,
\end{gather}
Substituting the two LZ matrices ~\eqref{LZ}, and the propagation phases  \eqref{phases} into monodromy matrix~\eqref{mon1} and  Bloch condition~\eqref{Bloch} yields the explicit forbidden-band equation
\begin{gather}
\label{band01}
    \begin{split}
    &\Big|\cos(\Phi_0-\Phi_1)\sqrt{P_0P_1}+\\&\sqrt{1+P_0}
    \sqrt{1+P_1}\cos(\Phi_0+\Phi_1+\phi_0+\phi_1)\Big|>1.
    \end{split}
\end{gather}
Here, $P_{0,1} = \exp(-2\pi\delta\ve |a_{0,1}| )$ are the LZ matrices at transition points $x_0$ and $x_1$ respectively. Coefficients $a_{0,1}$ are the respective Taylor expansion constants of the field $\phi(z)$ near the roots $x_{0,1}$ (see Eq.~\ref{sub1}). 
$\phi_{1,2}$ are the corresponding Stokes phases. 
Equation~\eqref{band01} separates naturally into two ingredients: the LZ probabilities \(P_{0,1}\), which control the band widths, and the total dynamical phase \(\Phi_0+\Phi_1\), which primarily fixes the band positions.

In the full richness, the band equation~\eqref{band01} can be analyzed only numerically. However, it can also be studied analytically in several important limits thanks to the initial smallness of $\delta\ve\ll\mu$. 
For both \(P_{0,1}\ll1\) and \(1-P_{0,1}\ll1\), the location of the band center is controlled, to leading order, by the second term in~\eqref{band01}, i.e. the total phase \(\Phi_0+\Phi_1\). In the limit $\delta\ve/\mu$ this phase can be evaluated analytically (Appendix~\ref{app:band}) as
\begin{align}
\label{full_phase}
    \Phi_0+\Phi_1 
    &= \mu\left[L_0 +L_{\rm eff}\frac{\delta\ve}{\mu}
    \ln\frac{B}{\sqrt{\frac{\delta\ve}{\mu}}}+...\right]\\
    L_{\rm eff} &= |a_0|+|a_1|\label{leff}\\
    L_0&=\int\limits_0^L\frac{\phi(x)}{1+\phi^2(x)}\,dx\label{lo}
\end{align}
where the accuracy is $o(\delta\ve/\mu)$ and \(B=O(1)\) is a field-dependent dimensionless coefficient (see Appendix ~\ref{app:phase}), while \(L_{\rm eff}\) and \(L_0\) are characteristic length scales whose meaning becomes more transparent below.
Solving the phase condition~\eqref{band01} for the band centers gives
\begin{gather}
\label{band-edges}
    \delta\ve_n(\mu) \approx \frac{\pi n - \mu L_0}{L_{\rm eff}}\frac{1}{\ln\mu L_{\rm eff}}.
\end{gather}
Equation \eqref{band-edges} agrees well with both the numerical solution of the band equation ~\eqref{band01} and the direct numerical solution of the original Dirac equation ~\eqref{Dirac2}, as shown in Fig.~\ref{fig:pLZ01a}.
\begin{figure}
    \centering
    \includegraphics[width=1.0\linewidth]{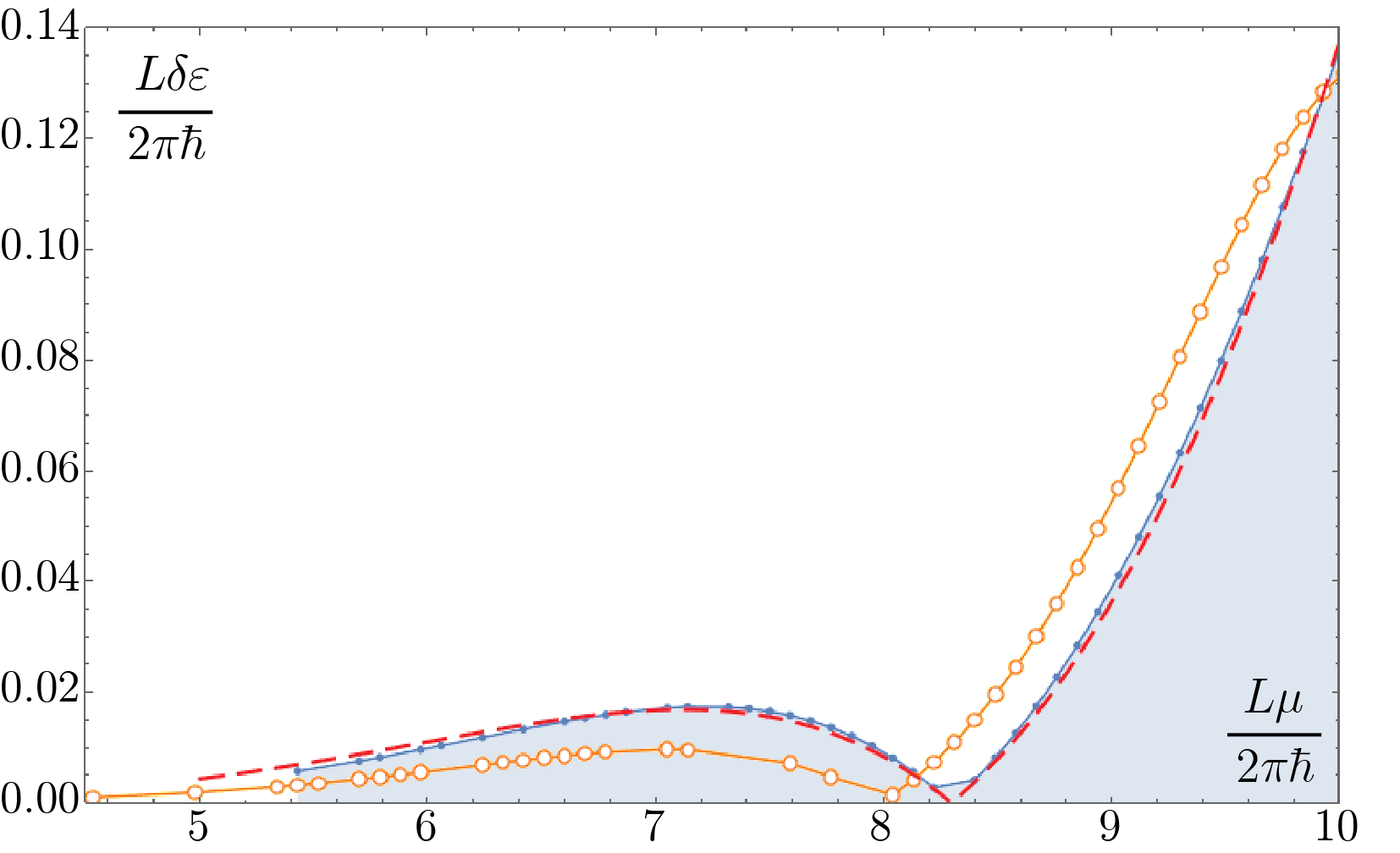}
    \caption{\justifying The $n = 8$ band gap for $\phi(x) = 0.5+\cos(2\pi x/L)$, calculated from semiclassical expression~\eqref{band01} (orange dots), direct numerical integration of the Dirac equation~\eqref{Dirac2} (blue dots) and the analytical approximations~\eqref{band-edges02} and ~\eqref{gap-sym} (red dashed line).}
    \label{fig:pLZ01b}
\end{figure}
\subsubsection{Band gaps}
For small energy offsets, \(\delta\varepsilon |a|\ll1\), Eq.~\eqref{band01} yields the following expressions for the gap widths:
\begin{align}
\label{band-edges01}
   \Delta_n(\mu) &= \frac{2}{L_{\rm eff}}\begin{cases}
       \arccos\left[\sin^2\frac{\mu\Delta L}{2}\right],\\
       \arccos\left[\cos^2\frac{\mu\Delta L}{2}\right],
   \end{cases}\ \ \delta\ve L\ll1,\\
   \Delta L & = \int\limits_{x_1}^{x_0}\frac{|\phi(x)|}{1+\phi^2(x)}\,dx-
   \int\limits_{x_0-L}^{x_1}\frac{|\phi(x)|}{1+\phi^2(x)}\,dx\label{DeltaL}
\end{align}
where the upper (lower) line in ~\eqref{band-edges01} corresponds to even (odd) $n$. 
$\Delta L$ is a geometric length determined solely by the deformation profile.  
We clearly see the oscillatory nature of band gaps as the functions of magnetic field $\mu$.

For larger energies $\delta\ve|a_i|\gg1$ the bands become exponentially small, as one can read directly from the band equation~\eqref{band01}. We obtain for the band gaps:
\begin{gather}
\label{band-edges02}
    \Delta_n(\mu) = \frac{[P_1+P_2\pm2\sqrt{P_1P_2}\cos\mu\Delta L]^{1/2}}{L_{\rm eff}\ln\mu L_{\rm eff}},\ \ \delta\ve L\gg 1
\end{gather}
where $\pm$ corresponds to even and odd levels respectively. For a symmetric potential $\phi(x)=\phi(L-x)$ one has $|a_0|=|a_1|=L_{\rm eff}/2$ and~\eqref{band-edges02} simplifies to
\begin{gather}
\label{gap-sym}
    \Delta_n(\mu) = \frac{2e^{-\pi L_{\rm eff}\delta\ve_n(\mu)/2}}{L_{\rm eff}\ln\mu L_{\rm eff}}
    \begin{cases}
        \left|\cos\frac{\mu\Delta L}{2}\right|,\ &\ n\ \ \hbox{is even}\\
        \\
        \left|\sin\frac{\mu\Delta L}{2}\right|,\ &\ n\ \ \hbox{is odd}
    \end{cases}.
\end{gather}
Equations~\eqref{band-edges01} and~\eqref{band-edges02} make the oscillatory dependence of the gap widths on the magnetic field explicit.
For the model profile $\phi(x)=1/2+\cos(2\pi x/L)$, this behavior is illustrated in Fig.~\ref{fig:pLZ01b}. 
Eqs.~\eqref{band-edges01},~\eqref{band-edges02} present the main intermediate results of this paper.
Having established the strong-field oscillations for a sign-changing profile, we now turn to the analytically different case in which \(\phi(x)\) has no real zeros.


\subsection{Sign-definite edge deformation}
The simplest realization of a sign-definite deformation is an overall bend of the TI edge whose magnitude exceeds that of the superimposed periodic undulation, as illustrated in Fig.~\ref{fig:bend2}. We focus on regular deformation profiles, for which the scattering is controlled by the branch points of the semiclassical momentum \(\pi(z)\), now located in the complex plane, rather than by possible poles of the deformation profile \(\phi(z)\).
\subsubsection{Exact equation near the turning point}
As in the sign-changing case, we expand around a zero \(z_0\) of \(\phi(z)\), which now lies in the complex plane. This time, unlike ~\eqref{sub1}, the respective Taylor expansion is written with explicit imaginary unity $i$.
\begin{gather}
    \phi(z) = i\frac{z - z_0}{a} + ... ,
\label{sub2}
\end{gather}
The scale \(a\) is generally complex. We restrict ourselves to \(|\arg a|\lesssim1\), so that varying its phase does not change the topology of the Stokes pattern discussed below.

The conditions under which \(|\arg a|\) remains small are discussed in Appendix~\ref{app:analyt01}. As before, \(|a|\) is of the order of the deformation period \(L\).

Substituting expansion~\eqref{sub2} into Eq.~\eqref{Dirac2} and retaining the leading terms in \(1/(\mu a)\), we obtain an equation closely analogous to Eq.~\eqref{exact_eq1}:
\begin{gather}
\begin{split}
    &\psi''_1 - 2\mu \frac{\zeta}{a} \psi'_1 + \mu\left[ 2\delta \ve - \frac{1}{a}  \right] \psi_1 = 0 \\
    &\zeta = z - z_0
\label{eq2}
\end{split}
\end{gather}
Unlike the sign-changing case, a generic sign-definite profile has only one zero per period, so the conventional two-transition LZS interference mechanism appears to be absent. Nevertheless, the complex orientation of the local deformation (due to the predominantly imaginary value of $\phi'(z_0)\equiv i/a$ in~\eqref{sub2}) produces a qualitatively different anti-Stokes topology. As we show below, the analytic structure of the TI Hamiltonian restores coherent interference through the interplay of the two nearly coalescing branch points with the nearby pole of the semiclassical momentum.

\subsubsection{Analysis of the anti-Stokes lines}
\begin{figure}[t]
    \includegraphics[width=1\linewidth]{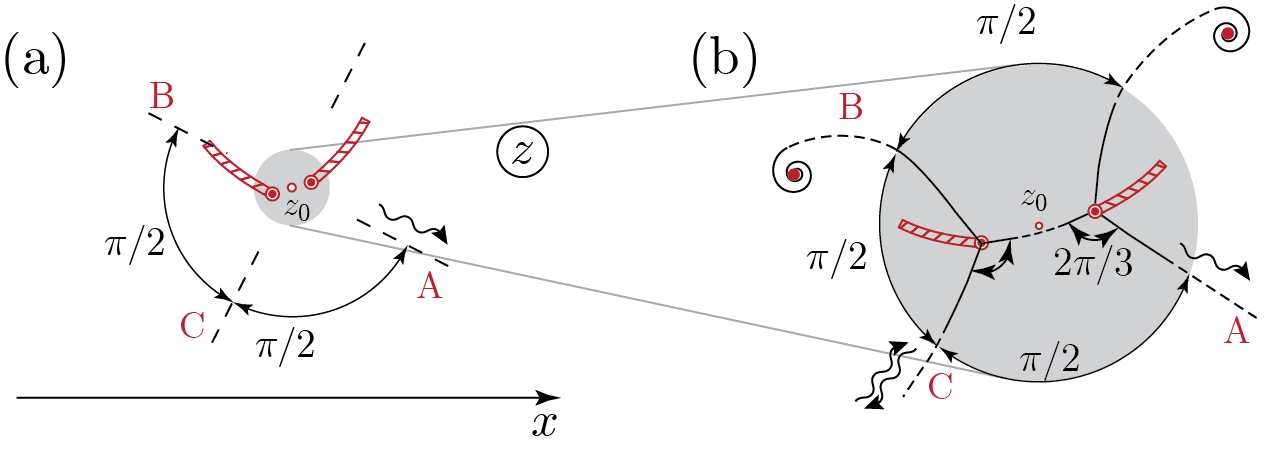}
    \caption{\justifying
    Sign-definite periodic deformation of the TI edge.\\
    (a) Large-scale anti-Stokes pattern associated with the pair of nearly coalescing branch points near the complex zero \(z_0\) of \(\phi(z)\). The transmitted wave propagates along line A, while the two left-going branches B and C make the continuation apparently ambiguous.\\
    (b) Local topology resolving this ambiguity. The two almost coalescent branch points generate six anti-Stokes branches: two connect the branch points, two spiral toward the poles of \(\pi(z)\), and the remaining two (A and C) form the asymptotic transmission and incident/reflected channels. 
    }
    \label{fig:antistokes2}
\end{figure}
Locally, the anti-Stokes structure appears almost identical to that of the sign-changing case~\eqref{moment1}. There is, however, a crucial global difference.

\begin{gather}
\label{moment2}
    \pi^2(\zeta) = \mu^2 \left(\frac{2\delta\ve}{\mu}-\frac{\zeta^2}{a^2}\right),\quad \left|\frac{\zeta}{a}\right|\ll1.
\end{gather}
As usual, in the limit $|\zeta/a|\gg \delta\ve/\mu$ we obtain the asymptotic directions for the anti-Stokes lines:
\begin{gather}\label{antiStokes-sign-definite}
    \begin{split}
       {\rm Im}\,S(z) &= {\rm Im}\int\limits_{\pm a\sqrt{2\delta\ve/\mu}}^z\pi(\zeta)d\zeta \underset{|\frac{\zeta}{a}|\gg \frac{\delta\ve}{\mu}}{\longrightarrow} {\rm Im} \frac{i\mu\zeta^2}{2a} = 0\Rightarrow\\
         \arg\zeta &= \frac{\pi}{4}+\frac{\arg a}{2}+\frac{\pi n}{2}.
    \end{split}
\end{gather}
Equation~\eqref{antiStokes-sign-definite}, as before, yields 4 directions separated by $\pi/2$.
At first sight, however, the anti-Stokes line pattern presents a rather puzzling picture, as one can judge from Fig.~\ref{fig:antistokes2}(a). The transmitted wave propagates along anti-Stokes line \(A\). Unlike in the sign-changing case, it is not immediately clear whether continuation along line \(B\) or \(C\) produces the incident and reflected waves: both extend in the required leftward direction.

To resolve this ambiguity, one must examine the anti-Stokes lines near the two branch points. As in the real-zero case, the strong-field limit produces two nearly coalescing branch points,
$\zeta_{\pm} = \pm a\sqrt{2\delta\ve/\mu}$,
each of which emits three local anti-Stokes lines separated by \(2\pi/3\). Thus six local branches emerge from the pair, see Fig.~\ref{fig:antistokes2}(b).  Since only four asymptotic directions remain at large \(|\zeta|\), two of these branches must join the two turning points directly.

\begin{figure}[t]
    \centering
    \includegraphics[width=1\linewidth]{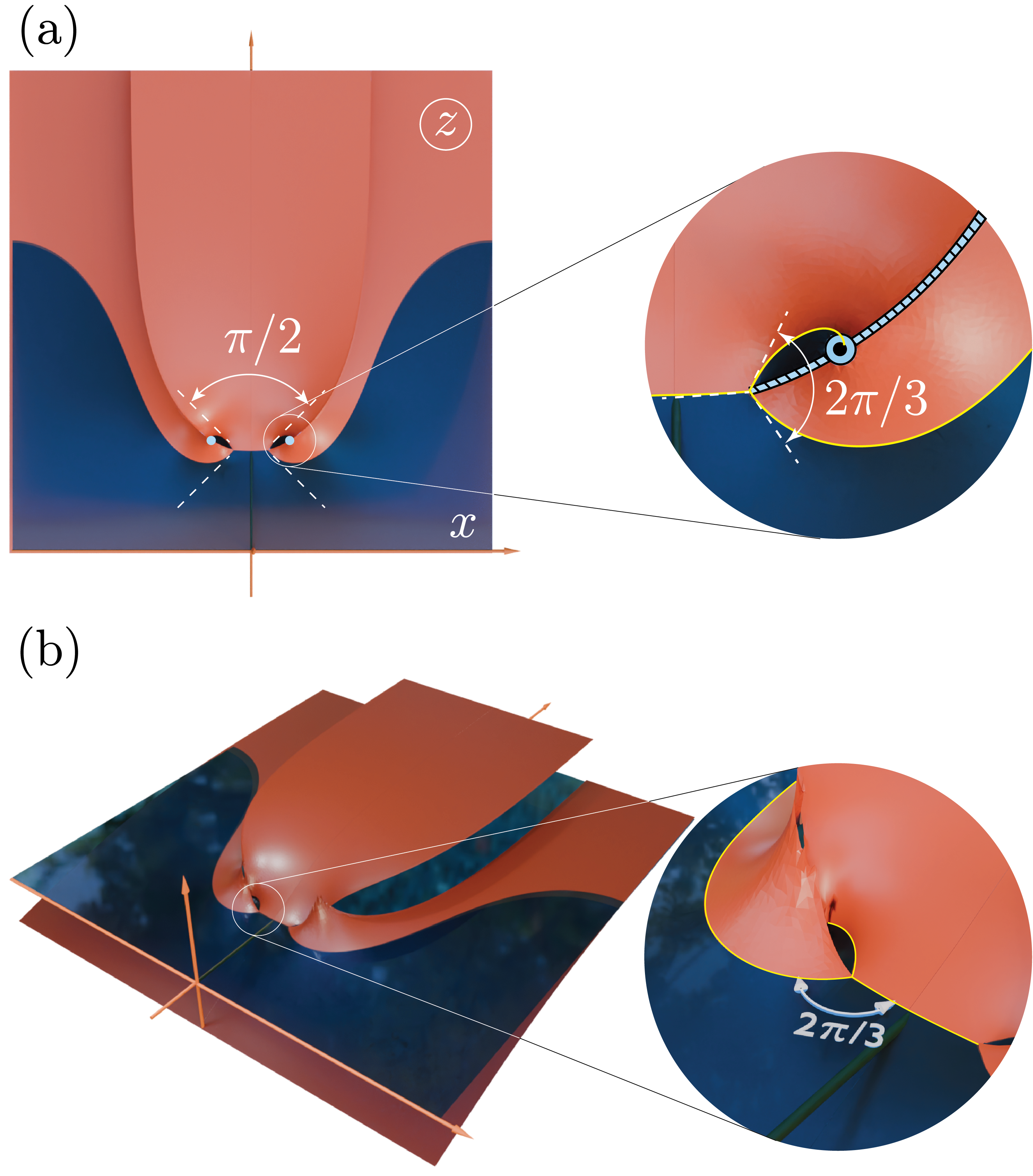}
    \caption{\justifying Anti-Stokes pattern for a sign-definite deformation. (a) Anti-Stokes lines emanating from the branch points of the semiclassical momentum \(\pi(z)\) for a representative deformation profile $\phi(x) = 2+\cos x $. (b) Three-dimensional visualization of the corresponding semiclassical phase surface. The anti-Stokes lines are shown as the intersection of the phase surface ${\rm Im}\,S(z)$ with the horizontal plane ${\rm Im}\,S(z) = {\rm Im}\,S(z_{+})$, where $z_+$ is the branch point of $\pi(z)$.}
    \label{fig:antistokes1}
\end{figure}

The fate of another pair of anti-Stokes lines becomes clear once we analyze the behavior of the semiclassical momentum and semiclassical phase $S(z)$ near the poles: $\phi^2(z_1)+1=0$. The Laurent expansion of the momentum near the pole $z_1$
yields the leading expression for the phase function:
\begin{gather}
   \begin{split}
        \pi(z) &= \frac{c_{-1}}{z-z_1}+\dots,\ z\rightarrow z_1\ \Rightarrow \\
        S(z) &= \int^z\pi(z)\,dz =c_{-1}\ln(z-z_1) +\mathcal{O}(1).
   \end{split}
\end{gather}
Parameterizing $z-z_1 = \rho e^{i\vf}$ we obtain the logarithmic spiral for the anti-Stokes line:
\begin{gather}
\label{spiral}
 \ln\rho + \vf\cot(\arg\,c_{-1}) = \rm const.
\end{gather}
As shown in Appendix~\ref{lemma} for a broad class of regular periodic profiles the semiclassical momentum \(\pi(z)\) possesses at least two simple poles within the upper half of each periodicity strip. The corresponding pair of anti-Stokes lines terminates at these poles, approaching them as logarithmic spirals [Fig.~\ref{fig:antistokes2}(b)]. The remaining two anti-Stokes lines are therefore uniquely identified as the transmitted and incident/reflected directions.
Unlike the previous case of real zero of $\phi(x)$,  they are separated by an angle \(\pi/2\).
We draw the exact picture of anti-Stokes line for a sample potential in Fig.~\ref{fig:antistokes1}.
\subsubsection{Reflection coefficient}
Constructing the exact solution of Eq.~\eqref{eq2} and matching its asymptotics to the semiclassical expressions~\eqref{semi0}, we obtain the reflection coefficient (see Appendix~\ref{app:def} for all the details):
\begin{gather}
\label{ans1}
\begin{split}
    R&= 4\left|\Gamma\left[\delta\ve a+\frac{1}{2}\right]\frac{1}{\sqrt{\pi}}\left\{\frac{e}{a\delta\ve}\right\}^{a\delta\ve}\right|^2
    \left|\cos^2\pi a\delta\ve\right| 
    e^{-4{\rm Im}S_0}\\
    S_0 &=   \int_{x_0}^{z_0}\pi(z)\,dz, 
\end{split}
\end{gather}
Here, the semiclassical phase $S_0$ is computed from an arbitrary point $x_0$ on the real axis to the complex zero of the deformation profile $\phi$: $z_0$.

Reflection coefficient~\eqref{ans1} again exhibits pronounced oscillations as a function of the magnetic field on the scale $|a\delta\ve|\gtrsim1$.
The prefactor containing the $\Gamma$ function has an evident Stirling asymptotics, which considerably simplifies the result in this regime.
Since the semiclassical approximation requires \(\mu |a|\gg1\), the limit \(|a\delta\varepsilon|\gtrsim1\) lies well within the range of applicability of the present analysis.
Using the Stirling identity for the Euler $\Gamma$ -  function $\Gamma(x+1/2)\approx \sqrt{2\pi}(x/e)^x,\ x\gtrsim1$ we obtain the reflection coefficient in the form:
\begin{gather}
  \label{ref02}
  R = 4\left|\cos^2\pi a\delta\ve\right|e^{-4{\rm Im}S_0},\quad \left|a\delta\ve\right|\gtrsim 1,
\end{gather}
where $S_0$ is defined by Eq.~\eqref{ans1}.  
Equation ~\eqref{ref02} will be the central result for determining the band-gap widths.
\subsubsection{Band positions and gap widths}
The centers of the forbidden bands are determined by Eq.~\eqref{action0}. Although the corresponding phase integral depends on the detailed geometry of the deformation, its energy dependence can be extracted analytically in the strong-field limit. Expanding the integrand in \(\delta\varepsilon/\mu\), we obtain the following equation for the band centers:
\begin{gather}
\label{band_eq}
\begin{split}
    \delta\ve_n &=  \frac{\pi n -\mu L_0}{L_0+L_1}\\
    L_1 &= \int_0^L\frac{dx}{\phi(x)}\frac{1}{1+\phi^2(x)}.
\end{split}
\end{gather}
The integer \(n\) is chosen such that \(|\delta\varepsilon_n|\ll\mu\), while \(L_0\) is defined in Eq.~\eqref{lo}.
Next, using Dykhne relation~\eqref{dykhne2}, and setting $\delta\ve=0$ in the traverse time in the pre-exponential term, we obtain the gap width
\begin{gather}
\label{band-gap03}
 \begin{split}
    \Delta &= \frac{4}{T_0}|\cos\pi a\delta\ve|e^{-2{\rm Im}S_0},\\
    T_0 & = \int_0^L \frac{dx}{\phi}.
\end{split}
\end{gather}
As in the LZS-interferometer case, the phase \(S\) admits a universal small-\(\delta\varepsilon/\mu\) expansion valid for an arbitrary smooth deformation profile:
\begin{gather}
\label{action2}
\begin{split}
    {\rm Im}S &=\mu L_{\rm eff}+\delta\ve \,{\rm Re}\,a \ln\frac{B}{\sqrt{\frac{\delta\ve}{\mu}}}\\
    L_{\rm eff} &= {\rm Im}\int\limits_{x_0}^{z_0}\frac{\phi(z)\, dz}{\phi^2(z)+1}
\end{split}
\end{gather}
The omitted terms are of relative order \(O(\delta\varepsilon/\mu)\), while \(B=O(1)\) is a numerical coefficient determined by the detailed shape of the deformation (see Appendix~\ref{app:phase5}). Equation~\eqref{action2} isolates the universal energy- and field-dependent part of the reflection amplitude. The resulting band gap is therefore
\begin{gather}
\label{band1}
    \begin{split}
    \Delta_n(\mu) &= \frac{4}{T_0}\left[\frac{1}{B^2}\frac{\delta\ve_n}{\mu}\right]^{\delta\ve_n {\rm Re }a}|\cos\pi a\delta\ve_n| e^{-2 \mu L_{\rm eff}}, 
    \end{split}
\end{gather}
Here \(\delta\varepsilon_n(\mu)\) is the center of the \(n\)th forbidden band given by Eq.~\eqref{band_eq}, and \(T_0\) is defined in Eq.~\eqref{band-gap03}. Equation ~\eqref{band1} shows that, as in the case of the sign-changing deformation, the gap widths oscillate with magnetic field.
\begin{figure}[t!]
    \centering
    \includegraphics[width=1.0\linewidth]{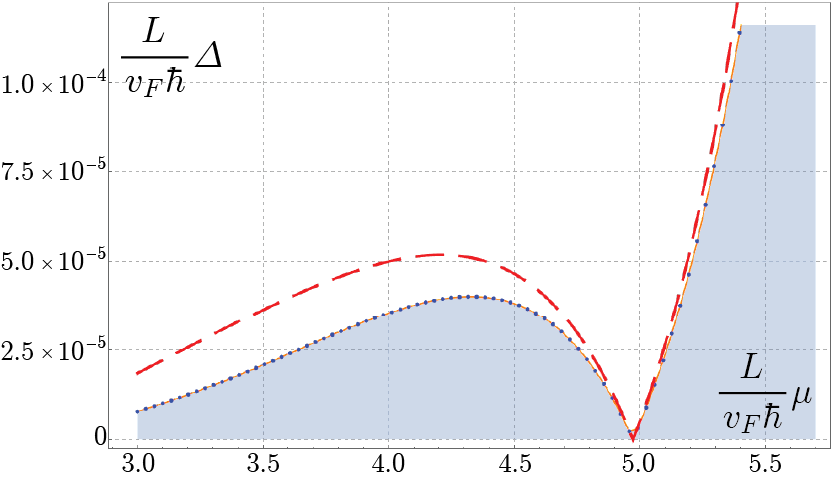}
    \caption{\justifying The band gap as a function of magnetic field $\mu$ for the potential $\phi(x)=2+\cos(2\pi x/L)$ for the band $n=5$. The analytic curve is given by~\eqref{band1} (dashed red line). The numerical solution  of the exact differential equation~\eqref{Dirac2} is given by the dotted line.}
    \label{fig:plot01}
\end{figure}
The comparison of semiclassical result~\eqref{band1} and numerical solution of the original Dirac equation~\eqref{Dirac2} is presented in Fig.~\ref{fig:plot01} for the model potential $\phi(x) = 2+\cos(2\pi x/L)$. As we see from Fig.~\ref{fig:plot01}, despite the astonishing complexity of the Dirac's equation~\eqref{Dirac2}, the semiclassical PkHD construction reproduces the band position on an 
$\mathcal{O}(1)$ energy scale with an absolute accuracy of order $10^{-4}$, a level of precision that is highly unusual for semiclassical methods. This accuracy is sufficient to resolve an exponentially narrow  band gap of width $10^{-4}$, whose magnitude is then reproduced to within $~\sim30 \%$, demonstrating that both the leading exponential and the pre-exponential structure are reliably captured.

Equation~\eqref{band_eq} shows that the centers of the forbidden bands exhibit an unusual strictly linear dependence on the magnetic field. At the same time, Eq.~\eqref{band1} reveals an important distinction from S--dH oscillations: whereas the latter are periodic in the inverse magnetic field, the present oscillations are periodic in the magnetic field itself.

Finally, the strong-field result~\eqref{band1} can be cast in a particularly transparent form in terms of the dimensionless magnetic field \(h\):
\begin{gather}
    \Delta(h) \propto\left|\cos[\#_1 h+\#_2]\right|e^{\#_3 h\ln h-\#_4h},\ \ h = \frac{\mu_B gL}{\hbar v_F}\mathcal{H}
\end{gather}
where \(\#_{1,2,3,4}=O(1)\) are dimensionless coefficients determined by the geometry of the deformation.

To summarize, the strong-field spectra of the both deformation geometries are remarkably similar.
In both cases the band positions are  linear in magnetic field \(\mu\) [Eqs.~\eqref{band-edges} and ~\eqref{band_eq}],
while the gap widths contain an oscillatory factor multiplied by a smooth exponential envelope [Eqs.~\eqref{gap-sym} and ~\eqref{band1}]. This robustness is physically natural: adding a smooth overall bend to the TI sheet merely moves the zeros of deformation field $\phi$ away from the real axis but should not eliminate the interference physics.
Thus, despite quite different complex-plane scattering geometries, both types of deformation exhibit satisfyingly similar patterns of magnetic quantum oscillations. This completes the strong-field picture. We now turn to the weak-field limit, where the failure of Eq.~\eqref{band_general} discussed above reveals a qualitatively different—and considerably more subtle—physical mechanism.

\section{Weak magnetic field. General considerations.}
We consider the weak-field regime
\begin{gather}
\label{weak}
 \mu\ll\ve   
\end{gather}
As discussed below Eq.~\ref{band_general}, the band gap must close in the absence of a magnetic field due to TR symmetry.  We approach this problem from two complementary viewpoints: 

(i) semiclassical regime, valid for  $\ve L\gg1$, 

(ii) perturbation theory in $\mu L \ll 1$, valid for arbitrary  $\ve L$.  In their common domain of applicability, the two approaches must coincide.

\subsection{Semiclassical regime}

Although a perturbative  expansion in \(\mu\) is tempting, it restricts the analysis to \(\mu L\ll1\). In the semiclassical regime \(\varepsilon L\gg1\), however, the weak-field condition \(\mu\ll\varepsilon\) remains compatible with \(\mu L\gtrsim1\), allowing us to access a much broader parametric range.

For small $\mu\ll\ve$  the branch point $z_0$ of the momentum  
$\phi(z_0) = i\sqrt{1 - \mu^2/\ve^2}$ (see Eq. \eqref{semi0}) is close to its pole $\phi(z_p)=i$.
The resulting scattering no longer follows the conventional isolated-turning-point WKB scenario. Instead, the reflected wave is controlled by the combined analytic structure of the nearly coalescing branch point and pole.
A closely related analytic structure (coalescing branch point and pole) occurs in scattering from the familiar potential \(-U_0/\cosh^2(x/a)\), which becomes reflectionless for particular values of \(U_0\), independently of the incident energy.

Let us briefly outline the main features of the treatment. 
The essential modification is that a local expansion around the branch point is no longer uniform: its domain of validity terminates at the nearby pole. We therefore Laurent-expand the deformation profile around the pole \(z_p\) of $\pi(z)$,  
\begin{gather}
\label{a-def}
    \phi(z) = i + \frac{z - z_p}{a} + ..,
\end{gather}
where \(a\), as before, is of the order of the deformation period and is generally complex. Expanding the original Dirac equation~\eqref{Dirac2} near \(z_p\), as in the previous cases, we obtain
\begin{gather}
\begin{split}
    &2i\zeta \psi'' + [3i - 2\mu a]\psi' - \ve^2 a\psi = 0, \\
    &\zeta = z - z_p,
\label{eq3}
\end{split}
\end{gather}

We then follow the standard PKh construction: local equation~\eqref{eq3} is solved exactly, the relevant anti-Stokes lines are identified, its asymptotics are analytically continued in the complex plane, and finally matched to the semiclassical wave functions.
The complete derivation is given in Ref.~\cite{dotdaev2024}. Here we only quote the resulting scattering amplitude:
\begin{gather}
\label{reflect1} 
r=2\sinh(\pi \mu a)  \exp\!\left[ 2i\!\!\int_{x_{0}}^{z_{+}}
\!\!\frac{
\sqrt{\varepsilon^{2}\!\left(\varphi^{2}+1\right)-\mu^{2}}
}
{\varphi^{2}+1}\,dz\right].
\end{gather}
Expanding Eq.~\eqref{reflect1} in $\mu/\ve$ and combining the result  with band-gap formula~\eqref{dykhne2} we obtain the width of the gap
\begin{gather}
\label{res_fin1}
\begin{split}
    &\Delta_n = \frac{4}{T_n} |\sinh\pi\mu a| \exp \left( - 2 {\rm Im} S_0+ \pi\mu {\rm Re}\, a \right),
\end{split}
\end{gather}
where $S_0$ is the action introduced in~\eqref{ans1} evaluated at $\mu = 0$. 
For deformation profiles with appropriate discrete symmetries, several branch points may lie at the same distance from the real axis. The reflection amplitude is then obtained by coherently summing the contributions from all such points \(z_i\):
\begin{gather}
    r = \sum\limits_{z_i} 2 \sinh\pi\mu a_i \exp\left[2iS_{z_i}\right],\ \ S_{z_i} = \int\limits_{x_0}^{z_i}
    \pi(z)\,dz.
\end{gather}

\subsubsection{On the possibility of quantum oscillations}
Equation~\eqref{res_fin1} has a peculiar feature. The parameter \(a\), defined by the local expansion~\eqref{a-def}, is generally complex and depends on the deformation profile. It is therefore instructive to consider the class of profiles for which \(a\) is purely imaginary,
\begin{gather}
    {\rm Re}\,a = 0.
    \label{condition}
\end{gather}
Of course, the exact condition~\eqref{condition} need not be realized experimentally. However, some relatively broad classes of periodic deformations satisfies the approximate condition ${\rm Re}\,a\ll {\rm Im}\,a$ (see Appendix~\ref{app:twoharmonic}), so that Eq.~\eqref{condition} can be approached to good accuracy in a realistic experimental setting.

In this regime, Eq.~\eqref{res_fin1} predicts pronounced quantum oscillations of the band gaps as a function of magnetic field $\mu$. Since the oscillatory prefactor in Eq.~\eqref{reflect1} is independent of the carrier energy, these oscillations should be relatively insensitive to thermal averaging. In particular, the gaps close at discrete values of the magnetic field. The underlying picture, however, turns out to be considerably more subtle and reveals a rich global structure of the relevant Riemann surface of the semiclassical action $S(z)$. To understand it, we must analyze the anti-Stokes lines governing the analytic continuation of the semiclassical solutions in the complex plane.

Before tackling this more involved problem, however, it is useful to examine the perturbative regime \(\mu L\ll1\). Although the scattering amplitude is then perturbative in \(\mu\), the semiclassical phase emerges in this approach as well and already retains the essential complex-plane geometry—its saddle-point structure and singularities—that reappears in the full semiclassical action \(S(z)\) for \(\mu L\gtrsim1\). The perturbative problem therefore provides a convenient testing ground for understanding the analytic continuation and anti-Stokes topology of the general case.

\subsection{Perturbative treatment}
\subsubsection{General expressions}
As shown in the Appendix~\ref{app:perturb}, the perturbative treatment is valid in the regime
\begin{gather}
    \label{perturbCrit}
    \mu L\ll1 
\end{gather}
In this regime, the Hamiltonian~\eqref{ham1} can be considerably simplified by the unitary transformation
\begin{gather}
\begin{split}
\label{unitary}
  \hat{P} &= e^{i\theta(x)\sigma_x/2}\quad
  \tan{\theta(x)} = \frac{1}{\phi(x)}.
\end{split}
\end{gather}
Under this transformation, the kinetic part of the Hamiltonian becomes
\begin{gather}
\begin{split}
    &\tilde{H} = P^{-1}\hat{H}_0(x)P= \frac{1}{2}(v(x) \hat{p}_x + \hat{p}_x v(x)) \sigma_z, \\
    &v(x) = \sqrt{1 + \phi(x)^2}
\end{split}
\end{gather}
whereas the Zeeman term transforms as
\begin{gather}
    \tilde{U} = P^{-1}UP= \frac{\mu}{v(x)} (\phi(x)\sigma_z - \sigma_y) 
\end{gather}
The band-gap widths and the positions of their centers are derived in the same Appendix~\ref{app:perturb}. Notably, the semiclassical Dykhne relations~\eqref{action0}  and ~\eqref{dykhne2} preserve their form even in the perturbative regime, with the reflection amplitude for the periodic deformation given by the following integral
\begin{gather}
\label{perturb1}
    r = \mu\int\limits_{0}^L e^{2i\ve_n\tau(x)}\frac{dx}{1+\phi^2(x)},\quad\tau(x) = \int\limits_0^x\frac{dx}{\sqrt{1+\phi^2(x)}}.
\end{gather}
Here, as before, $\ve_n$ is the center of the $n$th forbidden band, determined by
\begin{gather}
\label{center}
    \ve_n = \frac{\pi n}{T_0},\quad T_0=\int\limits_0^L\frac{dx}{\sqrt{1+\phi^2(x)}}.
\end{gather}
Combining Eqs.~\eqref{perturb1} and \eqref{center} with Eq.~\eqref{dykhne2} , we obtain the perturbative result for the band gap,
\begin{gather}
    \label{band_perturb}
    \Delta = \frac{2}{T}\left|\mu \int_0^L  e^{2i\ve_n\tau(x)}\frac{dx}{1+\phi^2(x)}\right|.
\end{gather}
The perturbative result~\eqref{perturb1} and the semiclassical expression~\eqref{reflect1} must coincide in their common domain of applicability, \(\mu L\ll1\) and \(\varepsilon_n L\gg1\). The connection, however, is less straightforward than one might expect. In the limit \(\varepsilon_n L\gg1\), the perturbative integral can be evaluated by the steepest-descent method. A peculiar feature of Eq.~\eqref{perturb1}  is that its phase function \(\tau(z)\) has no finite saddle points in the complex plane,
$\tau^\prime(z) = 1/\sqrt{\phi^2(z)+1}\neq0$ since \(\phi(z)\) is assumed to be regular within the periodicity strip. Therefore, the only finite singularities of $\tau(z)$ are the branch points. The steepest-descent directions issuing from these branch points, however, point toward the real axis. Consequently, the required contour deformation necessarily involves several sheets of the Riemann surface of \(\tau(z)\).

\subsubsection{Single harmonic profile}
\begin{widetext}

\begin{figure}
    \centering
    \includegraphics[width=1.0\linewidth]{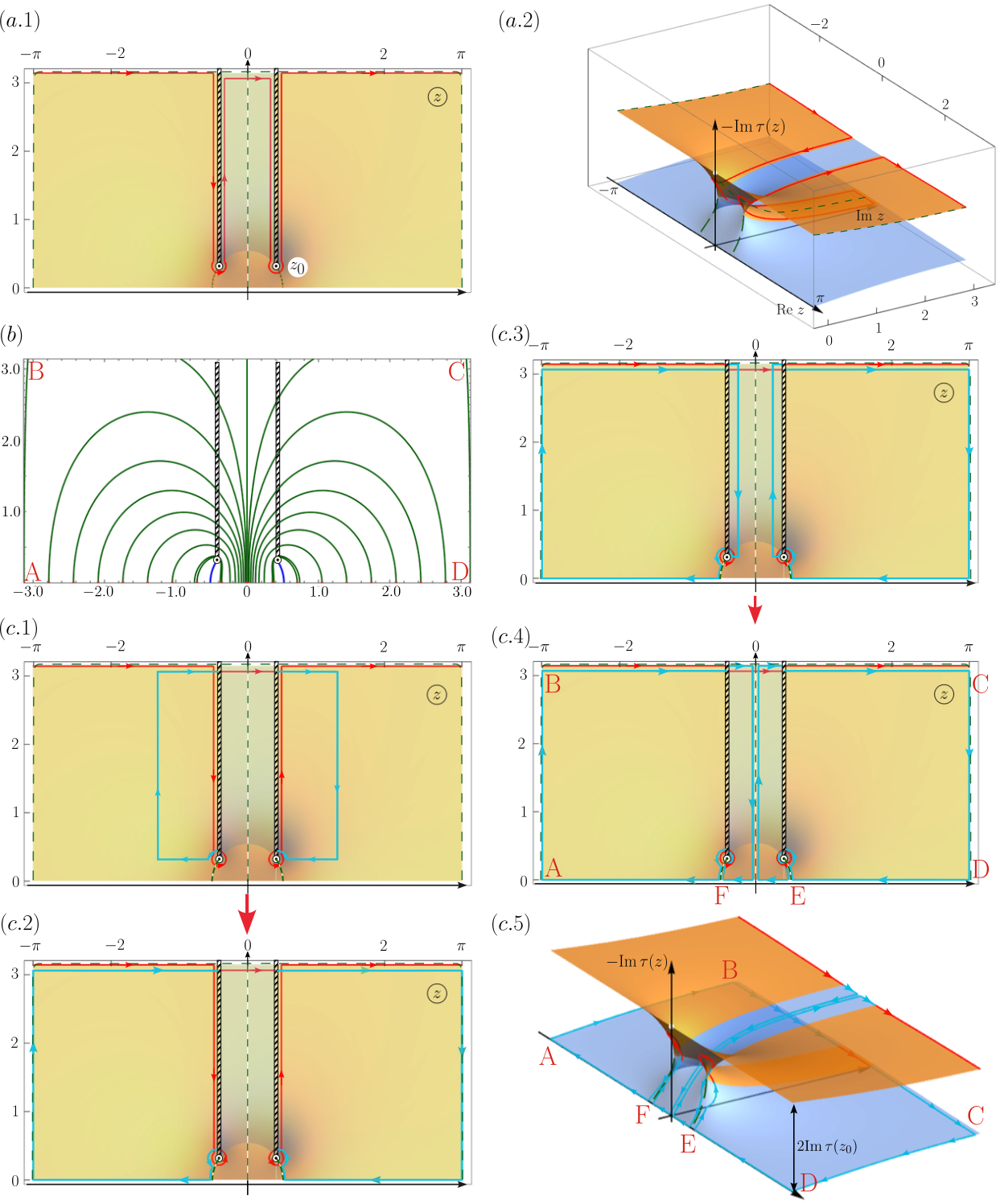}
    \caption{\justifying
    (a) The initial deformation of the contour for the sample single harmonic profile function ~\eqref{sample_pot}. Dark green dashed lines are the lines steepest descent. Two symmetric branch cuts emanate from branch points $z_0$ and $-z_0^*$. To the right, the placement of the contour on the surface $-{\rm Im}\,\tau(z)$ is shown.
    (b) The phase portrait of the steepest descent lines of function ${\rm Im}\,\tau(z)$.
    (c1-c4) The deformation of the contour for the single harmonic potential is shown along the steepest descent curves. (c.5) Final placement of the contour on the surface of ${\rm Im}\,\tau(z)$. The choice of $\pi$ as the upper boundary of all the plots is, of course, arbitrary. The actual steepest-descent contour extends to $+i\infty$.}
    \label{fig:steepest}
\end{figure}
\end{widetext}

We first consider the simplest case, in which a single Fourier harmonic dominates the deformation profile. The topology of the corresponding steepest-descent contours can then be understood almost completely and is illustrated in Fig.~\ref{fig:steepest}(a.1, a.2, b). To make the geometry explicit, we use the representative profile
\begin{gather}
\label{sample_pot}
\phi(x) = a_0 + a_1 \cos x,\ \ a_0 \approx 4.39,\ \ a_1\approx-4.57.
\end{gather}
\paragraph{The analytical structure of $\tau(z)$.}
For a periodic function whose only finite singularities are branch points, there exists a steepest-descent line connecting the real axis to \(i\infty\) between each neighboring pair of branch points. For the single-harmonic profile, these lines are vertical rays. The periodicity strip can always be chosen such that they form its boundaries ([lines AB and CD in Fig.~\ref{fig:steepest}(a)]. The steepest-descent curves infinitesimally displaced from these boundaries are initially almost vertical but, upon approaching \(i\infty\), bend by  \(\pi/2\) toward the corresponding branch cuts. In addition, a steepest-descent curve emanates downward from each branch point of \(\tau(z)\). The crucial feature of the scattering integral~\eqref{perturb1} is that the branch points of the phase \(\tau(z)\) are simultaneously poles of the pre-exponential factor of the integrand.

We now deform the integration contour onto the steepest-descent paths. In the present problem this deformation is highly unusual, since the contour must be continued across the Riemann surface of the multi valued \(\tau(z)\).

\paragraph{The contour deformation.}
It is convenient to begin the contour deformation by shifting the original contour upward toward \(+i\infty\). The vertical boundary segments cancel by periodicity of the deformation profile. Two Hankel-type loops encircling the branch points \(z_0\) and \(-z_0^*\) are then formed, as shown in Fig.~\ref{fig:steepest}(a). The subsequent steps of the deformation are illustrated in Figs.~\ref{fig:steepest}(c1-c4), while the final placement of the contour on the surface \(\operatorname{Im}\tau(z)\) is shown in Fig.~\ref{fig:steepest}(c.5).

Strictly speaking, this deformation is not yet complete. The contour in Fig.~\ref{fig:steepest}(c.5) still contains the horizontal segment AD, which is not a steepest-descent path. This can be remedied by pushing the segment into the lower half-plane until the contour lies entirely along steepest-descent directions. The resulting construction, however, is unnecessarily convoluted for our purposes: 
the structure of the contour already makes the origin of the leading contribution clear. The two steepest-descent segments approaching and leaving each branch point cancel exactly, because the branch point is encircled twice on the two-sheeted Riemann surface. The surviving leading contribution therefore comes entirely from a small contour winding twice around the branch point and scales as $\sim e^{-2\ve_n{\rm Im}\tau_0}$,  $\tau_0\equiv\tau(z_0)$. The contributions from the vertical segments cancel by periodicity, while the horizontal segment \(AD\), which lies on the lower Riemann sheet, is exponentially subleading,
$ e^{-4\varepsilon_n\operatorname{Im}\tau_0}$.  The remaining segment BC also gives no contribution, since its pre-exponential factor vanishes as \(\phi(i\infty)\to\infty\).

\paragraph{The analytical answer.}
The remaining local contour around each branch point is of the residue type, since the branch point coincides with a pole of the integrand. Evaluating these contributions in Eq.~\eqref{perturb1}, we obtain
\begin{gather}
\label{perturb_semi}
    r = 2\pi a\mu e^{2i\ve_n\tau(z_{+})},\ \ \ve a\gg1.
\end{gather}
This result coincides exactly with the \(\mu a\ll1\) limit of the semiclassical expression~\eqref{reflect1}, since
$\sinh\pi\mu a\rightarrow\pi\mu a$. 
The centers of the forbidden bands can likewise be obtained by expanding the Dykhne's quantization condition~\eqref{action0} in \(\mu/\varepsilon\ll1\), which gives
\begin{gather}
\label{band-analyt}
\begin{split}
    \ve_n &= \frac{\pi n}{L_0}+\frac{\mu^2 L_0L_1}{2(\pi n)^2}+o\left(\frac{1}{n^2}\right),\\
        L_{0}&=\int\limits_0^{L}\frac{dx}{\sqrt{1+\phi^2(x)}},\quad
        L_{1}=\int\limits_0^{L}\frac{dx}{[1+\phi^2(x)]^{3/2}}
\end{split}
\end{gather}
A more accurate value of \(\varepsilon_n\) can be obtained by evaluating the left-hand side of Eq.~\eqref{action0} numerically and solving the resulting equation for \(\varepsilon_n\).

For deformation profiles possessing an appropriate symmetry, such as evenness or oddness, the corresponding generalization of the steepest-descent result~\eqref{perturb_semi} takes the form
\begin{gather}
\label{perturb-steepest2}
    r = 2\pi \mu\sum\limits_{z_i}\sigma a_i e^{2i\ve_n\tau(z_i)},
\end{gather}
where \(a_i^{-1}=\phi'(z_i)\), \(\sigma=\pm1\), with \(i\sigma=\phi(z_i)\), and the sum runs over all roots \(z_i\) lying at the same minimal distance from the real axis.

\begin{figure}
    \centering
    \includegraphics[width=1.0\linewidth]{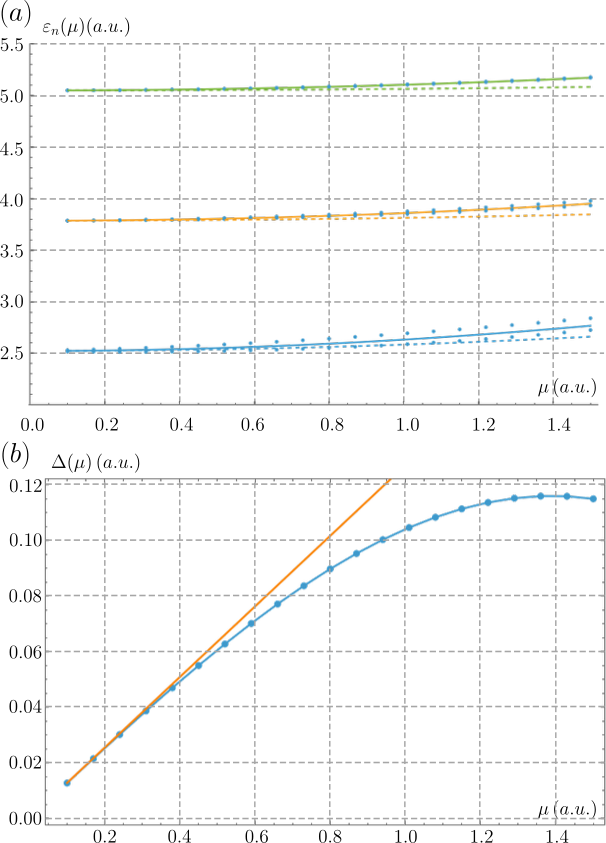}
    \caption{\justifying (a) Dots - bands (upper and lower edges) computed by full numerical integration of the Dirac equation~\eqref{Dirac2} for the single harmonic potential~\eqref{sample_pot} for $n = 2,3,4$. Solid lines - band centers, computed from semiclassical equation~\eqref{action0} by l.h.s. numerical integration. Dashed lines are given by analytical expression~\eqref{band-analyt}. \\
    (b) The bandgap for the $n=2$ band. Dots represent full numerical integration of Dirac equation~\eqref{Dirac2}; Solid linear in $\mu$ orange line is given by Eq.~\ref{perturb_semi}.}
    \label{fig:smallMu}
\end{figure}

Figure~\ref{fig:smallMu} compares direct numerical integration of the original Dirac equation~\eqref{Dirac2}, numerical evaluation of the semiclassical relation~\eqref{action0}, and its approximate analytical form~\eqref{band-analyt}.

The agreement between the steepest-descent perturbative result~\eqref{perturb_semi} and the nonperturbative semiclassical expression~\eqref{reflect1} is so compelling that little room for doubt seems to remain. A surprise emerges, however, when Eqs.~\eqref{reflect1}, ~\eqref{perturb1}, and ~\eqref{perturb_semi} are confronted with numerical results for a multi-harmonic deformation.

\section{Perturbation theory for a multi harmonic profile.}
\subsection{Introduction}
For a multiharmonic deformation $\phi(z)=\sum_n a_n e^{ik_n z},$ or, more generally, for any profile analytic within a periodicity strip, one would normally expect the semiclassical scattering amplitude—and, in the large-\(n\) limit, the perturbative amplitude~\eqref{perturb1}—to be controlled by the relevant roots of \(\phi^2(z)+1=0\) within that strip. The usual semiclassical intuition is that the dominant contribution should come from the roots closest to the real axis, since these normally correspond to the least exponentially suppressed scattering processes. For all cases considered so far, this geometrical rule has worked reliably.

\begin{figure}
    \centering
    \includegraphics[width=1.0\linewidth]{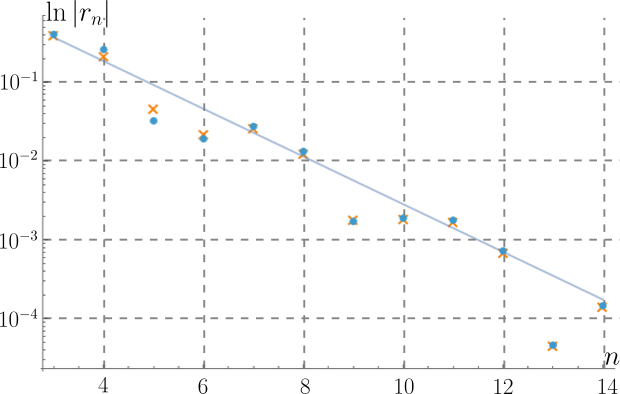}
    \caption{\justifying Blue dots represent the scattering amplitude computed by numerical evaluation of the integral in the r.h.s. of~\eqref{perturb_semi}. Orange crosses represent the analytical approximation~\eqref{perturb-steepest2}. Blue line is simply $-2 \ve_n{\rm Im\,}\tau_0$, where $\tau_0$ is defined in ~\eqref{perturb-semi2}.}
    \label{fig:function02}
\end{figure}

For additional confidence that everything works as expected, Fig.~\ref{fig:function02} compares the numerically evaluated scattering amplitude~\eqref{perturb1} with its steepest-descent approximation~\eqref{perturb-steepest2} for the simplest multiharmonic generalization of the single-harmonic profile, the two-harmonic deformation
\begin{gather}
\label{phi1}
    \phi_1(z) = a_1 \cos z + a_2\cos 2z,
\end{gather}
where $\ a_1\approx 1.17,\ a_2 = 1.0$.
As before, owing to the evenness of the deformation profile, two symmetry-related branch points closest to the real axis contribute to the large-\(n\) asymptotics of the scattering amplitude~\eqref{perturb1}, yielding
\begin{gather}
\label{perturb-semi2}
\begin{split}
    r_n &= 4\pi|a_0|\mu e^{-2\ve_n{\rm Im}\tau_0}\cos \left(2\ve_n{\rm Re}\tau_0+\arg a_0\right),\\
    \tau_0 &= \int_0^{z_0} \frac{dz}{\sqrt{1+\phi^2(z)}} 
\end{split}
\end{gather}
where $a_0^{-1} \equiv \phi'(z_0)$.

\begin{figure}
    \centering
    \includegraphics[width=1.0\linewidth]{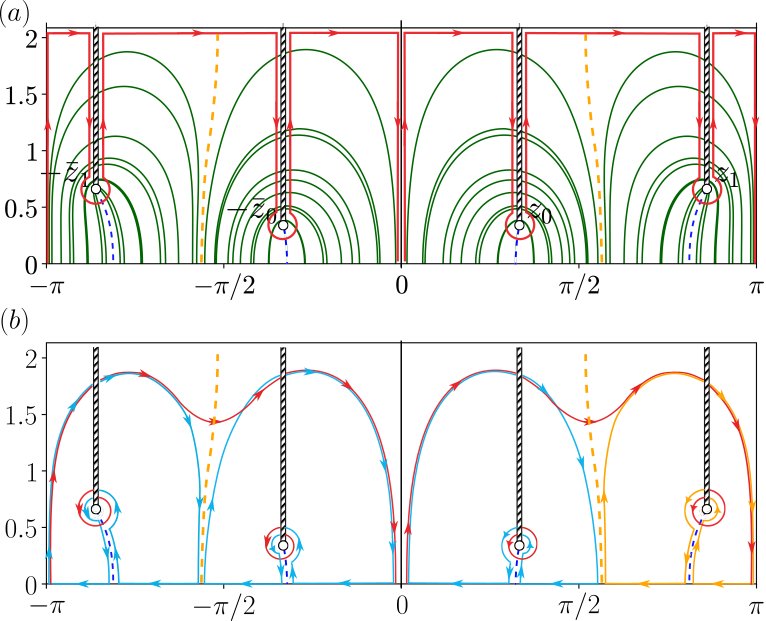}
    \caption{\justifying (a) The initial integration contour (red) and portrait of the steepest descent lines (green) for the function ${\rm Im}\,\tau(z)$ for the profile deformation $\phi_1(z)$ defined in Eq.~\ref{phi1}. The dashed steepest descent orange line is the separatrix.\\
    (b) The placement of the integration contour along the steepest descent lines of ${\rm Im}\,\tau(z)$.  
    }
    \label{fig:steepest3}
\end{figure}

To construct the steepest-descent approximation, we use a contour deformation analogous to that of Fig.~\ref{fig:steepest}(c.1--c.4), which isolates the contribution of each branch point. A crucial ingredient is the existence of a separatrix between every neighboring pair of branch points: a steepest-descent line extending to \(+i\infty\) onto which the final integration contour can be placed. For the single-harmonic profile these separatrices are the lines AB and CD in Fig.~\ref{fig:steepest}(b), together with the vertical boundaries \(\operatorname{Re}z=0\). For the profile~\eqref{phi1}, the corresponding separatrices are shown by the dashed orange lines in Fig.~\ref{fig:steepest3}, together with \(\operatorname{Re}z=0,\pm\pi\).

Fig.~\ref{fig:function02} compares Eq.~\eqref{perturb-semi2} with the numerical evaluation of Eq.~\eqref{perturb1} on a logarithmic scale. The plotted absolute slope is governed by the leading exponential \(\exp({-2\varepsilon_n\operatorname{Im}\tau_0})\), while the oscillatory modulation originates from interference between the two symmetry-related branch points (cosine term in Eq.~\ref{perturb-semi2}). The agreement is excellent.

For clarity and for the use in the next section, we present the portrait of the steepest descent lines and the respective contour deformation in Fig.~\ref{fig:steepest3}. 


\subsection{The paradox}
\begin{figure}
    \centering
    \includegraphics[width=1.0\linewidth]{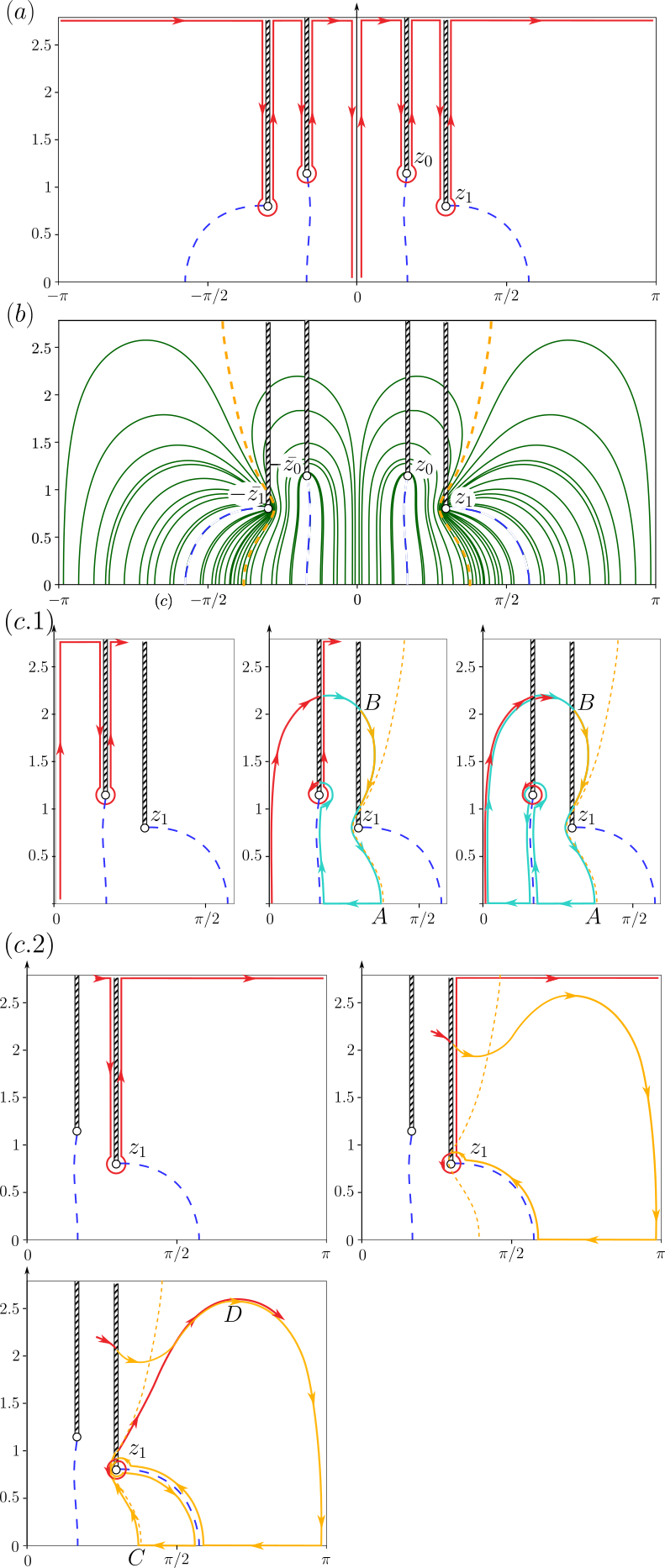}
    \caption{\justifying (a) Initial deformation of the contour for the profile function $\phi_2(z)$ (see Eq.~\ref{paradoxField}).\\ 
    (b) The portrait of the steepest descent lines for the function ${\rm Im}\,\tau(z)$ for the same deformation profile.\\ (c.1-c.2) Deformation of the right half of the symmetric integration contour along the steepest descent lines of the function ${\rm Im}\,\tau(z)$. The red curves lie on the initial Riemann sheet, while the turquoise curves lie on different Riemann sheets.}
    \label{fig:steepest2}
\end{figure}

We now turn to the most intriguing multiharmonic case: a profile \(\phi_2(z)\) tuned to satisfy condition~\eqref{condition}. As discussed above, such deformations are of particular interest because the semiclassical result~\eqref{reflect1} predicts magnetic quantum oscillations of the band gaps. As a representative example, we consider
\begin{gather}
\label{paradoxField}
    \phi_2(z) = \phi_0(z)+\cos 2z.
\end{gather}
The corresponding steepest-descent pattern of \(-\operatorname{Im}\tau(z)\) is shown in Fig.~\ref{fig:steepest2}(b). Condition~\eqref{condition} qualitatively changes both the arrangement of the branch points and the topology of the steepest-descent lines. 
In particular, the steepest-descent line emanating from the relevant branch point \(z_0\) now initially runs horizontally and reaches the real axis after a horizontal displacement \(\Delta x_1\sim\operatorname{Im}z_1\sim L\). Consequently, neighboring branch points are displaced away from the boundaries of the periodicity strip and approach one another in the horizontal direction, $|{\rm Re}(z_1-z_0)|\ll L$. 
The steepest-descent separatrix between \(z_0\) and \(z_1\) is therefore strongly squeezed, as shown by the dashed orange line in Fig.~\ref{fig:steepest2}(b). As we shall see, this seemingly innocuous geometrical rearrangement has dramatic consequences for the scattering amplitude.

\begin{figure}
    \centering
    \includegraphics[width=1.0\linewidth]{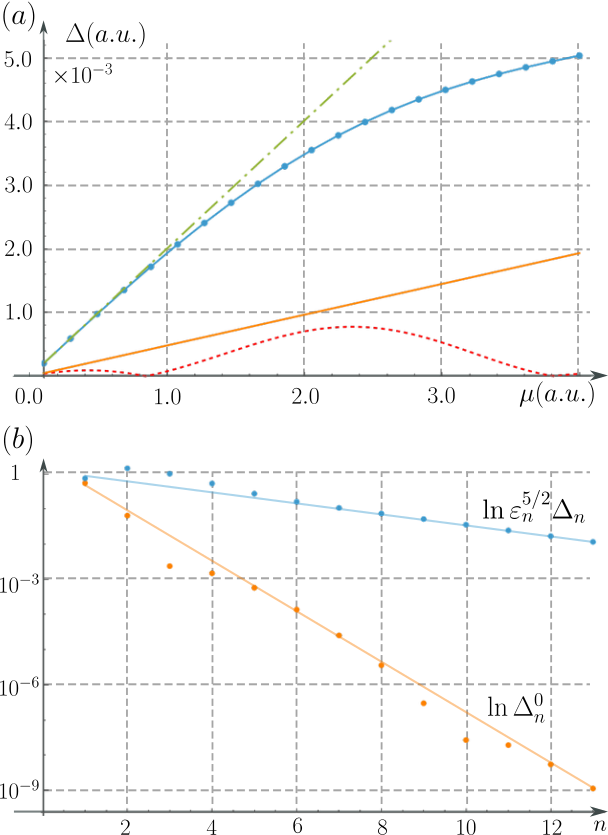}
    \caption{\justifying 
    (a) Bandgap for the $n=5$ band for the deformation profile $\phi_2(z)$ given by Eq.~\ref{paradoxField}. Connected blue dots show the result of direct numerical integration of the Dirac equation~\eqref{Dirac2}; the green dot-dashed line is obtained by numerical evaluation of the perturbative expression~\eqref{perturb1}; the solid orange line, linear in \(\mu\), corresponds to the steepest-descent result~\ref{perturb_semi}; and the dashed red line shows the semiclassical result~\ref{reflect1}.\\
    (b) Blue dots show \(\ln(\varepsilon_n^{5/2}\Delta_n)\), obtained by numerical evaluation of the scattering integral~\eqref{perturb1} as a function of the forbidden-band index $n$. Orange dots show \(\ln\Delta_n^{0}\), obtained from the steepest-descent approximation~\eqref{perturb_semi}, including the two branch points closest to the real axis and symmetric with respect to the imaginary axis. Straight lines are included to emphasize the approximately linear behavior of both dependencies. The slope of the blue line is determined by \(\operatorname{Im}\tau_{i\infty}\). 
    }
    \label{fig:smallMuParadox}
\end{figure}

Nevertheless, the integration contour can still be deformed onto the steepest-descent network, as illustrated in Fig.~\eqref{fig:steepest}(c.1-c.2). Despite its unusual appearance, the resulting contour seems to retain the same essential structure as before: each relevant branch point is encircled twice, and the scattering amplitude appears to reduce once again to the corresponding residue contributions from the branch points closest to the real axis.
Combining Eqs.~\ref{dykhne2} and ~\eqref{perturb_semi} we obtain:
\begin{gather}
    \Delta^0_n = \frac{4}{L_0} \left|\sin(\pi|a|\mu)\,\sin \left(2\ve_n{\rm Re}\tau_0\right)\right|e^{-2\ve_n{\rm Im}\tau_0}, 
\end{gather}
where $\tau_0 = \int_0^{z_{1}}[1+\phi_2^2(z)]^{-1/2}dz$ and $L_0$ is defined in~\eqref{band-analyt}.
Quite unexpectedly, however, the otherwise convincing agreement between the analytical results and the numerical calculations now breaks down completely, as seen in Fig.~\ref{fig:smallMuParadox}(a-b). Fig.~\ref{fig:smallMuParadox} compares the steepest-descent result~\eqref{perturb-steepest2}, the semiclassical expression~\eqref{reflect1}, and direct numerical integration of the original Dirac equation~\eqref{Dirac2}. The discrepancy reaches several orders of magnitude and is especially clear in the logarithmic plot of Fig.~\ref{fig:smallMuParadox}(b). 
Moreover, the numerical result shows no trace of the predicted oscillations.
At the same time, Fig.~\eqref{fig:smallMu}(a) still shows excellent agreement between the band gap obtained from direct numerical solution of the Dirac equation~\eqref{Dirac2} and from the numerical evaluation of the perturbative expression~\eqref{perturb1}.
Thus, both the semiclassical expression~\eqref{reflect1} and the steepest-descent approximation~\eqref{perturb_semi} fail badly, whereas the full perturbative expression remains perfectly accurate.

A clue comes from the numerical data. Plotting \(\ln(\varepsilon_n^{5/2}\Delta_n)\), rather than simply \(\ln\Delta_n\), reveals an almost linear dependence on \(n\). This suggests that the steepest-descent exponential behavior
$\ln |r|\sim -2\varepsilon_n\,\operatorname{Im}\tau(z_*) $
is accompanied by an additional pre-exponential factor scaling as \(\varepsilon_n^{-5/2}\). To uncover the analytic origin of this behavior, we must answer two questions:\\
(i) Why the previous steepest descent approach  stopped working?\\
(ii) Which region contribute to the integral?

\subsubsection{The breakdown of the ordinary steepest descent approximation}
The steepest-descent portrait in Fig.~\ref{fig:steepest2} already hints at the origin of the problem. The separatrix contour shown by the dashed orange line is squeezed between the two nearby branch points \(z_0\) and \(z_1\) and passes dangerously close to \(z_1\). Moreover, as discussed above, each branch point is simultaneously a first-order pole of the pre-exponential factor. The pole can therefore no longer be ignored when analyzing the steepest descent behavior in the vicinity of the branch point.

To expose the consequences, we treat the exponential and pre-exponential factors on equal footing and expand their logarithm near the branch point, \(z=z_1+\zeta\), \(|\zeta|\ll L\) [see Eq.\ref{a-def}]:
\begin{gather}
\label{exponent}
\begin{split}
    f(z) &= 2i\ve_n\tau(z)-\ln[\phi^2(z)+1]\\
    &=2i\ve_n\tau_0 +\underbrace{\left[2i\ve_n\sqrt{2|a|\zeta}-\ln\frac{\zeta}{|a|}\right]}_{\rm I+II}+...
\end{split}
\end{gather}
At first sight, the two terms in brackets appear to possess the usual semiclassical hierarchy, since their ratio is of order \(\varepsilon_n a\gg1\). However, the rescaling $ s=2\varepsilon_n\sqrt{2a\zeta}$ eliminates the large parameter altogether. To leading order,
\begin{gather}
    \label{exponent2}
    f(s) = {\rm const}+is-2\ln s,
\end{gather}
which has a saddle at \(s=-2i\), corresponding to \(\zeta=-1/(2\ve_n^2 a)\), parametrically close to the branch point. Thus, as condition~\eqref{condition} is approached, the usual separation between the exponential and pre-exponential factors breaks down in the branch-point region. Consequently, the previous argument that allowed us to discard the integrals along the separatrices also ceases to apply.

This observation explains why the previous steepest-descent construction becomes unreliable, but it does not yet solve the problem. Figure~\ref{fig:smallMuParadox}(b) shows that the dominant scattering contribution does not originate from the immediate vicinity of the branch point. Some other region of the separatrix must control the leading asymptotics. We are therefore led to a second question: where is the true accumulation region of the scattering integral?
\subsubsection{Hidden saddle. Alternative steepest descent expansion for the general harmonic profile.}
\paragraph{Preliminary remarks.}
The nearly linear behavior of the numerical curve in Fig.~\ref{fig:smallMuParadox} strongly suggests that some form of steepest-descent asymptotics must still apply: the scattering integral is governed by the rapidly varying exponential \(\exp[2i\varepsilon_n\tau(z)]\). Yet no ordinary finite stationary point of the phase exists, since $\tau'(z)=1/{\sqrt{\phi^2(z)+1}}$
never vanishes at a regular finite point. This leaves an unusual possibility: the dominant contribution may be associated with complex infinity. Indeed, a nonconstant entire periodic deformation cannot remain bounded in the complex plane, since otherwise Liouville's theorem would make it a constant. For the finite harmonic profiles considered here, this unboundedness manifests itself as \(|\phi(z)|\to\infty\) for \(z\to i\infty\), and consequently \(\tau'(z)\to0\). We parameterize the approach to complex infinity as \(z=b+iy\), \(y\to\infty\), where \(b\) lies within the periodicity strip. Thus, the variation of the phase becomes arbitrarily slow near complex infinity, making this region a natural candidate for the origin of the missing asymptotic contribution.

Which value of \(b\) is relevant depends on the deformation profile. At \(z\to i\infty\), \(\tau(z)\) approaches different constants in the sectors separated by the branch cuts; the dominant contribution is associated with the sector having the smallest \(\operatorname{Im}\tau_\infty\). For the two-harmonic profile \(\phi_2(z)\), the steepest-descent line approaching \(\pi+i\infty\) gives $\operatorname{Im}\tau_\infty=0.140766, $
precisely reproducing the slope of the numerical curve in Fig.~\ref{fig:smallMuParadox}(b). As we shall show below, this agreement is not accidental: the local structure of the integrand near complex infinity indeed determines the scattering asymptotics.

\paragraph{The saddles at infinity.}
To determine this contribution, we examine the analytic structure of the scattering integral~\eqref{perturb1} near \(z\to i\infty\). Although the discussion begins with the two-harmonic profile, the same construction extends with only minor changes to an arbitrary finite harmonic expansion. For simplicity, we take the deformation to be \(2\pi\)-periodic.

The numerical results already indicate that the pre-exponential factor plays an essential role in the asymptotics. We therefore absorb it into the exponent in the following analysis. This produces a stationary point of the full integrand and helps identify the region in which the integral is accumulated. However, no large parameter controls the expansion around this point, so it cannot be used for a conventional saddle-point approximation. Keeping the pre-exponential factor in the exponent therefore does not simplify the actual evaluation of the integral, but it reveals its underlying analytic structure.
For a general finite harmonic profile, the leading behavior at \(z\to i\infty\) can be written as
\begin{gather}
\label{largeZ}
    \phi_2(z) \underset{z\rightarrow i\infty}{=} Ae^{-i\gamma z}+...,\ \gamma\in \mathbb{N}.
\end{gather}
Including the logarithm of the pre-exponential factor, the saddle-point condition becomes
\begin{gather}
\label{saddleLarge}
    \frac{2i\ve}{\sqrt{\phi^2+1}}-\frac{2\phi\phi^\prime}{\phi^2+1} = 0.
\end{gather}
In the semiclassical limit \(\varepsilon\to\infty\), the stationary points move toward \(i\infty\), where Eq.~\ref{saddleLarge} simplifies to
\begin{gather}
\label{saddles}
    \phi^\prime = i\ve\ \rightarrow z_n =\frac{\pi}{\gamma}+\frac{\pi n}{\gamma}+i\ln\frac{\ve}{\gamma A}. 
\end{gather}
The positions~\eqref{saddles} are universal in the sense that, to leading order, they depend only on the highest harmonic of the deformation profile. There is, however, an important distinction between the previous \textit{nonparadoxical} profile and the one considered here. Comparing Figs.~\ref{fig:steepest3}(a) and~\ref{fig:steepest2}(a), we see that the rearrangement of the branch points, i.e., the roots of \(\phi^2(z)+1=0\), qualitatively changes the analytic structure of the integration domain. For the present profile, an analytic, branch-cut-free vertical strip opens between two adjacent stationary points, with real parts \(\pi/\gamma\) and \(3\pi/\gamma\). The integration contour can therefore be pushed upward to \(+i\infty\), made to pass through the left stationary point, and then returned to the real axis through the right one. Thus, the two vertical legs of the contour pass through two distinct stationary points. Since these stationary points themselves approach \(i\infty\) in the semiclassical limit, the dominant contribution to the integral originates from the vicinity of complex infinity. 

\begin{figure}
    \centering
    \includegraphics[width=1.0\linewidth]{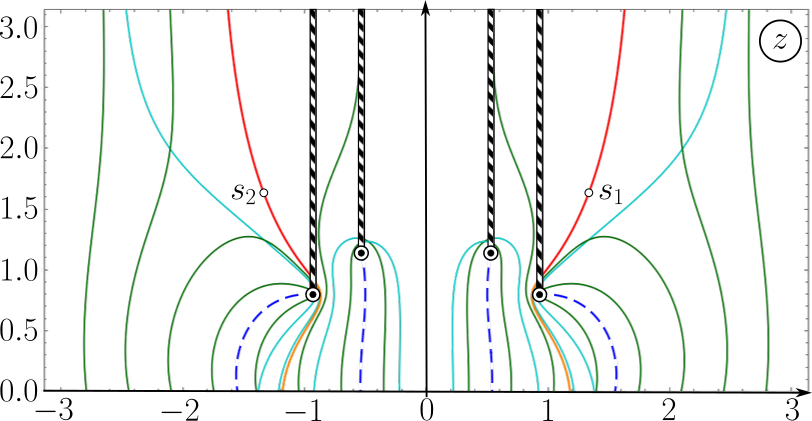}
    \caption{\justifying Phase portrait of the steepest-descent lines of the full exponent $f(z)=2i\varepsilon\tau(z)-\ln[\phi_2^2(z)+1] $ for the deformation profile \(\phi_2(z)\) defined in Eq.~\eqref{paradoxField}. Green curves belong to the original Riemann sheet of \(f(z)\), while cyan curves represent their continuation onto the lower sheet. The points \(s_{1,2}\) denote the stationary points. The red-orange curves are the steepest-descent lines passing through them; their orange segments lie on the lower Riemann sheet. The plot is shown for \(\varepsilon=26\).}
    \label{fig:steepestFull}
\end{figure}

The stationary points are shown in the phase portrait of the full exponent \(f(z)\) in Fig.~\ref{fig:steepestFull}.
Its second derivative determines the local convergence of the integral. At the stationary point one finds
$f^{\prime\prime} = 2\gamma^2$ and
$f^{\prime\prime\prime} = 2i\gamma^3$. The first relation shows that the relevant region has a width of order \(1/\gamma\), with the corresponding steepest-descent direction being vertical. The second, however, reveals the absence of any large parameter controlling the local expansion. Consequently, although the stationary point identifies the region in which the integral is accumulated, a conventional saddle-point approximation cannot be used.
\paragraph{The placement of the contour.}
An important question is whether the original integration contour can indeed be deformed so as to pass through both stationary points. Comparing Figs.~\ref{fig:steepestFull}  and ~\ref{fig:steepest2}(b), we see that the steepest-descent lines passing through them are precisely the former separatrices obtained when the pre-exponential factor was neglected. Once the pre-exponential factor is incorporated into the exponent, each such separatrix turns into a steepest-descent line passing through a stationary point of the full integrand. Therefore, the contour constructed previously in Fig.~\ref{fig:steepest2}(c1-c2) remains perfectly suitable for the present analysis.

\paragraph{The contribution from the saddle.}
As argued above, the integral is accumulated within an \(\mathcal{O}(1/\gamma)\) neighborhood of the stationary points \(s_{1,2}\), where the asymptotic form~\eqref{largeZ} is valid. To expose the structure of this contribution, we parameterize the steepest-descent contour passing through the right stationary point by
$z = \pi/\gamma+iy$: $s = e^{-\gamma y}$. 
Along the contour, 
\begin{gather}
    \begin{split}
    &\tau(z)=\tau_{i\infty}+\int\limits_{i\infty}^z \frac{ds}{\sqrt{\phi^2(z)+1}}\\
    &\underset{z\rightarrow i\infty}{=}
    \tau_{i\infty}+\frac{1}{A}\int\limits_{i\infty}^z e^{i\gamma s}\,ds+...
    =\tau_{i\infty}+\frac{e^{i\gamma z}}{i\gamma A}+...\\
    &\tau_{i\infty} -\frac{s}{i\gamma A}
    \end{split}
\end{gather}
Then the integral becomes:
\begin{gather}
\label{eq-saddle}
    \int\limits_{\rm right\ steepest\atop descent} = -\frac{i e^{2i\ve\tau_{i\infty}}}{\gamma A^2}\int\limits_0^{s_0}  e^{-\frac{2\ve s}{\gamma A}}sds
\end{gather}
where the saddle is passed in the downwards direction. 
The lower integration limit \(s_0\) corresponds to the point at which the asymptotic representation~\eqref{largeZ} ceases to be valid: $s\ll 1$. From Eq.~\eqref{eq-saddle}, however, the integral is seen to accumulate in the interval \(s\sim\gamma/\varepsilon\ll1\). We may therefore extend the upper limit to infinity without affecting the leading result. Although the integral cannot be evaluated by a local saddle-point approximation, it can be calculated exactly, yielding
\begin{gather}
\label{contribRight}
    \int\limits_{\rm right\ steepest\atop descent} = -\frac{i\gamma}{4\ve^2}e^{2i\ve\tau_{i\infty}}.
\end{gather}
However, the leading contributions from the left and right saddles~\eqref{contribRight} cancel exactly, since the corresponding steepest-descent contours pass through the two saddles in opposite directions. This cancellation is universal: within the leading asymptotic approximation~\eqref{largeZ} for the deformation profile, Eq.~\eqref{contribRight} gives equal contributions of opposite sign. A nonzero scattering amplitude can therefore arise only from subleading terms in the asymptotic expansion of \(\phi(z)\).

\paragraph{Next-to-leading order contribution.}
To obtain the first nonvanishing contribution, we now
consider a finite harmonic profile whose two highest
harmonics are nonzero and have consecutive harmonic
numbers, $\gamma$ and $\gamma-1$, with $\gamma\geq2$.
Near $z\to i\infty$, its asymptotic expansion reads
\begin{gather}
\label{subleading}
    \begin{split}
    \phi_2(z) &= A e^{-i\gamma z}+b e^{-i(\gamma-1)z}+...\\
    &=-Ae^{\gamma y}\left[1+\frac{b}{A}e^{\frac{i\pi}{\gamma}-y}+...\right]
    \end{split}
\end{gather}
The correction term is of order \((\gamma/\varepsilon)^{1/\gamma}\ll1\). The integral along the right steepest-descent contour then takes the form
\begin{gather}
    \begin{split}
    &\int_{\rm right\ steepest\atop contour} = -e^{2i\ve\tau_{i\infty}} \frac{2bi}{A^3\gamma}e^{i\pi/\gamma}\\
    &\times\int\limits_0^{\infty}\left(\frac{\ve s}{A(\gamma+1)}-1\right)s^{1+1/\gamma}\exp\left(-\frac{2\ve}{\gamma A}s\right)ds
    \end{split}
\end{gather}
Evaluating the integral in terms of $\Gamma$-functions, we obtain
\begin{gather}
    \begin{split}
    &\int_{\rm right\atop left} = \pm\frac{bi}{4\ve^2A\gamma}\Gamma\left(\frac{1}{\gamma}\right)\left(\frac{\gamma A}{2\ve}\right)^{1/\gamma}e^{\pm\frac{i\pi}{\gamma}}e^{-2\ve {\rm Im}\tau_{i\infty}}(-1)^n
    \end{split}
\end{gather}
The full scattering amplitude~\eqref{perturb1} is therefore
\begin{gather}
\label{amp_correct_gen}
    r(\ve_n) = (-1)^n\frac{b\mu}{2\ve^2A\gamma}\sin\frac{\pi}{\gamma}\Gamma\left(\frac{1}{\gamma}\right)\left(\frac{\gamma A}{2\ve}\right)^\frac{1}{\gamma}e^{-2\ve{\rm Im\tau_{i\infty}}}.
\end{gather}
Equation~\eqref{amp_correct_gen} is the key result for the scattering amplitude. To compare it with our numerical results for the profile~\eqref{paradoxField}, we set \(\gamma=2\), which yields the particularly compact expression
\begin{gather}
    \label{amp_correct}
    r(\ve_n) = (-1)^n\frac{b\mu}{4}\sqrt{\frac{\pi}{A}}\frac{1}{\ve^{5/2}}e^{-2\ve_n{\rm Im\tau_{i\infty}}}.
\end{gather}
Its pre-exponential dependence, \(\varepsilon^{-5/2}\), precisely reproduces the scaling inferred earlier from the numerical data. 
\begin{figure}
    \centering
    \includegraphics[width=1.0\linewidth]{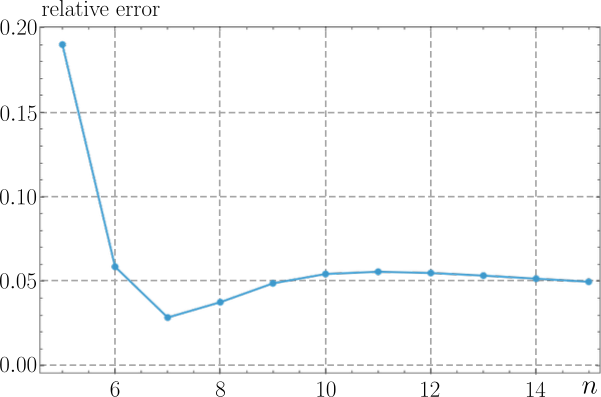}
    \caption{\justifying Comparison of the analytical steepest-descent result~\eqref{amp_correct} with the direct numerical evaluation of the perturbative integral~\eqref{perturb1}. The vertical axis shows the relative error.}
    \label{fig:org01}
\end{figure}
Figure~\ref{fig:org01} compares Eq.~\eqref{amp_correct} with the direct numerical evaluation of the perturbative integral~\eqref{perturb1}. The relative error falls below \(6\%\) already for \(n\ge6\).

Equations~\eqref{amp_correct_gen} and~\eqref{amp_correct}
thus complete our perturbative analysis of the scattering
mechanism controlled by complex infinity. The decisive
feature is not the detailed harmonic content of the
deformation profile, but the global analytic structure of
the scattering integral. In particular, a sufficient
geometric condition for complex infinity to provide the
leading contribution is the existence of an analytic,
branch-cut-free strip of width $2\pi/\gamma$, where
$\gamma$ is the highest harmonic number in the
trigonometric expansion. Such a strip allows the integration
contour to be deformed toward complex infinity, where the
leading contributions cancel and the first nonvanishing
term is determined by the subleading asymptotics of the
deformation profile.

For the class of profiles described by
Eq.~\eqref{subleading}, this mechanism yields
Eq.~\eqref{amp_correct_gen}. The adjacent harmonic with
coefficient $b$ determines the leading nonzero
pre-exponential factor. If this harmonic is absent, the
same mechanism can still operate, but the first nonvanishing
contribution must be obtained from the next available term
in the asymptotic expansion.

\section{SEMICLASSICAL SCATTERING AT COMPLEX INFINITY}
\subsection{Preliminaries. Anti-Stokes lines at $z=i\infty$}
We now return to weak-field scattering beyond the perturbative regime, \(\mu\ll\varepsilon\) but \(\mu L\gtrsim1\). As before, for pedagogical clarity, we use the two-harmonic deformation profile as a representative example of the simplest nontrivial multiharmonic case. The construction extends to a general multiharmonic profile, and we indicate the necessary modifications where they arise. 
Near $z=i\infty$ the deformation profile is dominated by its highest harmonic ~\eqref{largeZ}. 
The semiclassical action therefore behaves as
\begin{gather}
\label{action_fin}
    \int\limits_{i\infty}^z \pi(z)\,dz \underset{z\rightarrow i\infty}{=} \ve\int\limits_{i\infty}^z\frac{dz}{\phi(z)} \approx\frac{\ve}{iA\gamma}e^{i\gamma z}
\end{gather}
Choosing \(A>0\) without loss of generality, Eq.~\eqref{action_fin} yields the asymptotic anti-Stokes and Stokes lines
\begin{gather}
\begin{split}
{\rm Re}\,z_{n} &= \frac{\pi}{2\gamma}+\frac{\pi n }{\gamma}\quad -\hbox{anti-Stokes lines},\\
    {\rm Re}\,z&=\frac{\pi n}{\gamma}\quad -\hbox{Stokes lines.}
\end{split}
\end{gather}
The semiclassical description requires the phase accumulated from complex infinity to be large. At the same time, the asymptotic form of \(\phi(z)\) requires \(e^{i\gamma z}\) to remain small. The overlap region is therefore
\begin{gather}
\label{condition2}
    \frac{\gamma}{\ve}\ll |e^{i\gamma z}|\ll 1
\end{gather}
For \(\gamma=2\), four anti-Stokes lines lie within one \(2\pi\)-periodicity strip, at \(\operatorname{Re}z=\pi/4+\pi n/2\).

\subsection{Dirac equation near $i\infty$}
For the analysis near complex infinity, it is convenient to use the same unitary transformation $P$ ~\eqref{unitary} introduced in the perturbative treatment, supplemented by the ansatz
\begin{gather}
    \psi = \frac{1}{\sqrt{2v(x)}}\exp\left(-i \mu\int\frac{dx\phi(x)}{1+\phi^2}\right)P[\sigma_z+\sigma_x]\chi(x).
\end{gather}
Then, the transformed Dirac equation becomes:
\begin{gather}
\label{dirac3}
    \begin{split}
    -\de_\tau^2\chi_{\sigma}+(m^2+\sigma\de_\tau m)\chi_\sigma &= \ve^2\chi_\sigma,\\
    m &= \frac{\mu}{v(\tau)}.
    \end{split}
\end{gather}
where, as before, \(d\tau=dx/v(x)\) and \(\sigma=\pm1\). Equation~\eqref{dirac3} is particularly convenient for analyzing the large-\(|\phi|\) region near complex infinity.

\begin{figure}
    \centering
    \includegraphics[width=1.0\linewidth]{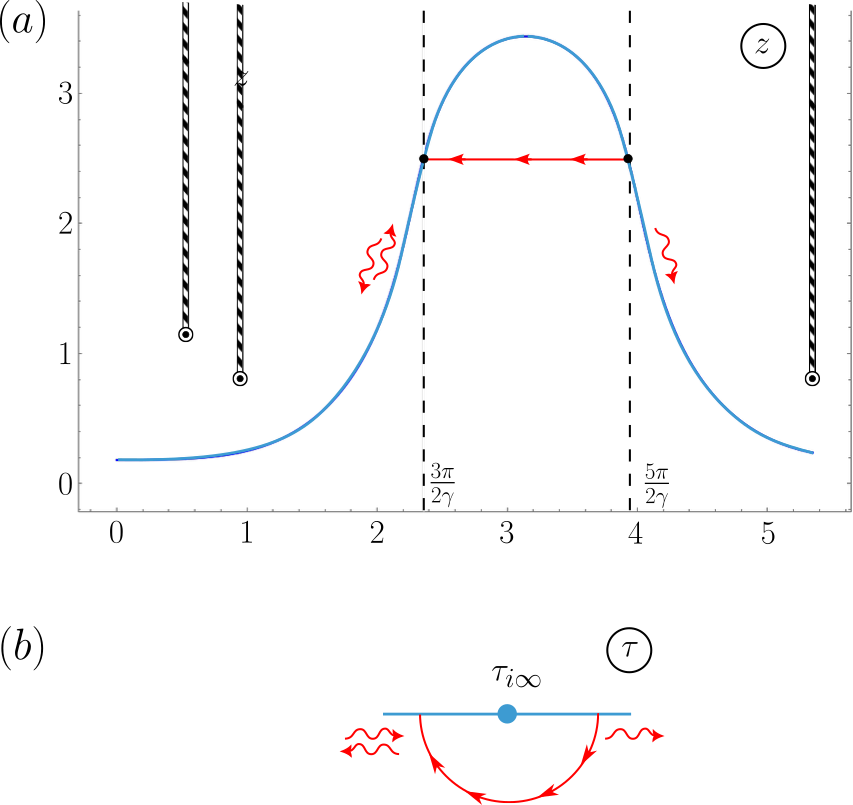}
    \caption{\justifying (a) Two anti Stokes lines passing through points $2.5+3\pi/2\gamma$ and $2.5+5\pi/2\gamma$. As one clearly sees they meet at Stokes-line at $2\pi/\gamma +i\infty$. The horizontal arrow indicates the translation between the two lines near \(i\infty\). (b) Mapping of the same anti-Stokes lines onto the \(\tau\) plane near complex infinity. The translation in the \(z\) plane becomes a rotation by exactly \(\pi\) in the \(\tau\) plane.}
    \label{fig:anti-Sotkes-fin}
\end{figure}

The perturbative analysis has already provided an important clue as to where the scattering amplitude is accumulated. The relevant contributions were associated with the asymptotic regions lying on the lines \(\operatorname{Re}z=\pi/\gamma\) and \(2\pi/\gamma\). As before, the scattered wave arises through analytic continuation of the transmitted solution from one anti-Stokes line to another. The relevant anti-Stokes pattern for the deformation profile \(\phi_2(z)\) is shown in Fig.~\ref{fig:anti-Sotkes-fin}(a). 
The two anti-Stokes lines approaching the asymptotic directions \(\operatorname{Re}z=3\pi/(2\gamma)\) and \(5\pi/(2\gamma)\), although nearly vertical at large \(\operatorname{Im}z\), meet at complex infinity at \(\operatorname{Re}z=2\pi/\gamma\). To clarify the geometry of this scattering process, let us briefly examine the mapping from the \(z\) plane to the \(\tau\) plane. From Eq.~\eqref{perturb1}, near \(i\infty\) we have
\begin{gather}
\label{tauPhi01}
    \tau(z) \approx \tau_{i\infty}+\int_{i\infty}^z\frac{1}{\phi(z)}\approx\tau_{i\infty}+\frac{e^{i\gamma z}}{Ai\gamma}
\end{gather}
and the change of anti-Stokes lines is mapped into $\tau$ - domain as:
\begin{gather}
    \begin{split}
    z(\vf) &= \frac{5\pi+\vf}{2\gamma}+iy,\ \ \vf\in[0,-\pi]\\
    \tau(\vf) &= \frac{e^{-\gamma y}}{A\gamma}e^{i\vf}
    \end{split}
\end{gather}
Thus, a translation between the two asymptotic anti-Stokes lines in the \(z\) plane becomes a clockwise rotation in the \(\tau\) plane. In the present case the rotation angle is exactly \(\pi\), as illustrated in Fig.~\ref{fig:anti-Sotkes-fin}(b).

\subsection{Scattering near $z=i\infty$}
\subsubsection{Leading exponential approximation}
We begin the analysis of the exact Dirac equation~\eqref{dirac3} near complex infinity by retaining only the leading exponential of the deformation profile $\phi(z) =  A e^{i\gamma z}$. Using Eq.~\eqref{tauPhi01} we observe that $\tau$ is related to $\phi(z)$ via a simple transformation:
\begin{gather}
    \tau-\tau_{i\infty} = \frac{1}{i\gamma\phi(z)}
\end{gather}
The Dirac equation~\eqref{dirac3} consequently reduces to
\begin{gather}
\label{weber}
    \begin{split}
    -\de_t^2\chi_\sigma -\gamma^2\mu^2t^2\chi_\sigma &= (\ve^2-\mu i\gamma\sigma)\chi_{\sigma},\\
t &= \tau-\tau_{i\infty}
    \end{split}
\end{gather}
Equation~\eqref{weber} is a Weber equation. However, in the weak-field regime \(\mu/\varepsilon\ll1\), its \(t\)-dependent term is parametrically small. In the matching region relevant here [see Eqs.~\eqref{condition2}, ~\eqref{tauPhi01}], \(|t|\sim1/\varepsilon\), the corresponding asymptotic does not undergo a Stokes transition, and no reflected wave is generated. A nonzero Stokes jump appears only at much larger distances, \(|t|\gg1/\mu\) (see Appendix~\ref{app:weber_stokes}), where the leading-exponential approximation of the deformation profile is no longer valid. Thus, retaining only the highest harmonic is insufficient to describe the scattering. This is precisely analogous to the perturbative calculation, where the leading exponential of \(\phi(z)\) likewise produced a vanishing scattering amplitude.

\subsubsection{Adding subleading exponential}
We therefore retain the first nonvanishing subleading
exponential in the asymptotic expansion~\eqref{subleading}.
For the class of profiles considered here, this term is the
adjacent harmonic with coefficient $b$. Although subleading
in $\phi(z)$, it generates the dominant $t$-dependent term
in the differential equation: its characteristic scale,
$\sim(\mu/L)(\varepsilon L)^{-1/\gamma}$, parametrically
exceeds that of the quadratic Weber term,
$\sim(\mu/\varepsilon)^2L^{-2}$. The latter was retained in Eq.~\eqref{weber} only because, within the leading-exponential approximation, it is the first \(t\)-dependent term that could in principle generate scattering. We thus obtain
\begin{gather}
\label{diracFinal}
    -\chi_\sigma^{\prime\prime} - \frac{b}{A}\mu\sigma e^{i\pi\gamma/2}(\gamma A)^{1/\gamma}t^{1/\gamma}\chi_\sigma = (\ve^2 -\mu i \gamma\sigma)\chi_\sigma.
\end{gather}
Here the quadratic Weber term has been discarded in comparison with the subleading contribution proportional to \(t^{1/\gamma}\). On the right-hand side, the term \(\mu\gamma\) can likewise be neglected relative to \(\varepsilon^2\) in the limit \(\mu\ll\varepsilon\); retaining it would only generate a relative correction of order \(\mu/\varepsilon\).

The resulting equation can be treated perturbatively in the \(t^{1/\gamma}\) term. 
At first sight, the appearance of perturbation theory inside a semiclassical construction may seem unusual. Near complex infinity, however, a new small parameter emerges naturally, \(1/\phi\ll1\). Since \(1/\phi\propto t^{1/\gamma}\), the local equation~\eqref{diracFinal} can be treated perturbatively in \(t^{1/\gamma}\), independently of the semiclassical expansion itself. Simple dimensional analysis shows that the corresponding dimensionless expansion parameter is $(\mu/\ve)(\ve L)^{-1/\gamma-1}$.
It is most convenient to work directly in the variable \(t=\tau-\tau_{i\infty}\). 
In the absence of the \(t^{1/\gamma}\) term, the solutions are simply the two plane waves $\chi = \exp(\pm i\ve\tau)$ and no scattering occurs. To construct the perturbative solution, we use the method of variation of constants; an equivalent Green-function representation offers no particular advantage in the present complex-plane setting.
\begin{gather}
\label{variation}
    \begin{split}
    \chi(t) &= C_1(t)e^{i\ve t}+C_2(t)e^{-i\ve t}\\
    0 &=C^\prime_1(t)\chi_1(t)+ C^\prime_2(t)\chi_2(t)
    \end{split}
\end{gather}
We start from the solution describing the transmitted wave only, \(\chi^{(0)}(t)=e^{i\varepsilon t}\). Thus, the unperturbed values of the coefficients are \(C_1=1\) and \(C_2=0\). The boundary conditions for the perturbed coefficients are therefore
\begin{gather}
    \label{boundary2}
    C_1(-\infty) = 1,\ \ C_2(\infty) = 0. 
\end{gather}
Using~\eqref{boundary2} we obtain the solution for $C_2$ in the form
\begin{gather}  
\label{C2}
        C_2(t) =\frac{\mu b}{2i\ve A}e^{i\pi\gamma/2}(\gamma A)^{1/\gamma}\int_{+\infty}^t s^{1/\gamma}e^{2i\ve s}ds
\end{gather}
Since we are interested in the reflection amplitude, we take the limit \(t\to-\infty\) in Eq.~\eqref{C2}. The local reflection amplitude generated near complex infinity is then \(r_{i\infty}=C_2(-\infty)\), which reduces the problem to the evaluation of a definite integral. The remaining integral requires a consistent choice of the branch of \(t^{1/\gamma}\) and analytic continuation between the two anti-Stokes lines. The corresponding contour deformation is described in Appendix~\ref{finali} and yields
\begin{gather}
\label{amplitude-final}
    r_{i\infty} = -\frac{b\mu}{2\ve^2 A\gamma}\Gamma\left(\frac{1}{\gamma}\right)\left(\frac{\gamma A}{2\ve}\right)^{\frac{1}{\gamma}}\sin\frac{\pi}{\gamma}
\end{gather}
which, up to an inessential phase factor, coincides exactly with the perturbative result~\eqref{amp_correct_gen}. We then match this local solution near \(\tau_{i\infty}\) to the semiclassical solutions in the overlap region~\eqref{condition2}. Using Eqs.~\eqref{semi0}, \eqref{tauPhi01}, and \eqref{dirac3}, one readily finds that \(t=\tau-\tau_{i\infty}\) is mapped onto
\begin{gather}
    \ve(\tau-\tau_{i\infty}) \underset{z\rightarrow i\infty}{=} \int\limits_{i\infty}^z\pi(s)\,ds
\end{gather}
Therefore, the solution with reflection becomes
\begin{gather}
    \begin{split}
    \psi(\tau) &= e^{i\ve(\tau-\tau_{i\infty})}+r_{i\infty}e^{-i\ve(\tau-\tau_{i\infty})},\\
    \psi(\tau) &= e^{-i\int_0^{i\infty}\pi dz}\big[e^{i\int_0^{z}\pi dz}+r_{i\infty} e^{2i\int_0^{i\infty}\pi dz}e^{-i\int_0^z\pi dz}\big]
    \end{split}
\end{gather}
Consequently, the modulus of the full semiclassical reflection amplitude for the deformation profile \(\phi(z)\) is
\begin{gather}
\label{res3}
    |r_2| = |r_{i\infty}|\exp\left[-2\,{\rm Im}\!\!\!\!\int\limits_{\pi}^{\pi+i\infty}
    \frac{\sqrt{\ve^2(1+\phi^2)-\mu^2}}{1+\phi^2(z)}dz\right] 
\end{gather}
which reduces to the perturbative result~\eqref{amp_correct_gen} in the limit \(\mu\to0\). As before, the corresponding band gap is then restored using Dykhne's relation~\eqref{dykhne2}, \(\Delta_n=2|r_2|/T\).

Equation~\eqref{res3} is one of the key results of this work. In this section, we have extended the Pokrovsky--Khalatnikov construction to scattering controlled by complex infinity. In this unusual situation, the conventional Stokes analysis changes its character substantially. Instead of using exact Airy- or Weber-type solutions to connect different asymptotics across the anti-Stokes network, we exploit the additional small parameter \(1/\phi(z)\) near complex infinity and solve the local scattering problem perturbatively. The resulting solution is then matched to the standard semiclassical wave functions in the usual way.

\section{Discussion}
In this work, we study transport along a periodically deformed edge of a two-dimensional topological insulator with strong spin-orbit coupling. Time-reversal symmetry is broken in the minimal way, by adding a Zeeman term. We determine the resulting band structure and its evolution with magnetic field. Together with our previous study of an isolated edge deformation, the present work completes a natural semiclassical program for this class of edge-scattering problems.

Our results predict a novel type of magnetic quantum oscillations: the widths of the forbidden gaps oscillate as a function of magnetic field. We consider both strong fields, where the Zeeman energy \(\mu\) is of the order of the quasiparticle energy \(\varepsilon\), and weak fields, \(\mu\ll\varepsilon\). The two regimes are governed by qualitatively different features of the complex-analytic structure of the Dirac equation and of the corresponding scattering amplitude.

In the strong-field regime, \(\mu\sim\varepsilon\), we considered two classes of deformation profiles: sign-changing and sign-definite. The real-space deformations are closely related, while the corresponding complex-plane scattering structures are qualitatively different.
For a sign-changing deformation, the problem maps naturally onto a Landau--Zener--Stückelberg interferometer. Each real zero of the deformation profile is accompanied by a pair of nearly coalescing complex turning points, while two consecutive zeros provide the two spatially separated Landau--Zener transitions required for interference. As a result, the forbidden-band widths exhibit magnetic quantum oscillations and may vanish at discrete values of the field.

Quantitatively, Eq.~\eqref{band1} shows that these oscillations lead to a complete collapse of the \(n\)th band gap at a discrete sequence of magnetic fields

\begin{gather}
\label{magnetic1}
    H^{(n)}_k = \frac{2v_F\hbar}{\mu_B g L}\left[\#_1 n-\#_2\left(k+\frac{1}{2}\right)\right]. 
\end{gather} 
Here, \(\#_1=\pi L/L_1\) and \(\#_2=L_2L/(L_1{\rm Re}\,a)\) are dimensionless factors determined by the geometry of the edge deformation.

Equation~\eqref{magnetic1} reveals an unusual property of these oscillations: unlike conventional magnetic quantum oscillations in normal metals, which are periodic in inverse magnetic field, they are periodic in the magnetic field itself. This field dependence provides a clear experimental signature that separates them from conventional MQO observed in various topological-insulator systems~\cite{Taskin2009,Wang2010,Ramazashvili2015,Zhao2019,LaBarre2022,Huang2022,Alisultanov2023}. The oscillation period also provides access to characteristics of the edge states, such as the Fermi velocity and the effective \(g\)-factor.

\subsubsection{Possible experimental realization}
A natural question is whether periodic out-of-plane
deformations of the type considered here can be realized
experimentally. Ref.~\cite{niu2021direct} provides an
encouraging indication: monolayer WTe$_2$ was suspended
and characterized in a free-standing geometry, revealing
pronounced anisotropic out-of-plane ripples with a
characteristic lateral scale $L\approx12$ nm.
Although the observed corrugation is not strictly periodic, this experiment demonstrates that free-standing WTe$_2$ can naturally support smooth out-of-plane deformations on the nanometer scale required in our model.
For the edge states of WTe$_2$, the $g$-factor can reach $g\approx7.5$~\cite{fei2017edge}, while the Fermi velocity  obtained from~\cite{jia2022tuning} is $\hbar v_F\approx1.0\ {\rm eV}\cdot{\rm \AA}$.

Using these parameters in Eq.~\eqref{magnetic1}, the characteristic magnetic-field scale for the naturally occurring ripples with $L\approx12$ nm is rather high, $H\sim19$ T. Since this scale is inversely proportional to the deformation period, increasing $L$ to $1\ \mu{\rm m}$ reduces the required field by almost two orders of magnitude. For the $12$-nm undulation reported in Ref.~\cite{niu2021direct}, our results, e.g., Fig.~\ref{fig:pLZ01b}, give a typical band gap $\Delta\sim0.2$ meV, making the band structure resolvable at temperatures $T\ll1.5$ K.

The most direct experimental signature should appear in edge transport. By varying the magnetic field at sufficiently low temperatures, the Fermi energy can be driven into and out of a forbidden band. Inside a gap, propagation through a sufficiently long periodically deformed edge segment is exponentially suppressed, whereas conducting edge modes are restored when the gap closes. This should produce pronounced oscillations of the edge conductance with magnetic field. Experimentally, two independent control parameters are available: the magnetic field and the chemical potential, the latter being tunable by gate voltage.

\subsubsection{Experimental validity of the approximations}

Throughout this work we employ two complementary approximations: semiclassics and perturbation theory in the magnetic field. The semiclassical treatment requires the electronic wavelength to be shorter than the characteristic deformation scale. The experiment of Ref.~\cite{niu2021direct} did not determine the Fermi energy of the edge states, so we estimate the relevant energy range using characteristic scales reported in other experiments on monolayer WTe$_2$.

To our knowledge, the Fermi energy of the edge states measured from the Kramers point has not been determined directly. Nevertheless, an order-of-magnitude estimate is possible. Bulk spectroscopy~\cite{cucchi2019microfocus} and direct measurements of the chemical potential in monolayer WTe$_2$ indicate characteristic low-energy scales of several tens of meV. In Ref.~\cite{cucchi2019microfocus}, the chemical potential was found approximately \(9\) meV above the conduction-band minimum, while Ref.~\cite{sun2022evidence} demonstrated electrostatic tuning over a range of approximately \(40\) meV. Taking a representative edge-state energy \(\ve\sim10\!-\!40\) meV and using \(\hbar v_F\simeq1.0 \,{\rm eV}\,{\rm \AA}\), we obtain $ \lambda=\hbar v_F/{\ve}\simeq12.5\!-\!3.1\ {\rm nm}$. For the naturally occurring deformation scale \(L\simeq12\) nm reported in Ref.~\cite{niu2021direct}, the semiclassical condition is therefore only marginal at the lower end of this energy range, but becomes progressively better satisfied as \(\ve\) increases. For larger engineered deformation periods, such as those discussed above, the semiclassical regime is reached much more comfortably. The condition for the perturbative expansion~\eqref{perturbCrit} can also be readily satisfied. For \(L\simeq12\) nm and \(H=1\) T, one finds
$ \mu L/(\hbar v_F)\simeq0.04, $
well within the perturbative regime.

Finally, in the weak-field limit we have constructed a unified description of band-gap formation for a broad class of periodic deformation profiles using both perturbation theory and semiclassical analysis. A particularly unusual result is that the dominant reflection need not be associated with the turning points closest to the real axis. For a certain class of profiles, it is instead controlled by the asymptotic region \(z\to i\infty\). We developed an analytic treatment of this contribution and obtained complete agreement between the semiclassical and perturbative results in their common domain of validity, as well as excellent agreement with direct numerical calculations. This construction extends the Pokrovsky--Khalatnikov treatment of over-barrier scattering to a qualitatively new situation in which the relevant Stokes phenomenon is controlled by complex infinity.

\section*{ACKNOWLEDGMENTS}
Ya. I. R. warmly thanks Andrew G. Semenov for many insightful and illuminating discussions.
P.D.G. acknowledges the RSF grant No. 26-12-00345.

\section*{DATA AVAILABILITY}
All data are available upon reasonable request from the
corresponding author at yaroslav.rodionov@gmail.com

\section*{CONFLICT OF INTEREST}
The authors declare no conflicts of interest.

\appendix

\section{Dykhne's formalism}
\label{Dykhne}
Here, we summarize the main result of the work by Dykhne~\cite{dykhne1961}.
Consider a one-dimensional Schr\"{o}dinger  periodic problem with period \(L\), \(\phi(x+L)=\phi(x)\). 
Linear independent solutions read $f(x)$ and $f^*(x)$.
Translation by one period mixes the solutions. Therefore, we write
\begin{gather}
\label{app:translation}
\begin{split}
    &f(x + L) = D f(x) + r f^*(x), \\
    &f^*(x + L) = D^* f^*(x) + r^* f(x),
\end{split}
\end{gather}
where $r$ and $D$ are certain functions of energy $\ve$.
Conservation of the Wronskian implies
\begin{gather}
    \label{prprp}
    |D|^2-|r|^2=1. 
\end{gather}
Quasi-periodicity condition  $\psi(x + L) = \lambda\psi(x)$ together with~\eqref{prprp} yields the dispersion relation
\begin{gather}
\label{disp}
    \lambda^2 - 2\lambda {\rm Re} D + 1 = 0.
\end{gather}
The wave propagation condition is $|\lambda| = 1$. Consequently, the forbidden bands equation is the condition of the reality of roots of~\eqref{disp}
\begin{gather}
\label{app:disp2}
    |{\rm Re}\, D| \geq1\ \Rightarrow\ \left|\cos(\arg D)\right|\geq\frac{1}{\sqrt{1+|r|^2}}
\end{gather}
So far all the relations were exact. Now we employ WKB approximation. 
WKB wave function reads
\begin{gather}
 f(x)=\frac{1}{\sqrt{\pi(x)}}\exp\left( i\int^x \pi(x')\,dx'\right). 
\end{gather}
where $\pi(x)$ is the semiclassical momentum.
In the over-barrier semiclassical regime the reflection amplitude is exponentially small, \(|r|\ll1\), whereas \(D=O(1)\). Consequently, the forbidden regions determined by Eq.~\eqref{app:disp2} are confined to narrow neighborhoods of $\arg D = \pi n,\ n\in\mathbb{N}$.
which yields the band center equation
\begin{gather}
    \arg D(\ve_n) = \pi n,\ n\in\mathbb{N}
\end{gather}
Expanding l.h.s of~\eqref{app:disp2} in $\ve$ and its r.h.s. in $|r|$ we obtain the width of the forbidden band in the form
\begin{gather}
\label{app:Delta}
    \begin{split}
    \left|(\ve-\ve_n)\de_\ve \arg D(\ve)|_{\ve_n}\right| &\leq |r|\ \Rightarrow\\
    \Delta_n  &= \frac{2|r(\ve_n)|}{|\de_{\ve}\arg D(\ve)|_{\ve_n}}
    \end{split}
\end{gather}
In the leading WKB approximation, propagation of the \(f\) branch through one period in the absence of exponentially small interbranch mixing yields
\begin{gather}
 f(x+L)= \exp\!\left(i\int_x^{x+L}\pi(x')dx'\right)f(x).
\end{gather}
Since \(\pi(x+L)=\pi(x)\), the accumulated phase is independent of \(x\), and therefore, to leading WKB order,
\begin{gather}
\label{app:d}
\arg D(\varepsilon) = S_L(\varepsilon) +\mathcal{O}\left(\frac{v_F}{\ve L}\right),
\quad S_L(\varepsilon) = \int_0^L\pi(x,\varepsilon)\,dx,
\end{gather}
Using the last equation we obtain
\begin{gather}
\frac{\partial\,\arg D}{\partial\varepsilon} = \int_0^L \frac{\partial \pi(x,\varepsilon)}{\partial\varepsilon}\,dx \equiv T ,
\end{gather}
where \(T\) is the semiclassical traversal time over one spatial period. Substituting this result into Eq.~\eqref{app:Delta} yields
\begin{gather}
 \Delta_n=\frac{2|r|}{T},
\end{gather}
which reproduces Eqs.~\eqref{action0} and~\eqref{dykhne2} of the main text.

\section{Strong magnetic field. Sign-changing deformation}
\label{app:indef}
Before addressing the exact solution of Eq.~\eqref{exact_eq1}, it is instructive first to study the semiclassical solutions of the same problem to get the asymptotic appearance of the incident-reflected-transmitted triple.  
\subsection{Semiclassical solution}
For the local momentum obtained from Eq.~\eqref{moment1},
we choose the zero of the action at the center of the
turning-point pair, $\zeta=0$. The two semiclassical actions
are therefore
\begin{gather}
    S_{\gtrless}(\zeta) = \pm\frac{\mu \zeta^2}{2a}+\frac{s}{2} \pm s \ln\frac{2\zeta}{a}\sqrt{\frac{\mu}{2\delta\ve}},
\end{gather}
where $>$ and $<$ denote the right and left directions respectively and $s = \delta\ve a$.
The semiclassical wave functions are then obtained from relations~\eqref{semi0}:
\begin{gather}
\label{app:semi0}
\begin{split}
    \psi_{+>} (\zeta)&= \sqrt{2}\left[\frac{2\zeta}{a}\right]^{is-\frac{1}{2}}\left[\frac{2\delta\ve}{\mu}\right]^{-\frac{is}{2}+\frac{1}{2}} e^{\frac{is}{2}},\\
    \psi_{+<} (\zeta)&= \sqrt{2}\left[\frac{2\zeta}{a}\right]^{-is-\frac{1}{2}}\left[\frac{2\delta\ve}{\mu}\right]^{\frac{is}{2}+\frac{1}{2}} e^{-\frac{i\mu \zeta^2}{a}-\frac{is}{2}},\\
    \psi_{-<} (\zeta)&=
    \sqrt{2}\left[\frac{2\zeta}{a}\right]^{is-\frac{1}{2}}\left[\frac{2\delta\ve}{\mu}\right]^{-\frac{is}{2}-\frac{1}{2}} e^{\frac{is}{2}},\quad 
\end{split} 
\end{gather}

\subsection{Exact solution}
\begin{figure}
    \centering
    \includegraphics[width=1.0\linewidth]{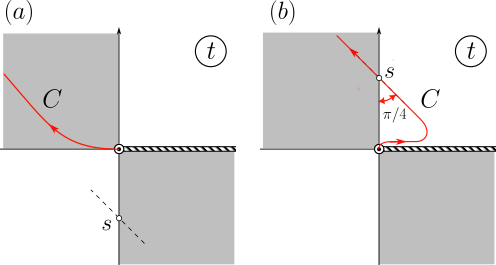}
    \caption{\justifying Integration contour and its deformation. Shaded regions are the ones where $V$ -- function from~\eqref{V_function}. (a) $C$ yields transmitted wave, the main contribution comes from the vicinity of the origin. (b) Transformation of the contour after rotation
    $\zeta\rightarrow\zeta e^{-i\pi}$. The contour is deformed along the steepest descent directions near the origin and near the saddle .}
    \label{fig:contour-first}
\end{figure}
We now match the semiclassical solutions~\eqref{app:semi0}
to the exact solution of Eq.~\eqref{exact_eq1}. Although the
Hermite equation is well understood, the present calculation
requires the analytic continuation of a particular solution
between the right and left anti-Stokes lines, together with
its normalization relative to the semiclassical states. For
this purpose, we use a Laplace contour-integral representation
following the method described in Ref.~\cite{LandauIII}. The exact solution of the differential equation of type:
\begin{gather}\label{LD1}
  \sum\limits_{m=0}^n (a_m+b_m \zeta)\frac{d^m\theta}{d\zeta^m}=0
\end{gather}
is given by the following contour integral:
\begin{gather}\label{LD2}
  \begin{split}
     \theta(\zeta)  = \int_C e^{\zeta t}Z(t)\,dt,\quad
      Z(t) = \frac{1}{Q}\exp\int \frac{P}{Q}\,dt.
  \end{split}
\end{gather}
where contour $C$ is such, that function $V(t) = e^{\zeta t}QZ$ assumes identical values at its end points and the polynomials $P$ and $Q$ are defined as:
\begin{gather}\label{LD3}
  P(t) = \sum\limits_{m=0}^n a_m t^m,\quad
  Q(t) = \sum\limits_{m=0}^n b_m t^m,
\end{gather}
Applying Eqs.~\eqref{LD2} and \eqref{LD3} to the local
equation~\eqref{exact_eq1}, we recover the contour-integral
solution~\eqref{sol1} given in the main text. The corresponding
boundary function is
\begin{gather}\label{V_function}
   V = e^{\zeta t -\frac{ait^2}{4\mu}} t^{-si+\frac{1}{2}}
\end{gather}
We choose the integration contour so that the boundary
function $V$ vanishes at its endpoints. Among the admissible
contours, we select the one whose asymptotics matches the
transmitted semiclassical solution $\psi_{+>}$ in
Eq.~\eqref{app:semi0}. This contour is shown in
Fig.~\ref{fig:contour-first}(a).
The multivalued powers in Eqs.~\eqref{sol1} and
\eqref{V_function} are defined using the same branch of
$\ln t$. We choose $\arg t=0$ on the positive real axis,
with the branch cut and the integration contour specified
in Fig.~\ref{fig:contour-first}.

\subsection{Matching of semiclassical and exact solutions}
To evaluate the asymptotics of the exact solution, we employ
the method of steepest descent. For large positive $\zeta$,
the steepest descent curve emanating from $t = 0$ is presented in Fig.~\ref{fig:contour-first}(a). The leading
contribution then comes from the vicinity of the endpoint
$t=0$, where the exponent can be expanded and only the term
linear in $t$ retained. As one can see from Fig.~\ref{fig:contour-first}(a) the saddle remains inaccessible.

We next analytically continue the solution to the negative
real axis through a clockwise rotation by $-\pi$. During this
continuation, the saddle-point contribution becomes accessible.
The contour deformation shown in
Fig.~\ref{fig:contour-first}(b) then yields two contributions:
one from the endpoint $t=0$ and the other from the saddle
point. Their sum gives the asymptotic behavior of the exact
solution on the negative real axis.
\begin{gather}
\label{app:exact2}
    \psi(\zeta) =\begin{cases}
        - i \zeta^{is-\frac{1}{2}}e^{\pi s}\Gamma\left(\frac{1}{2}-is\right),\ &\ \zeta>0\\
        -|\zeta|^{is-\frac{1}{2}}e^{\pi s}\Gamma\left(\frac{1}{2}-is\right)+\\-ie^{-\frac{i\mu \zeta^2}{a}}\left[\frac{2|\zeta|}{a}\right]^{-is-\frac{1}{2}}\!\!\!\!\sqrt{\frac{4\pi}{a}}e^{\frac{\pi s}{2}}\mu^{-is},\ \ &\ \zeta<0
    \end{cases}
\end{gather}
We now fix the overall normalization of the contour-integral
solution~\eqref{sol1}. Matching its positive-$\zeta$ asymptotics
in Eq.~\eqref{app:exact2} to the transmitted semiclassical
state $\psi_{+>}$ in Eq.~\eqref{app:semi0}, we obtain
\begin{gather}
    A = -i\left[\frac{2}{a}\right]^{is-\frac{1}{2}}
    \left[\frac{2\delta\ve}{\mu}\right]^{-is+\frac{1}{2}}
    \frac{\sqrt{2}}{\Gamma(\frac{1}{2}-is)}e^{-\frac{is}{2}-\frac{\pi s}{2}}.
\end{gather}
With the normalization fixed by the transmitted wave, we
return to the negative-$\zeta$ asymptotics in
Eq.~\eqref{app:exact2}. Matching the endpoint and saddle-point
contributions to the corresponding semiclassical states
$\psi_{+<}$ and $\psi_{-<}$ in Eq.~\eqref{app:semi0}, we obtain
the connection formula
\begin{gather}
    \psi_{+>}\rightarrow \psi_{+<}\frac{\sqrt{2\pi}\left(\frac{e}{s}\right)^{is}e^{-\frac{\pi s}{2}}}{\Gamma(\frac{1}{2}-is)} - i\psi_{-<}.
\end{gather}
This connection formula gives the first row of the
Landau--Zener transfer matrix~\eqref{LZ}, with the conventions
for the incoming and outgoing amplitudes adopted in the main
text. The remaining row follows from the corresponding
connection formula for the next independent solution, or,
equivalently, from complex conjugation and conservation of
the Wronskian. Thus the exact local solution reproduces the
transfer matrix used in the main-text band-structure analysis.

\subsection{The computation of the phase integral}
\label{app:phase}
Here we build an asymptotic expansion of the full action integral~\eqref{full_phase} $\Phi_1+\Phi_2\equiv \int_0^L\pi(x)\,dx$
at small values of energies $\delta\ve\ll\mu$. Plugging in semiclassical expressions we obtain:
\begin{gather}
    \int\limits_0^L\pi(x)\Big|_{\ve= \mu+\delta\ve}\,dx =\ve\int\limits_0^L\frac{\sqrt{\phi^2(x)+\frac{2\delta\ve}{\mu}}}{1+\phi^2(x)}\,dx\equiv\ve I(b^2),
\end{gather}
where we denoted $b^2 = 2\delta\ve/\mu$ for brevity. The integral can't be computed with simple Taylor expansion in $b^2$ since there are regions where $\phi(x)\sim b$. Those are precisely the $x_{0,1}$ points of LZ-transitions (roots of $\phi(x)$). As usual, to build a formal expansion in infinitesimal $b^2$ we isolate these points with $s$ - neighborhoods and pick $s$ in such a way that $s\ll a$, where $a =1/f^\prime(x_{0,1})$. This allows one to perform Taylor expansion of $\phi(x)$ near the roots for $x\in U_{s}(x_{0,1})$. On the other hand, we pick $s$ large enough such that: $\phi^2(x_{0,1}\pm s)\gg b^2$, i.e. $b |a_{1,2}|\ll s\ll |a_{1,2}|$. Now we split:
\begin{gather}
\label{app:int00}
    I(b^2) = \underset{\rm I}{\int\limits_0^{x_1-s}}+\underset{\rm II}{\int\limits_{x_1-s}^{x_1+s}}+\underset{\rm III}{\int\limits_{x_1+s}^{x_0-s}}+\underset{\rm IV}{\int\limits_{x_0-s}^{x_0+s}}+\underset{\rm V}{\int\limits_{x_0+s}^L}
\end{gather}
Here, roman numbers I-V denote integrals as well as the respective integration domains.  
In the first, third, and last integrals, it is possible to expand the integrand in $b^2$:
\begin{gather}
\begin{split}
    &{\rm I+III+V} =\!\!\!\!\!\!\!\underbrace{\underset{{\rm I+III+V}}{\int}\!\!\!\!\frac{|\phi(x)|}{1+\phi^2(x)}\,dx}_{Q_1}+
    \frac{b^2}{2}\!\!\!\!\underbrace{\int\limits_{\rm I+III+V}\!\!\!\!\frac{1}{|\phi(x)|}\frac{dx}{1+\phi^2(x)}}_{Q_2}
\end{split}
\end{gather}
Integral $Q_1$ is transformed as:
\begin{gather}
\label{app:int01}
\begin{split}
    Q_1 &= \int\limits_0^L\frac{|\phi(x)|}{1+\phi^2(x)}\,dx - \int\limits_{-s}^s\frac{|z|\,dz}{|a_1|} -
    \int\limits_{-s}^s\frac{|z|\,dz}{|a_2|}\\
    &=I(0)-s^2\left[\frac{1}{|a_1|}+\frac{1}{|a_0|}\right].
\end{split}
\end{gather}
where we added and subtracted the $s$-neighborhoods to the integration domain. When subtracting, we changed the exact expression for $\phi$ with its Taylor expansion utilizing the equality $s\ll|a_{1,2}|$. The $Q_2$ integral is transformed using the regularization subtraction:
\begin{gather}
\begin{split}
    Q_2 &= \!\!\!\!\int\limits_{\rm I+III+V}\frac{1}{|\phi(x)|}\left[\frac{1}{1+\phi^2(x)}-\frac{|a_1|}{|x-x_1|}-
    \frac{|a_0|}{|x-x_0|}\right]\,dx\\
    &+\int\limits_{\rm I+III+V}\left[\frac{|a_1|}{|x-x_1|}+\frac{|a_0|}{|x-x_0|}\right]\,dx
\end{split}
\end{gather}
Since the expression for $Q_2$ already enters the expression with $b^2$ prefactor, we may set $s\rightarrow 0$ wherever possible when computing the integral. This way we obtain:
\begin{gather}
\label{app:int02}
\begin{split}
    Q_2 &= C +|a_1|\ln\frac{x_1(L-x_1)}{s^2}
   + |a_0|\ln\frac{x_0(L-x_0)}{s^2}\\
   C &=
\int_0^L
\left[
\frac{1}{|\phi(x)|(1+\phi^2(x))}
-\sum_{i=0,1}\frac{|a_i|}{|x-x_i|}
\right]dx.
\end{split}
\end{gather}
Finally we compute the integrals II and IV by using the Taylor expansion of $\phi(x)$ near its roots:
\begin{gather}
    \begin{split}
    &{\rm II+IV} = \sum\limits_{i=0,1}\int\limits_{-s}^s\sqrt{\frac{z^2}{a_i^2}+b^2}\,dz\\
    &=\sum\limits_{i=0,1} \frac{s^2}{|a_i|}+
    \frac{|a_i|b^2}{2}+b^2 |a_i|\ln\frac{2s}{|a_i|b}.
    \end{split}
\end{gather}
Combining expressions~\eqref{app:int01},~\eqref{app:int02} and plugging them into~\eqref{app:int00}, we immediately notice that all $s$-dependent terms cancel out (as they should).
Next,  introducing $L_{\rm eff}$ from Eq.~\eqref{leff} and suitable constant
\begin{equation}
\ln \sqrt{2}B=\frac{1}{L_{\rm eff}}\left[
\frac{C}{2}+L_0
+\sum_{i=0,1}|a_i|
\ln\frac{2\sqrt{ex_i(L-x_i)}}{|a_i|}
\right].
\end{equation}
we obtain expression~\eqref{full_phase} presented in the main body of the paper.

\subsection{Analysis of the band equation}
\label{app:band}
\subsubsection{Small energies $\delta\ve|a_{0,1}|\ll1$}
In this case, one can safely put $P_1\approx P_2\approx 1$
and get the band equation in the form:
\begin{gather}
\label{app:band02}
\begin{split}
    &|\cos(\mu \Delta L)+2\cos(\Phi_1+\Phi_2+\vf_1+\vf_2)|>1,\\
    &\Delta L = \int\limits_{x_1}^{x_0}\frac{|\phi(x)|\,dx}{1+\phi^2(x)}-\int\limits_{x_0-2\pi}^{x_1}\frac{|\phi(x)|\,dx}{1+\phi^2(x)}
\end{split}
\end{gather}
We seek for the band edges. Changing inequality sign in~\eqref{app:band02} with equivalence and tackling the modulus, we obtain
\begin{gather}
\label{app:bandedges}
    \begin{split}
        2\cos\left[\Phi_1+\Phi_2+\vf_1+\vf_2\right] =\pm 1-\cos\mu\Delta L
    \end{split}
\end{gather}
The Stokes-phases $\phi_{i}$ need to be taken in the leading order in $s\equiv\delta\ve |a_i|\ll1$:
\begin{gather}
\label{app:stokes}
    \vf_1+\vf_2 \rightarrow -\sum\limits_{i=0,1}\delta\ve|a_i|\ln \delta\ve|a_i|
\end{gather}
Comparing expression for the full phase~\eqref{full_phase} and Stokes phases~\eqref{app:stokes} we see that despite their functional similarity, the energy dependent term in the full phase contains additional factor $\ln\sqrt{\mu(|a_0|+|a_1|)}\gg1$ which is absent in the Stokes phase terms. Therefore we may discard the Stokes phase all together. 

Using expression~\eqref{app:bandedges} for the band edges and the phase~\eqref{full_phase}, we quickly capture the approximate positions for the center of bands~\eqref{band-edges} as well as the width of the forbidden bands themselves. This explains negative and approximately constant slope of forbidden bands depicted in plot~\eqref{fig:pLZ01a}. 

\subsubsection{Large energies $\delta\ve a\gg1$}
In this case, we exploit the smallness of the LZ-transition amplitude $P_{1,2}$.
That means the band gaps become exponentially small. The r.h.s. of band equation~\eqref{band01} is dominated by the second term. The Stokes phases vanish in the limit $\delta\ve a\gg 1$.  
Introducing the phase shift $\Delta S$ according to:
\begin{gather}
 \Phi_1+\Phi_2 = \pi n +\Delta S   
\end{gather}
 and expanding Eq.~\eqref{band01} in exponentially small $P_{1,2}$ we obtain.
\begin{gather}
    \Delta S = \pm[P_1+P_2\pm2\sqrt{P_1P_2}\cos\Delta\Phi]^{1/2}.
\end{gather}
where $\pm$ corresponds to  even and odd bands respectively. 
Using expression ~\eqref{full_phase}, we obtain the band gap~\eqref{band-edges02} in the main part.

\section{Strong magnetic field. Sign-definite deformation}
\label{app:def}
\subsection{Zeroes of deformation profile}
\label{app:analyt01}
To get some understanding of the possible analytical structure of the deformation profile function let us track down the evolution of the roots of $\phi(x)$ as we apply the uniform bend to the edge $\phi(x)\rightarrow \phi(z)+c$. At some moment $c =  c_*$ becomes equal to the $\min\{\phi(x)\}$ where both roots $x_0$ and $x_1$ merge into saddle point of $\phi$: $x_*$.  As the growth of $c$ continues but $c$ remains small compared to the period of the deformation, the roots of the function can be accessed with Taylor expansion near the saddle: $\phi(z) = (1/2)\phi^{\prime\prime}(x_*)(z-x_*)^2+(c-c_*)$. Therefore, the derivative of $\phi(z)$ at  roots $z_{0}$ becomes purely imaginary: $\phi'(z_0) = \pm i\sqrt{2\phi^{\prime\prime}(x_*)(c-c_*)}$. This, of course, is only a local property which violates as soon as complex roots travel far enough from the real axis. Nevertheless, we expect that property to hold generally for not very large bending of the TI edge. In other words, the value of the derivative of the deformation potential remains predominantly imaginary for the potential overall bend $c\lesssim L$.

\subsection{Analytical structure of $\phi(z)$}
\label{lemma}
Here we prove the important lemma concerning the poles of the semiclassical momentum $\pi(z)$. As follows from the definition, the poles of $\pi(z)$ are the zeros of function $\phi^2(z)+1$, hence we prove the following\\
\textbf{Lemma}. Let \(\phi(z)\) be a nonconstant real trigonometric polynomial of degree \(N\), i.e. a finite Fourier series satisfying \(\phi(x)\in\mathbb R\) for real \(x\). Then, within the upper half of one periodicity strip, the equation $\phi^2(z)+1=0$ 
has exactly \(2N\) roots, counted with multiplicity. For generic Fourier coefficients these roots are simple.

\textit{Proof.}
We choose the coordinate such that the deformation period is $2\pi$ and write
\begin{gather}
    \phi(z)=\sum_{n=-N}^{N} c_n e^{inz},
    \qquad c_{-n}=c_n^* ,
\end{gather}
where the latter relation follows from the reality of $\phi(x)$ on the real axis.
Introducing
\begin{gather}
    w=e^{iz},
\end{gather}
maps the upper half-plane, ${\rm Im}\,z>0$, onto the interior of the unit circle,
$|w|<1$. The deformation profile becomes a Laurent polynomial,
\begin{gather}
    \phi(z)\equiv F(w)=\sum_{n=-N}^{N}c_n w^n .
\end{gather}

The equation $1+\phi^2(z)=0$ splits into two equations,
\begin{gather}
    \phi(z)=\pm i .
\end{gather}
Consider first $\phi(z)=i$ and define
\begin{gather}
    P_+(w)=w^N[F(w)-i].
\end{gather}
Since $c_{\pm N}\neq0$, $P_+(w)$ is an ordinary polynomial of degree $2N$,
and $P_+(0)=c_{-N}\neq0$.

On the unit circle, $w=e^{ix}$, we have
\begin{gather}
    P_+(e^{ix})=e^{iNx}\,[\phi(x)-i].
\end{gather}
As $x$ varies from $0$ to $2\pi$, the first factor winds $N$ times around the
origin. The second factor, however, has zero winding number: since $\phi(x)$
is real, the curve $\phi(x)-i$ lies entirely in the lower half-plane and never
crosses the origin. Hence the total change of the argument of $P_+$ is
$2\pi N$. By the argument principle, $P_+(w)$ therefore has exactly $N$
zeros inside the unit circle.

The same reasoning applied to
\begin{gather}
    P_-(w)=w^N[F(w)+i]
\end{gather}
shows that the equation $\phi(z)=-i$ also has exactly $N$ roots in $|w|<1$.
Thus, within the upper half of one $2\pi$-periodicity strip,
\begin{gather}
    1+\phi^2(z)=0
\end{gather}
has exactly $2N$ roots, counted with multiplicity.

Finally, a multiple root would require simultaneously
\begin{gather}
    \phi(z_p)=\pm i,
    \qquad
    \phi'(z_p)=0 .
\end{gather}
These two conditions require an additional fine tuning of the Fourier
coefficients. Therefore, for a generic deformation profile the roots are
simple, and the corresponding singularities of the semiclassical momentum
are simple poles.
\hfill$\square$

\subsection{Semiclassical action}
\label{app:def:semi}
To perform the matching of semiclassical solution with exact near the turning point we, as before, compute the semiclassical action. The action on both anti-Stokes lines (right and left) is given by identical analytical function.
It is convenient to choose the
right branch point as the reference point of the action. This
choice fixes the constant phases of the semiclassical
solutions; a different reference point leads to an equivalent
normalization of the same physical wave function. We obtain
\begin{gather}
\label{phase1}
    \begin{split}
       S &=\!\!\!\!\int\limits_{a\sqrt{\frac{2\delta\ve}{\mu}}}^\zeta \!\!\!\!\pi(\zeta)d\zeta 
         =\frac{i\mu\zeta^2}{2a}-\frac{ia\delta\ve}{2}-i\delta\ve a\ln\frac{\zeta}{a}\sqrt{\frac{2\mu}{\delta\ve}}
    \end{split}
\end{gather} 
For the two anti-Stokes lines used in the matching, the local
coordinate is parameterized as
\begin{gather}\label{app:anti-stokes1}
  \zeta=
    \begin{cases}
      |\zeta|e^{-\frac{i\pi}{4}+\frac{i\arg a}{2}},\   &\ \hbox{right line}\\
      |\zeta|e^{-\frac{3\pi i}{4}+\frac{i\arg a}{2}},\ &\ \hbox{left line}.
  \end{cases}
\end{gather}
Then, according to relations~\eqref{semi2} and carefully tracing the correct regular branches of $\pi(\zeta)$ we arrive at the following semiclassical wave functions:
\begin{gather}
\label{asymp02}
    \begin{split}
   \psi_{s+}(\zeta) &= e^{\frac{a\delta\ve}{2}-\frac{i\pi}{4}}\left[\frac{\zeta}{a}\right]^{\delta\ve a-\frac{1}{2}}\left[\frac{\delta\ve}{2\mu}\right]^{-\frac{\delta\ve a}{2}}\\
   \psi_{s-}(\zeta) &= e^{\frac{\mu\zeta^2}{a}-\frac{a\delta\ve}{2}-\frac{i\pi}{4}}\left[\frac{\zeta}{a}\right]^{-\delta\ve a-\frac{1}{2}}\left[\frac{\delta\ve}{2\mu}\right]^{\frac{\delta\ve a}{2}}
    \end{split}
\end{gather}
These expressions provide the asymptotic basis on both
anti-Stokes lines, with the branches understood as specified
above.
\subsection{Exact solution}
\label{app:def:exact}
\begin{figure}[t!]
	\centering
\includegraphics[width=1\columnwidth]{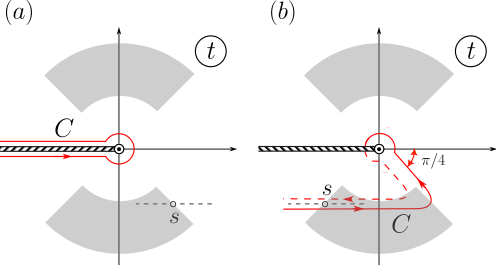}
    	\caption{\justifying
Integration contour and its steepest-descent deformation.
The shaded regions indicate the convergence sectors in which
the boundary function $V(t)$ vanishes as $|t|\to\infty$.
(a) The initial Hankel contour, whose leading asymptotics on
the right anti-Stokes line is determined by the endpoint
$t=0$.
(b) Deformation of the contour after analytic continuation
to the left anti-Stokes line. The saddle-point contribution
becomes accessible in addition to the endpoint contribution.
The dashed lines indicate the local steepest-descent
directions. For clarity, the contours are shown for
$\arg a=0$.}
\label{fig:cont01}
\end{figure}

We now determine the connection between the two semiclassical
bases by analytically continuing a single exact solution.
As in the sign-changing case, the local equation~\eqref{eq2}
admits a Laplace contour-integral representation.
\begin{gather}
    \begin{split}
    \psi(\zeta) &= A\int\limits_C e^{\zeta t - \frac{at^2}{4\mu}}t^{-s -\frac{1}{2}}\,dt,\\
    V(t) &= e^{\zeta t - \frac{at^2}{4\mu}}t^{-s +\frac{1}{2}}\
    \end{split}
\end{gather}
Here $s\equiv a\delta\ve$, and $V(t)$ is the boundary function
whose values at the endpoints of the contour must coincide.
The multivalued powers are defined using a fixed branch of
$\ln t$, with the branch cut along the negative real semiaxis.
We choose the Hankel contour shown in
Fig.~\ref{fig:cont01}(a), which satisfies the boundary
condition and matches the transmitted asymptotic solution
$\psi_{s+}$ on the right anti-Stokes line.

For large $\zeta$ along the right anti-Stokes line, the
contour can be deformed so that the leading contribution
comes from the vicinity of the branch point $t=0$. In this
region the exponent is expanded in $t$, and the quadratic
term does not contribute to the leading endpoint asymptotics.
The resulting Hankel integral gives
\begin{gather}
  \label{asymp03}
    \psi(\zeta)\Big|_{\rm right} = \frac{2\pi i A}{\Gamma\left(s+\frac{1}{2}\right)}\zeta^{s-\frac{1}{2}}.
\end{gather}
We fix the overall normalization by matching the endpoint
asymptotics~\eqref{asymp03} to the transmitted semiclassical
solution $\psi_{s+}$ in Eq.~\eqref{asymp02}, in their common
domain of validity. This yields
\begin{gather}
    A = \frac{\Gamma\left(s+\frac{1}{2}\right)}{2\pi i} e^{\frac{s}{2}-i\frac{\pi}{4}}
    \left[\frac{\delta\ve}{2\mu}\right]^{-\frac{s}{2}}\frac{1}{a^{s-\frac{1}{2}}}.
\end{gather}
We next continue the same exact solution from the right
anti-Stokes line to the left one, rotating $\zeta$ clockwise
from $\arg\zeta=-\pi/4+\arg a/2$ to $\arg\zeta=-3\pi/4+\arg a/2$ for
$\arg a=0$. The integration contour is deformed continuously,
with the branch of the integrand and the overall normalization
fixed by the preceding matching. As shown in
Fig.~\ref{fig:cont01}(b), the saddle-point contribution
becomes accessible during this continuation.

The endpoint contribution is the analytic continuation of
the transmitted branch already obtained on the right line,
whereas the saddle generates the additional reflected branch.
Matching both contributions to the semiclassical basis
\eqref{asymp02}, with the same normalization constant $A$, we
obtain
\begin{gather}
\label{match3}
\begin{split}
    \psi(\zeta)\Big|_{\rm left} = \psi_{s+}(\zeta)\\
    - 2i\cos(\pi s) \psi_{s-}(\zeta)\left[\left(\frac{e}{s}\right)^s\frac{\Gamma(s+\frac{1}{2})}{\sqrt{\pi}}\right]e^{-\pi i s}
\end{split}
\end{gather}
The phase factor $e^{-i\pi s}$ in Eq.~\eqref{match3} depends
on the reference point chosen for the semiclassical action
in Eq.~\eqref{phase1}. If the action is instead referred to
the zero of the deformation profile, the corresponding
constant phase is transferred to the normalization of the
semiclassical basis. This change of convention leaves the
physical reflection probability unchanged.

Using the full semiclassical solutions~\eqref{semi0} to
restore the normalization of the scattering states, we
obtain the reflection coefficient~\eqref{ans1} of the main
text. The Gamma-function factor and the factor
$\cos(\pi s)$ in Eq.~\eqref{match3} determine the
pre-exponential dependence, while the continued
semiclassical action supplies the exponential attenuation
expressed through $S_0$ in Eq.~\eqref{ans1}.

\subsection{The computation of the phase integral II}
\label{app:phase5}
Here we compute the phase integral entering ~\eqref{ans1} in the manner, analogous to Appendix~\ref{app:phase}.
It is going to be suitable to pick $x_0 = {\rm Re}\,z_0$ as a starting integration point. 
As before, we separate the integration line into two pieces surrounding the proximity of the root $z_0$ with $s$ - neighborhood. We have:    
\begin{gather}
    S = \ve\int\limits_{x_0}^{z_0}\frac{\sqrt{\phi^2(z)+\frac{2\delta\ve}{\mu}}}{1+\phi^2(z)}\,dz \equiv \ve I(b^2)
\end{gather}
where $b^2 = 2\delta\ve/\mu$.
Next we split:
\begin{gather}
\label{app:int2}
    I(b^2) = \int\limits_{x_0}^{z_0-s}+\int\limits_{z_0-s}^{z_0}\equiv {\rm I+II}.
\end{gather}
The first integral can be Taylor-expanded in $b^2$:
\begin{gather}
\begin{split}
    {\rm I} &= \int\limits_{x_0}^{z_0-s}\frac{\phi(z)dz}{1+\phi^2(z)}+\frac{b^2}{2}\int\limits_{x_0}^{z_0-s}\frac{dz}{\phi(z)}\frac{1}{1+\phi^2(z)} \\
    &= I(0) +\frac{is^2}{2a}+\frac{b^2}{2}{\rm III}.
\end{split}
\end{gather}
Here we used Taylor expansion of  $\phi(z)$ in the $s$-neighborhood of $z_0$ when computing the integral. 
Expression III has the pole at $s\rightarrow0$ and we add the regularization function (which we'd like to be real on a real axis):
\begin{align}
\label{app:int3a}
    {\rm III} &= -C +\int\limits_{x_0}^{z_0-s}\frac{dz}{\phi_0(z)},\\
    C &= \int_{x_0}^{z_0}\left[\frac{1}{\phi_0(z)}-\frac{1}{\phi(z)}\frac{1}{\phi^2(z)+1}\right]\,dz,\\
    \frac{1}{\phi_0(z)} &= \frac{a}{i}\frac{1}{z-z_0} - \frac{a^*}{i}\frac{1}{z-z^*_0}. \label{app:int3b}
\end{align}
Next, the integral in~\eqref{app:int3a} is computed exactly as:
\begin{gather}
\label{app:int3c}
    \int\limits_{x_0}^{z_0-s}\frac{dz}{\phi_0(z)} = \frac{a}{i}\ln\frac{s}{i{\rm Im}z_0}-\frac{a^*}{i}\ln2.
\end{gather}
Integral II in~\eqref{app:int2} is worked out via the Taylor expansion of the field $\phi(z)$ in the $s$-neighborhood of the root:
\begin{gather}
\label{app:int3}
    {\rm II} = \!\!\!\!\int\limits_{z_0-s}^{z_0}\!\!\sqrt{b^2-\frac{(z-z_0)^2}{a^2}}\,dz = -\frac{is^2}{2a}+\frac{b^2i}{4a}+\frac{ab^2}{2i}\ln\frac{b}{-2is}. 
\end{gather}
Combining~\eqref{app:int3}, ~\eqref{app:int3c} and~\eqref{app:int3a} we obtain for the~\eqref{app:int2}:
\begin{gather}
\begin{split}
&I(b^2) = I(0)\\
&+\frac{b^2}{2}
\left[
-C+\frac{a}{i}
\left(
\ln\frac{ab}{2\operatorname{Im}z_0}
-\frac12
\right)
-\frac{a^*}{i}\ln 2
\right]
+o(b^2).
\end{split}
\end{gather}
Finally, we are ready to obtain the full expression for the phase constant $B$ entering formula~\eqref{action2} in the paper:
\begin{gather}
\begin{split}
&\ln \sqrt{2}B=
\frac{1}{\operatorname{Re}a}
\bigg\{
-\operatorname{Im}C
-\operatorname{Re}
\left[
a\ln\left(\frac{a}{2\operatorname{Im}z_0}\right)
\right]
+\\
&\operatorname{Re}a
\left(
\frac12+\ln 2
\right)
+L_{\rm eff}\bigg\}.
\end{split}
\end{gather}

\section{Weak magnetic field}

\subsection{Two-harmonic deformation with an imaginary
turning-point derivative}
\label{app:twoharmonic}
To understand, how in principle the multi-harmonic potential satisfying condition~\eqref{condition} can look like geometrically, we construct a simple 
class of two harmonic periodic deformation profiles.
We require that the turning points [$\phi^2(z_0)+1=0$] closest to the real axis have a
predominantly imaginary derivative.
Such profiles provide a useful realization of the special
weak-field regime discussed in the main text.

Consider the even, real-valued two-harmonic profile
\begin{equation}
    \phi(x)=a_0+a_1\cos x+a_2\cos 2x,
    \qquad a_0,a_1,a_2\in\mathbb R.
    \label{app:twoharmonic-profile}
\end{equation}
This function has exactly 2 inequivalent roots in the upper complex plane inside its periodicity strip.  
It is suitable to seek a turning point at the line
\begin{gather}
    z_0=\frac{\pi}{4}+iy,\qquad y>0,
    \label{app:twoharmonic-z0}\\
    \phi(z_0)=i,\qquad
    \operatorname{Re}\phi'(z_0)=0.
    \label{app:twoharmonic-conditions}
\end{gather}
The position of the turning point is thus fixed by the
parameter $y$, while the three real coefficients are
determined by the three real conditions in
Eq.~\eqref{app:twoharmonic-conditions}.
Conditions~\eqref{app:twoharmonic-conditions} reduce to
\begin{align}
    a_0+\frac{a_1}{\sqrt2}\cosh y&=0,
    \nonumber\\
    \frac{a_1}{\sqrt2}+2a_2\cosh y
        &=-\frac{1}{\sinh y},
    \nonumber\\
    \frac{a_1}{\sqrt2}\cosh y
        +2a_2\cosh 2y&=0.
    \label{app:twoharmonic-linear}
\end{align}
Solving these equations gives
\begin{equation}
\begin{aligned}
    a_0(y)&=
        \frac{\cosh y\cosh 2y}{\sinh^3y},\\
    a_1(y)&=
        -\frac{\sqrt2\cosh 2y}{\sinh^3y},\\
    a_2(y)&=
        \frac{\cosh y}{2\sinh^3y}.
\end{aligned}
\label{app:twoharmonic-coefficients}
\end{equation}
Direct substitution yields a purely imaginary derivative
\begin{equation}
    \phi'(z_0)=
    \frac{i\cosh2y}{\sinh^2y}.
    \label{app:twoharmonic-derivative}
\end{equation}
It remains to show that $z_0$ belongs to the lower of the
two inequivalent pairs of turning points. Introducing
$w=\cos z$, Eq.~\eqref{app:twoharmonic-profile} becomes
\begin{equation}
    \phi(z)=
    2a_2w^2+a_1w+a_0-a_2.
\end{equation}
For the coefficients~\eqref{app:twoharmonic-coefficients},
the two roots of $\phi(z)=i$ are
\begin{align}
    w_0&=\frac{\cosh y-i\sinh y}{\sqrt2},
    \nonumber\\
    w_1&=
    \frac{1+3\sinh^2y}{\sqrt2\cosh y}
    +\frac{i\sinh y}{\sqrt2}.
    \label{app:twoharmonic-wroots}
\end{align}
The first root corresponds to $z_0$ in
Eq.~\eqref{app:twoharmonic-z0}. The second root has an
upper-half-plane preimage $z_1$ with
$\operatorname{Im}z_1>y$. Indeed, the two values in
Eq.~\eqref{app:twoharmonic-wroots} have equal absolute
imaginary parts, whereas
\begin{equation}
    \operatorname{Re}w_1-\operatorname{Re}w_0
    =
    \frac{\sqrt2\sinh^2y}{\cosh y}>0.
    \label{app:twoharmonic-order}
\end{equation}
For $w=u+iv=\cos(x+iY)$, the coordinates satisfy
\begin{equation}
    \frac{u^2}{\cosh^2Y}
    +\frac{v^2}{\sinh^2Y}=1.
\end{equation}
At fixed $|v|$, the upper-half-plane height $Y$ increases
with $|u|$. Equation~\eqref{app:twoharmonic-order} therefore
establishes the claimed ordering.

Since the coefficients are real and $\phi$ is even, the
remaining roots are obtained by complex conjugation and
reflection. The construction does not require a large second harmonic.
In fact,
\begin{equation}
    \frac{|a_2|}{|a_1|}
    =
    \frac{\cosh y}{2\sqrt2\cosh 2y},
    \label{app:twoharmonic-ratio}
\end{equation}
which decreases as $1/(4\sqrt2\sinh y)\sim e^{-y}/(2\sqrt{2})$ for large $y$.

Finally, as discussed in the main body, the exact cancellation of the real part of
$\phi'(z_0)$ is not essential. Since the turning points are
simple and their heights are strictly ordered, sufficiently
small perturbations of the coefficients preserve the smallness of the right part of~\eqref{condition}.
For example, turning all hyperbolic functions into exponential we obtain the following approximate formula for the two harmonic deformation potential
\begin{equation}
    \phi(x)= 2 - 4\sqrt{2}e^{-y}\cos x+2 e^{-2y}\cos 2x,\ \ y\gg1.    
    \label{app:twoharmonic-example}
\end{equation}
For example, for $y=3$ we obtain ${\rm Im}\,\phi'(z_0)/{\rm Re}\,\phi'(z_0)\approx 17$.
This way, the condition
$\operatorname{Im}\phi'(z_0)\gg
|\operatorname{Re}\phi'(z_0)|$ can therefore be realized in
a finite neighborhood of the profiles
\eqref{app:twoharmonic-coefficients}.

\subsection{Perturbation approach}
\label{app:perturb}
After unitary transform~\eqref{unitary} the Schrodinger equation for the edge states in the absence of the Zeeman term takes the form
\begin{gather}
\label{shr}
    -\frac{i}{2}\left(v(x)\frac{\partial}{\partial x} + \frac{\partial}{\partial x}v(x)\right) \psi^{(0)}_{1,2}(x) = \pm \psi^{(0)}_{1,2}(x) \ve.
\end{gather}

With substitution $\psi^{(0)}_{1,2}(x) \rightarrow \psi^{(0)}_{1,2}(x)/\sqrt{v(x)}$ and after taking the derivatives, Eq.~\eqref{shr} becomes
\begin{gather}
\label{eqap}
    -i v(x) \psi^{(0)'}_{1,2}(x) = \pm \ve \psi^{(0)}_{1,2}(x).
\end{gather}
The exact solution of Eq.~\eqref{eqap} is
\begin{gather}
\label{sol0}
    \psi^{(0)}_{1,2}(x) = \frac{e^{\pm i \ve \tau(x)}}{\sqrt{v(x)}}, \; \tau(x) = \int_0^x \frac{ds}{v(s)}. 
\end{gather} 
Let us define
\begin{gather}
    T = \tau(L) = \int_0^L \frac{ds}{v(s)}
\end{gather}
as the time required for a particle to traverse a distance equal to the period of the potential.

Now, we assume that $\mu \ll \ve$ and add external magnetic field as a small perturbation $\hat{V} = -\mu \sigma_z$. 
The term with $\phi(x) \sigma_z$ renormalizes the momentum $p_x$ only and doesn't bring any essential physical effect. We will focus on the influence exerted by the second term. As we omitted tildes for convenience, Hamiltonian of the system in external magnetic field after the unitary transformation reads as 
\begin{gather}
\label{ham_v}
\begin{split}
    \hat{H}(x) &=  \hat{H}_0(x) + \hat{V}(x) \\
    &= \frac{1}{2}(v(x) \hat{p}_x + \hat{p}_x v(x)) \sigma_z + \frac{\mu}{v(x)} (\phi(x)\sigma_z - \sigma_y).
\end{split}
\end{gather} 
Schrodinger equation is 
\begin{gather}
    (\hat{H}_0(x) + \hat{V}(x))\psi_{1,2}(x) = \pm \ve \psi_{1,2}(x)
\end{gather}

Solution for $\psi_{1,2}(x)$ is expressed according to perturbation theory:
\begin{gather}
    \psi_{1,2}(x) = \psi^{(0)}_{1,2}(x) + \psi^{(1)}_{1,2}(x) + \psi^{(2)}_{1,2}(x) + O(\mu)^3
\end{gather}

Zero order solution is known (look at Eq.~\eqref{sol0}). First order solution could be found using Green function~$G(x, x') = (\ve - \hat{H}_0)^{-1}$ of the fieldless system:
\begin{gather}
\begin{split}
\label{deriv}
    &(\hat{H}_0 + \hat{V})\psi_{1,2} = \pm \ve \psi_{1,2} \\
    &\hat{V}\psi_{1,2} = (\pm \ve - \hat{H}_0)\psi_{1,2} \\
    &\psi_{1,2}(x) = \psi^{(0)}_{1,2}(x) + \int_0^L \hat{G}(x, x')\hat{V}(x')\psi_{1,2}(x') \; dx'.
\end{split}
\end{gather}
In the last expression terms of the highest order should be omitted and now we have the equation for the first order solution:
\begin{gather}
\label{firstap}
    \psi^{(1)}_{1,2}(x) = \int_0^L \hat{G}(x, x')\hat{V}(x')\psi^{(0)}_{1,2}(x') \; dx'.
\end{gather}

The Green function is expressed as
\begin{gather}
\label{green}
\begin{split}
    &\hat{G}(x, x', \ve) = \sum_{\alpha} \frac{\psi_{\alpha}(x)\psi^{\dagger}_{\alpha}(x')}{\ve - \ve' + i0} \\ 
    &= \frac{-ie^{i\ve|\tau(x)-\tau(x')|}}{2\sqrt{v(x)v(x')}}(1+{\rm sgn}[\tau(x)-\tau(x')]\sigma_z)
\end{split}
\end{gather}
The expression~\eqref{green} for the Green's function together with exact equation~\eqref{deriv2} allows to construct the full perturbation series.  
We need to keep in mind that eigenvalue equation for the desired $2\times2$ monodromy matrix is of at least second order in perturbation parameter $\mu$. Therefore, from the very start we aim at second order in $\mu$ perturbation expansion. 
Integration of Eq.~\eqref{firstap} gives the first order corrections:
\begin{gather}
\begin{split}
\label{correct1}
    \psi^{(1)}_1 &= -\mu \psi^{(0)}_2(x) [g(L) - g(x)], \\ 
    \psi^{(1)}_2 &= \mu \psi^{(0)}_1(x) g^*(x), \\
    g(x) &= \int^x_0 \frac{e^{2i\ve\tau(x')}}{v^2(x')} \; dx'.
\end{split}
\end{gather}
Here, we note that in a general situation (when parameter $\ve L$ is arbitrary and not necessarily large) $g(L)\sim L$. Therefore, the perturbative expansion parameter, as can be read from Eq.~\ref{correct1} is 
indeed given by inequality~\eqref{perturbCrit} presented in the main body of the paper.
The second order terms are then expressed from plugging in the first order corrections~\eqref{correct1} into exact eqution \eqref{deriv2}:
\begin{gather}
\begin{split}
\label{deriv2}
    \psi_{1,2}(x) &= \psi^{(0)}_{1,2}(x) + \int_0^L \hat{G}(x, x')\hat{V}(x')\psi_{1,2}(x') \; dx', \\
    \psi^{(2)}_{1,2}(x) &= \int_0^L \hat{G}(x, x')\hat{V}(x')\psi^{(1)}_{1,2}(x') \; dx'.
\end{split}
\end{gather}
Integration of Eq.~\eqref{deriv2} gives the necessary second order corrections:
\begin{gather}
\begin{split}
\label{corect2}
    \psi^{(2)}_1(x) &= - \mu^2 \psi^{(0)}_1(x) \int_0^x \frac{e^{-2i\ve\tau(x')}}{v^2(x')}[g(L)-g(x')] \; dx', \\
    \psi^{(2)}_2(x) &= - \mu^2 \psi^{(0)}_2(x) \int_x^L \frac{e^{2i\ve\tau(x')}}{v^2(x')}g^*(x') \; dx'
\end{split}
\end{gather}
Summing up corrections \eqref{sol0}, \eqref{correct1}, and \eqref{corect2} yields the solution with the required accuracy:
\begin{gather}
\begin{split}
   \psi_1(x) &= \psi^{(0)}_1(x)\left[ 1 - \mu^2 g_Lg^*(x) + \mu^2 J(x)\right] \\
   &- \mu \psi_2^{(0)}(x)[g_L-g(x)] + O(\mu^3), \\
   \psi_2(x) &= \psi^{(0)}_2(x) \left[ 1 - \mu^2 [J^*_L - J^*(x)] \right]  \\
   &+ \mu \psi^{(0)}_1(x) g^*(x) + O(\mu^3),
\end{split}
\end{gather}
where 
\begin{gather}
\label{gandJ}
\begin{split}
    J_L &\equiv J(L) = \int_0^L \frac{e^{-2i\ve\tau(x')}}{v^2(x')}g(x') \, dx' = \int_0^L g(x)dg^*(x), 
\end{split}
\end{gather}
and $g_L\equiv g(L)$. From the last identity we obtain the useful relation:
\begin{gather}
\label{app:aux01}
    J_L+J_L^* = |g_L|^2.
\end{gather}
The monodromy matrix $\hat{M}$ relates solutions $\psi_{1,2}(x)$ and $\psi_{1,2}(x+L)$ in the following way:
\begin{gather}
    \begin{pmatrix}
     \psi_1(x + L) \\
     \psi_2(x + L)
    \end{pmatrix} = \hat{M} 
    \begin{pmatrix}
     \psi_1(x) \\
     \psi_2(x)
    \end{pmatrix}
\end{gather}

Combining monodromy matrix is expressed as
\begin{gather}
\label{monodromy}
    \hat{M} =\\
    \begin{bmatrix}  e^{i \ve T} \!\left[ 1\!-\!\mu^2 |g_L|^2 \!+\! \mu^2 J_L\right] & \mu e^{i \ve T} g_L \\
\mu e^{i \ve T} g^*_L & \!\!\!\!e^{- i \ve T} \! \left[ 1\!+\!\mu^2 J^*_L \!+\! \mu^2 e^{2 i \ve T} |g_L|^2 \right] \end{bmatrix}.\notag
\end{gather}
Matrix~\eqref{monodromy} should be unitary with $\mathcal{O}(\mu^2)$ accuracy. Using the auxiliary identity in ~\eqref{app:aux01} one easily sees that the relation $\det M = 1$ is, in fact, exact fo~\eqref{monodromy} (therefore eigenvalues $\lambda_1\lambda_2 = 1$). The normalization condition for the wave functions then (they wave function should not grow exponentially) then requires that $\lambda_1 = \lambda_2^* = \exp(KL)$ where $K$ is quasi momentum. Therefore, we have for the band the standard condition $2\cos KL = {\rm tr}\,M$, where $K$ is a quasi momentum. 
As a result, the band gap equation reads
\begin{gather}
    |{\rm tr}\,M|\geq 2.
\end{gather}
Computing the trace we obtain:
\begin{gather}
\label{band2}
    \left|\cos\ve T+\frac{\mu^2}{2}\left(J_L e^{i\ve T}+c.c.\right)\right|\geq 1
\end{gather}
Since $\mu$ is small, the only way to satisfy the equation~\eqref{band2} is to put $\ve T$ close to $\pi n$. Therefore, setting $\ve T  = \pi n +\delta,\ \ \delta\ll1$ we obtain:
\begin{gather}
    \cos\delta +\frac{\mu^2}{2}\left(J_L e^{i\delta}+J^*_L e^{-i\delta}\right)
    \geq 1
\end{gather}
Discarding $e^{\pm i\delta}$  in the perturbative term (since, as we are going to see $\delta$ itself is proportional to perturbation $\mu$) and using~\eqref{app:aux01} once again we obtain:
\begin{gather}
    \mu^2 |g_L|^2\geq\delta^2 \Rightarrow |\delta|\leq \mu|g_L|
\end{gather}
which is nothing but result~\eqref{band_perturb} quoted in the man body of the paper.

\subsection{Absence of Stokes switching in the local Weber problem} \label{app:weber_stokes} 

\begin{figure}[t!]
    \centering
    \includegraphics[width=0.75\linewidth]{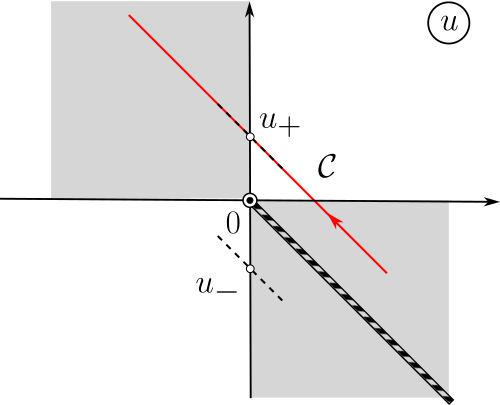}
    \caption{\justifying 
    The contour yielding the transmitted wave \(\exp(i\lambda/\sqrt{\mathcal E})\). The gray sectors indicate the directions in which the boundary term \eqref{eq:weber_boundary} vanishes as \(|u|\to\infty\). The points \(u_\pm\) are the saddle points ~\eqref{eq:fixed_saddles}. The dashed lines indicate the local steepest-descent directions near the saddles.}
    \label{fig:contours}
\end{figure}

In this Appendix we clarify a somewhat unusual property of the local Weber problem relevant to the asymptotic region considered in the main text. Namely, analytic continuation of the corresponding solution from positive to negative real \(s\) through the lower half-plane, \(\arg s:0\to-\pi\), does not switch on an additional reflected-wave contribution. This can be seen particularly transparently from a contour-integral representation of the Weber solution and the associated saddle-point structure. After an inessential rescaling of the independent variable, the local equation can be written in the dimensionless form 

\begin{equation} 
\psi''(s)+\left(s^2+\mathcal E\right)\psi(s)=0, \qquad \mathcal E\gg 1. \label{eq:weber_local} 
\end{equation} 
where $\mE = \ve^2/(\gamma\mu)$ and $s = \sqrt{\gamma\mu}t$.
We are interested in the asymptotic solutions of Eq.~\eqref{eq:weber_local} in the overlap region

\begin{gather}
\label{app:limit}
 1/\mE\ll s\ll \sqrt{\frac{\mu}{\gamma}}\sim 1.
 \end{gather}
A convenient way to obtain the required asymptotics is to construct a contour-integral representation of the solution. Using the standard substitution  $\psi(s)=e^{is^2/2}h(s)$, Eq.~\eqref{eq:weber_local} reduces to an equation with coefficients linear in \(s\),  
\begin{equation} 
    h''(s)+2is\,h'(s)+(\mathcal E+i)h(s)=0. 
    \label{eq:h_weber} 
\end{equation}  
Using Laplace method, we write the contour-integral solution of Eq.~\eqref{eq:h_weber}  
\begin{equation} 
    h(s)= \int_{\mathcal C}d\xi\, \exp\left(s\xi-\frac{i}{4}\xi^2\right) \xi^{-1/2-i\mathcal E/2}. 
    \label{eq:weber_contour} 
\end{equation}  
Here, the contour $\mathcal C$ is chosen such that the boundary term 

\begin{equation} 
V(\xi) = \exp\left(s\xi-\frac{i}{4}\xi^2\right) \xi^{1/2-i\mathcal E/2} \longrightarrow 0 
    \label{eq:weber_boundary} 
\end{equation}
at both ends of the contour. The branch of the multivalued power in Eq.~\eqref{eq:weber_contour} will be specified below.
To analyze Eq.~\eqref{eq:weber_contour} in the regime ~\eqref{app:limit}, we introduce the intermediate large parameter
\begin{gather}
\label{lambda}
 1\ll \lambda =  s\mE\ll \mE
\end{gather}
and the rescaling $\xi = \mE u$. The integral can then be evaluated by the method of steepest descent.
Apart from an irrelevant $\lambda$-independent prefactor, Eq.~\eqref{eq:weber_contour} then takes the form  
\begin{gather} 
    \begin{split}
h(s) &= {\rm const} \int_{\mathcal C}du\, u^{-1/2} e^{\Phi(u)}, \\
\Phi(u) &= \lambda u-\frac{i\mathcal E^2}{4}u^2 -\frac{i\mathcal E}{2}\ln u,\ \ 1\ll\lambda\ll \mE
\label{eq:weber_scaled_integral} 
\end{split}
\end{gather}  
The branches of \(\ln u\) and \(u^{-1/2}\) are fixed by the branch cut shown in Fig.~\ref{fig:contours} We choose \(\arg u=0\) on the positive real axis.
The saddle points are therefore determined by the quadratic equation  
\begin{equation} 
\mathcal E^2u^2+2i\lambda u+\mathcal E=0. 
\label{eq:weber_saddle_quadratic} 
\end{equation}  
Thus the two saddles can be found explicitly: 
\begin{equation} 
u_{\pm}(\lambda) = \frac{i}{\mathcal E^2} \left[ -\lambda\pm\sqrt{\lambda^2+\mathcal E^3} \right]. 
\label{eq:weber_saddles} 
\end{equation}  
Expanding in the limit ~\eqref{lambda} we
obtain
\begin{align} 
    u_{\pm} &= \pm\frac{i}{\sqrt{\mathcal E}} -\frac{i\lambda}{\mathcal E^2} \pm \frac{i\lambda^2}{2\mathcal E^{7/2}} +... 
    \label{eq:uminus_expansion} 
\end{align} 
Hence, to the leading order, the saddles remain fixed at 
\begin{equation} 
    u_\pm\simeq \pm\frac{i}{\sqrt{\mathcal E}}. 
    \label{eq:fixed_saddles} 
\end{equation}  
Although both saddles approach the origin as \(\mathcal E\to\infty\), they remain parametrically well separated on the scale of their saddle-point regions. Indeed,
\begin{gather}
\label{app:second}
\Phi''(u_{\pm}) = -i \mE^2+...
\end{gather}
so that the characteristic Gaussian width of either saddle-point region is
$\Delta u \sim 1/\mE$. On the other hand, $ |u_+-u_-|\simeq2/\sqrt{\mathcal E} \gg1/\mathcal E.$ The steepest descent direction is inferred from~\eqref{app:second} for both saddles to be $\vf = 3\pi/4$.
Substituting the saddle-point expansions~\eqref{eq:uminus_expansion} into \(\Phi(u)\), we obtain
\begin{gather}
\Phi(u_\pm) = \Phi_\pm^{(0)} \pm\frac{i\lambda}{\sqrt{\mathcal E}} -\frac{i\lambda^2}{2\mathcal E^2} +O\!\left(\frac{\lambda^3}{\mathcal E^{7/2}}\right),
\end{gather}
where \(\Phi_\pm^{(0)}\) are \(\lambda\)-independent constants determined by the chosen branch of the logarithm.
Combining this result with $ \exp\!\left(\frac{is^2}{2}\right) = \exp\!\left(\frac{i\lambda^2}{2\mathcal E^2}\right),$ we find that, to leading order, the two saddle-point contributions are
\begin{gather}
    \psi_\pm(s)\propto \exp\!\left(\pm\frac{i\lambda}{\sqrt{\mathcal E}}\right) = \exp(\pm is\sqrt{\mathcal E}) = \exp(\pm i\varepsilon t),
\end{gather}
i.e., precisely the two plane-wave solutions in the original variable \(t\).

Consider now the solution defined by the contour \(\mathcal C\) shown in Fig.~\ref{fig:contours}. Its leading asymptotic contribution comes from the saddle \(u_+\), and hence
\begin{gather}
\label{app:transmit}
    \psi(s) \propto e^{i s\sqrt{\mE}}.
\end{gather}
This is the transmitted branch.

We now rotate \(s\) through the lower half-plane by \(-\pi\). Accordingly, \(\lambda\) undergoes the same rotation. Remarkably, Eq. \eqref{eq:weber_saddles}  immediately shows that at the end of the continuation
\begin{gather}
 u_\pm(-\lambda)-u_\pm(\lambda) = \frac{2i\lambda}{\mathcal E^2}. 
\end{gather}
Thus both saddles undergo exactly the same displacement, while their separation remains unchanged. Since their characteristic Gaussian width is \(\Delta u\sim1/\mathcal E\),
\begin{gather}
 \frac{|u_\pm(-\lambda)-u_\pm(\lambda)|}{\Delta u} \sim \frac{\lambda}{\mathcal E} \ll1. 
 \end{gather}
Moreover, the convergence sectors at \(|u|\to\infty\) are determined by the quadratic term \(-i\mathcal E^2u^2/4\) and are therefore independent of \(\lambda\). The contour \(\mathcal C\) can consequently be continuously deformed during the rotation without crossing the branch cut and while remaining attached to the same saddle \(u_+\). Its homology class is unchanged, and no contribution from \(u_-\) is switched on. Therefore, the analytic continuation of the transmitted branch remains a single transmitted branch, and no reflected wave is generated.

\subsection{The scattering integral for $|\phi|\rightarrow\infty$}
\label{finali}
\begin{figure}[h]
    \centering
    \includegraphics[width=1.0\linewidth]{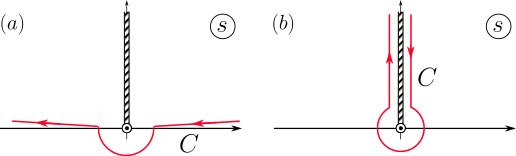}
    \caption{\justifying The transformation of the contour for the computation of the scattering integral~\eqref{C2} at $t\rightarrow-\infty$.}
    \label{app:fig:finale}
\end{figure}
Here, we present small technical details for the computation of the scattering integral~\eqref{C2} at $t\rightarrow-\infty$.  
The transition from positive to negative $t$ is performed via the rotation along the lower semicircle (see Fig.~\ref{fig:anti-Sotkes-fin})(b). To assure the convergence of the integral determining $C_2$,  both legs of the anti Stokes lines should be bent slightly upwards [see Fig.~\ref{app:fig:finale}(a)]. Bending then both legs of the contour along the vertical direction, we turn the integral into a Hankel representation for the Gamma-function. The respective arguments of $s$ are $\pi/2$ and $-3\pi/2$ [see Fig.~\ref{app:fig:finale}(b)]. 
Then we have:
\begin{gather}
\begin{split}
    &\int\limits_C s^{1/\gamma}e^{2i\ve s}\,ds\\
    &= \int\limits_{\infty}^0 e^\frac{i\pi}{2\gamma}e^{-2\ve \rho}\rho^{1/\gamma}\,d\rho+
    \int\limits_{0}^\infty e^\frac{-3\pi i}{2\gamma}e^{-2\ve \rho}\rho^{1/\gamma}\,d\rho\\
    &=-2i e^{-\frac{i\pi}{2\gamma}}\sin\frac{\pi}{\gamma}\frac{\Gamma\left(\frac{1}{\gamma}+1\right)}{(2\ve)^{\frac{1}{\gamma}+1}}.
\end{split}
\end{gather}

This yields final formula~\eqref{amplitude-final} in the main body.

\bibliography{ti.bib}

\end{document}